\documentclass[a4paper,prb,preprint,groupedaddress]{revtex4-2}

\usepackage{graphicx}
\usepackage[normalem]{ulem}
\usepackage[final]{changes}
\usepackage{siunitx}
\usepackage{upgreek}

\usepackage{xcolor}
\usepackage{ulem}

\newcommand\varpm{\mathbin{\vcenter{\hbox{%
				\oalign{\hfil$\scriptstyle+$\hfil\cr
					\noalign{\kern-.3ex}
					$\scriptscriptstyle({-})$\cr}%
}}}}
\newcommand\varmp{\mathbin{\vcenter{\hbox{%
				\oalign{$\scriptstyle({+})$\cr
					\noalign{\kern-.3ex}
					\hfil$\scriptscriptstyle-$\hfil\cr}%
}}}}

\usepackage{amsmath}
\usepackage{relsize}
\usepackage{float}

\begin{document}

\title{Unexpected Gaussian line shapes reveal electron-adsorbate interaction as dominant broadening mechanism in quantum corrals}

\author{Marco Weiss}
\author{Michael Schelchshorn}
\author{Fabian Stilp}
\author{Alfred J. Weymouth}
\author{Franz J. Giessibl}
\email{franz.giessibl@ur.de}
\affiliation{Institute of Experimental and Applied Physics, University of Regensburg, 93053 Regensburg, Germany}

\date{\today}

\begin{abstract}

Understanding the factors influencing the lifetime of electronic states in artificial quantum structures is of great significance for advancing quantum technologies. This study focuses on CO-based quantum corrals on Cu(111). Tunneling spectroscopy measurements reveal a strong correlation between the size of the quantum corral and spectral width, characterized by a predominant Gaussian line shape. We attribute this dominant Gaussian-shaped lifetime broadening to the interaction of surface state electrons with the corral boundary.
To further investigate this phenomenon, we constructed corrals of varying wall densities. 
Our findings indicate that elastic processes, such as tunneling, are more sensitive to wall density than coupling to the bulk.

\end{abstract}

\maketitle

\section{Introduction}
In a neutral atom, electronically excited states will \hyphenation{ty-pi-cal-ly} typically decay within a few nanoseconds~\cite{Verolainen1982}.
Because the probability of the electron remaining excited decreases exponentially with time, the observed spectral line shape of the decay is Lorentzian.
For certain realizations of artificial atoms (e.g. quantum dots) \cite{Kastner1993}, the observed spectral peaks are also well-described by Lorentzian functions \citep[e.g.][]{Wegscheider1998, Jahn2015, Pisoni2018}. 
The spectral line shapes and their widths are measurements used to determine not only the lifetimes of the confined states but also the lifetime limiting mechanisms within artificial atoms.
Understanding these mechanisms is vital for making quantum dots with long lifetimes. 
Long-lived confined states have been proposed, for applications in quantum sensing, quantum computing and for creating artificial quantum simulators~\citep[e.g.][]{Sierda2023, Gao2022, Pezzagna2021, Ladd2010, Cirlin2018, Ruf2019, Ates2009, Maurer2013}.

We investigated a realization of an artificial atom made by manipulation of adsorbates on a noble metal surface with scanning probe microscopy, called a quantum corral~\cite{Crommie1993}.
A quantum corral confines the quasi-free 2D electron gas present on the surface \cite{Zangwill}, resulting in a set of resonant eigenstates. 
While the original corral, consisting of Fe adatoms arranged in a circle on Cu(111), was published in 1993 \cite{Crommie1993}, quantum corrals have been constructed on more exotic surfaces including semiconductors \cite{Sierda2023}, a Rashba surface alloy \cite{Jolie2022a}, a proximity superconductor \cite{Schneider2023} and topological insulators \cite{Chen2019}. 
Interpreting the spectral peaks of these corral states requires a thorough understanding of the lifetime limitations.

There are three mechanisms which influence the lifetime of surface state electrons: electron-phonon scattering, electron-electron scattering and electron-defect scattering \citep[e.g.][]{McDougall1995, Burgi1999, Kliewer2000, Echenique2001,Fukui2001, Braun2002, Eiguren2002, Hellsing_2002, Vergniory2005, Chulkov2006, Hofmann2009}. 
The common property of electron-phonon scattering and electron-electron scattering is a temporally constant and spatially homogeneous scattering background for surface state electrons which can be expressed with an exponentially decreasing survival probability which results in a Lorentzian peak shape~\cite{Chulkov2006}. 
For studies of quantum corrals, electron-defect scattering describes the scattering at the corral walls.
While randomly-spaced defects should also result in a Lorentzian contribution to the line shape~\cite{Chulkov2006}, the defects that make up the corral walls are not randomly-spaced. Considering the corral walls as well-arranged defects, a denser wall not only limits the lateral transmission out of the corral (e.g. tunneling) but also increases scattering into bulk states.

Since the realization of the first quantum corral with Fe adatoms \cite{Crommie1993}, electron-wall interaction has been a topic of theoretical investigation. Using multiple scattering theory with an additional absorbing channel at the $\delta$-peak shaped scattering potentials, it was concluded that the widths of the corral states energy distributions (spectral peaks) are mainly dictated by the absorbing channel  (e.g. scattering to bulk states). An estimate indicated that approximately $25\%$ of an incident wave is transmitted and about $50\%$ is absorbed \cite{Heller1994}. 
Other studies also emphasize the significant role of scattering to bulk states in reproducing the energy width of corral states in simulations \cite{Crampin1994, Crampin1996}.
On the other hand, some theoretical considerations of quantum corrals are based entirely on elastic mechanisms. For instance, a corral was modeled with finite-height potential barriers representing the adsorbates, and it was concluded that reproducing the standing wave pattern in a corral does not require bulk coupling \cite{Harbury1996}. 
Furthermore, other theoretical considerations with Gaussian-shaped potentials can also reproduce the energetic behavior of a quantum corral without additional lossy channels \cite{Rahachou2004}.

In this article we probe the states of quantum corrals by means of STS (scanning tunneling spectroscopy) measurements.
The spatial behavior of states in a corral is well described by a hard-wall model. This enables us to characterize the energy distributions of corral states by fitting the states of the hard-wall potential well to the experimentally acquired local density of states. 
By comparing two corrals of different sizes, we show that the spectral widths are fundamentally influenced by the confinement, i.e. the corral wall. Moreover, for the states that can be well characterized, we find that the line shape of the energy distributions is not purely Lorentzian as it was previously reported~\cite{Crommie1993}, but rather better described by a Gaussian. 
The energy dependence of the widths of the energy distributions prompted us to propose a classical model which connects the lifetime of the corral states to the average path length of a confined surface state electron. 
To further understand the influence of the interaction of the electrons with the corral wall, we constructed corrals with various densities of adsorbates in the walls. 
Corrals with a denser wall host states with thinner energy distributions, allowing us to conclude that transmission (i.e. tunneling through the corral wall) is more sensitive to the wall density than coupling to the bulk.

\section{Constructing a quantum corral}

The first quantum corral, built with a scanning tunneling microscope (STM), was created by arranging 48 Fe adatoms in a circular shape on Cu(111) \cite{Crommie1993}. 
While this artificial structure showed discrete energy states, a detailed investigation with large voltage variation was difficult because the corral wall was not stable \cite{Crommie1993}. 
This can be attributed to the weak bonding of Fe on Cu(111) \cite{Berwanger2018}.
By using a more stable wall than the original quantum corral, we were able to investigate states far away from the Fermi level and to precisely measure their spectral widths and line shapes. 
Carbon monoxide (CO) binds 6 times stronger to Cu(111) than Fe \cite{Berwanger2018, Ternes2008}, making it a good candidate for creating stable artificial structures. 
Spectroscopic STM measurements on CO-based structures therefore show the desired stability and allow for a larger voltage window \cite{Slot2017,Gomes2012,Jolie2022,Freeney2020_1} to perform a detailed analysis of the corral's energy levels. 

\begin{figure}
	\includegraphics[width=0.7\columnwidth]{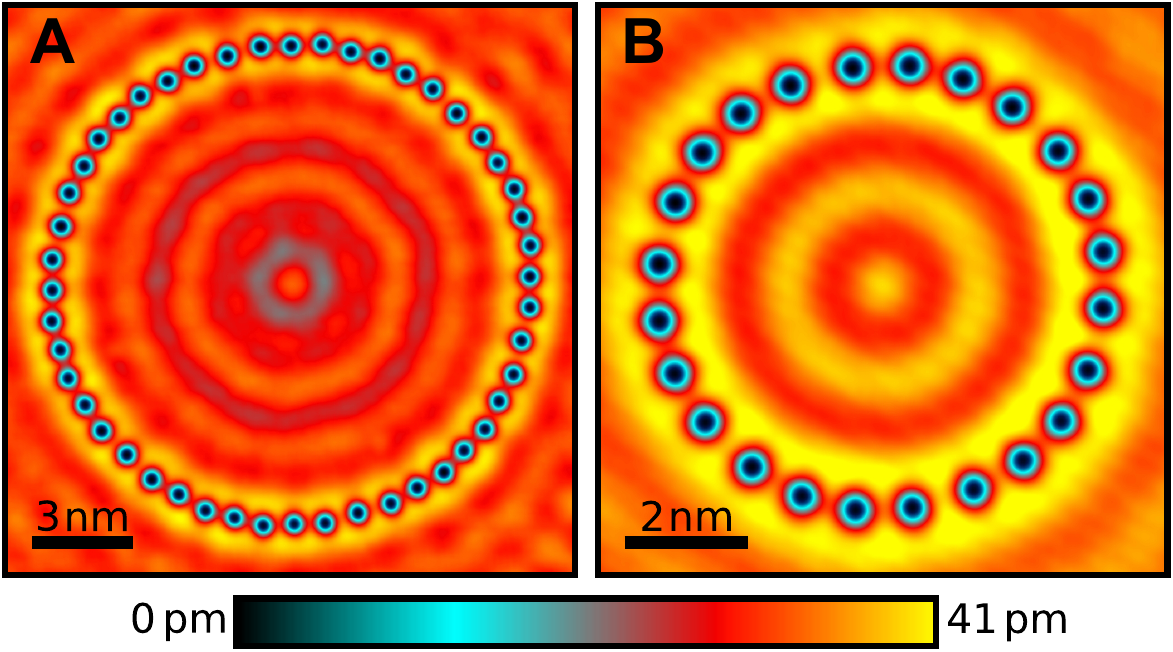}%
	\caption{\label{Fig1} \textbf{A}: Constant current STM image ($10\,$mV/$10\,$pA) of the big 48-CO quantum corral, which has a radius of $R_{48\mathrm{,large}} = 7.13\,$nm. \textbf{B}: STM image ($50\,$mV/$100\,$pA) of the small 24-CO quantum corral with a radius of $R_{24\mathrm{,small}} =3.57\,$nm. Both corrals have an average inter molecular distance of $938\,$pm.} 
\end{figure}

The spatial behavior of states in a ring-shaped quantum corral is almost identical to one of states in an infinitely high cylindrical potential well \cite{Crommie1993, Stilp2021a}, commonly known as the hard-wall model. 
Solving the Schrödinger equation for the hard-wall model yields the solutions $\Psi_{n,l}(r,\phi,z) = \psi_{n,l}(r) \cdot \psi_{l}(\phi) \cdot \psi(z)$, where $n$ is the main quantum number and $l$ the angular momentum quantum number. 
The radial distance from the center of the corral is denoted with $r$, $\phi$ is the azimuthal angle and $z$ is the direction perpendicular to the surface. 
Here $\psi_{n,l}(r)$ describes the radial dependent component of the wavefunction including Bessel functions of the first kind and $\psi_{l}(\phi) = 1/\sqrt{2\pi} \cdot \mathrm{exp}(il\phi)$ describes the angular dependence. 
The $z$-component of the Shockley surface state remains unaffected by the cylindrical potential well and above the surface, for $z>0$, $\psi(z) \propto \mathrm{exp}(-\kappa z)$ with the decay constant $\kappa$ \cite{Stilp2021a}.
The absolute square of each wavefunction, which can be measured with STM, gives the probability density $|\Psi_{n,l}(r,\phi,z_0)|^2 = |\psi_{n,l}(r)|^2 \cdot |\psi_{l}(\phi)|^2 \cdot |\psi(z_0)|^2  =|\psi_{n,l}(r)|^2 \cdot 1/2\pi \cdot C_z$. As measurements over each corral were performed at the same same tip-sample distances, $z_0$, and $\kappa$ is the same for every corral state \cite{Stilp2021a}, the z-component $|\psi(z_0)|^2$ can be expressed as a constant factor $C_z$.
For a more detailed discussion and calculations see the supplemental material of Stilp \textit{et al.} \cite{Stilp2021a}. 
In this work corral states are characterized by their main quantum number $n$ and their angular momentum quantum number $l$.

For the initial line shape analysis, measurements (see SM1 of the supplemental material \cite{SupplMat} for experimental details) were taken of two differently sized CO corrals: one with a radius of $R_{48\mathrm{,big}} = 7.13\,$nm (48 CO molecules) and another with a radius of $R_{24\mathrm{,small}} =3.57\,$nm (24 CO molecules). Both structures share an identical wall density with an average inter molecular distance of $938\,$pm. Detailed information on the construction plans of the two differently sized corrals is available in the supplemental material (SM2) \cite{SupplMat}. STM topography images of these corrals are presented in FIG. \ref{Fig1}A\&B.

\section{Line shape analysis}
In order to study the line shapes of the corral states, we performed scanning tunneling spectroscopy. However, performing stationary d$I$/d$V$ measurements (fixing $\vec{r}_\mathrm{tip}$ and sweeping the bias voltage $V_\mathrm{B}$) restricts analysis to the $l=0$ states, drastically limiting the amount of analyzed states. 
The reason for this is that one cannot find a position $\vec{r}_\mathrm{tip}$ where a particular $l\neq0$ state is energetically and spatially isolated. 
An example of this problem is shown in Figure S4 of the supplemental material \cite{SupplMat}.

\begin{figure*}
	\includegraphics[width=\columnwidth]{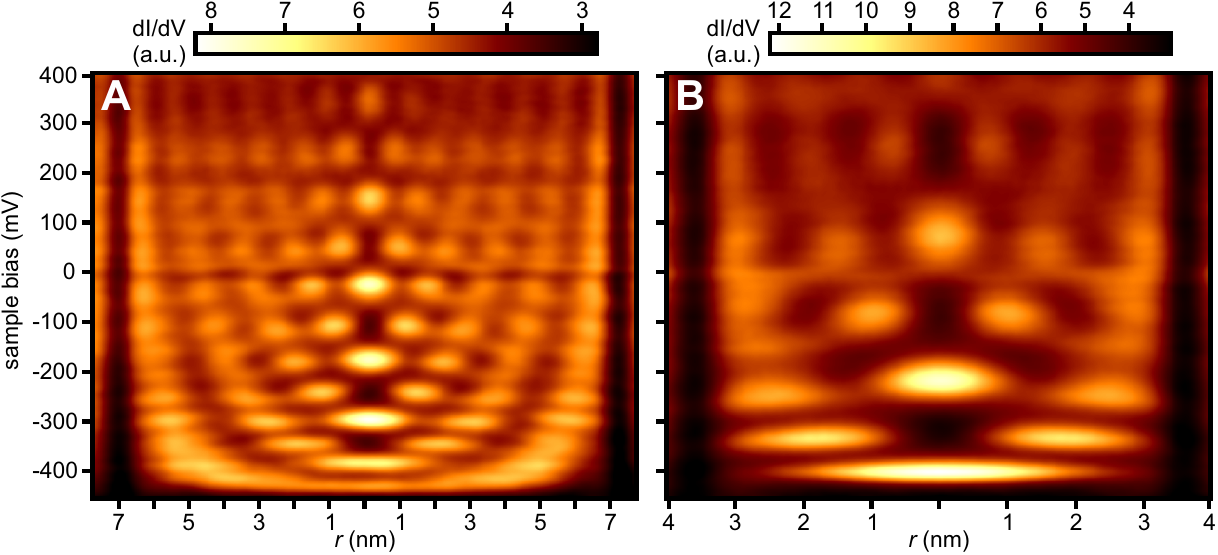}%
	\caption{\label{linescans} Differential conductance (d$I$/d$V$) measurements are shown as a function of radial distance $r$ from the center and sample bias for the (\textbf{A}) large 48-CO corral ($R_{48\mathrm{,big}} = 7.13\,$nm) and the (\textbf{B}) small 24-CO corral ($R_{24\mathrm{,small}} =3.57\,$nm). In A and B respectively the two outermost vertical, black stripes are originating from the CO molecules (corral walls). The bright appearing features can be assigned to the corral states.}
\end{figure*}

A complete description of the corral states includes their energetic and spatial characteristics. By using d$I$/d$V$ line scans across the whole diameter of the corral (constant height measurements at a fixed sample bias) one can record the spatial behavior of the local density of states (LDOS) at a single energy $eV_\mathrm{B}$, with $e$ being the elementary charge.
We performed line scans over a bias range from $-450\,$mV to $400\,$mV across both corrals with increments of $5\,$mV. 
Combining these d$I$/d$V$ line scans results in FIG. \ref{linescans}A\&B which represents a spatially (horizontal axis) and energetically (vertical axis) dependent local density of states (color coding) map of the quantum corral.

Each d$I$/d$V$ line scan (each horizontal line in FIG. \ref{linescans}A\&B) consists of a combination of several different states that, due to an energetic overlap, contribute at different magnitudes to the local density of states. 

In the initial exploration of the quantum corral, it was suggested that the standing wave pattern observed within the structure using STM can be effectively described with a linear combination of corral states~\cite{Crommie1993}.
We extend this concept to d$I$/d$V$ line scans measured at different bias voltages:
\begin{equation}\label{eq:2}
\frac{\mathrm{d}I}{\mathrm{d}V}(r, eV_\mathrm{B}) \propto \sum\limits_{n,l} \alpha_{n,l}(eV_\mathrm{B}) \cdot |\psi_{n,l}(r)|^2.
\end{equation}
In this equation the prefactor $\alpha_{n,l}(eV_\mathrm{B})$ describes how much the probability density  $|\psi_{n,l}(r)|^2$ contributes to the measured LDOS at a specific energy $eV_\mathrm{B}$. Therefore $\alpha_{n,l}(eV_\mathrm{B})$ is the energy distribution of the associated corral state.
Additional details about applying eq. (\ref{eq:2}) to determine $\alpha_{n,l}(eV_\mathrm{B})$ are given in the supplemental material (SM4) \cite{SupplMat}.

\begin{figure}
	\includegraphics[width=0.7\columnwidth]{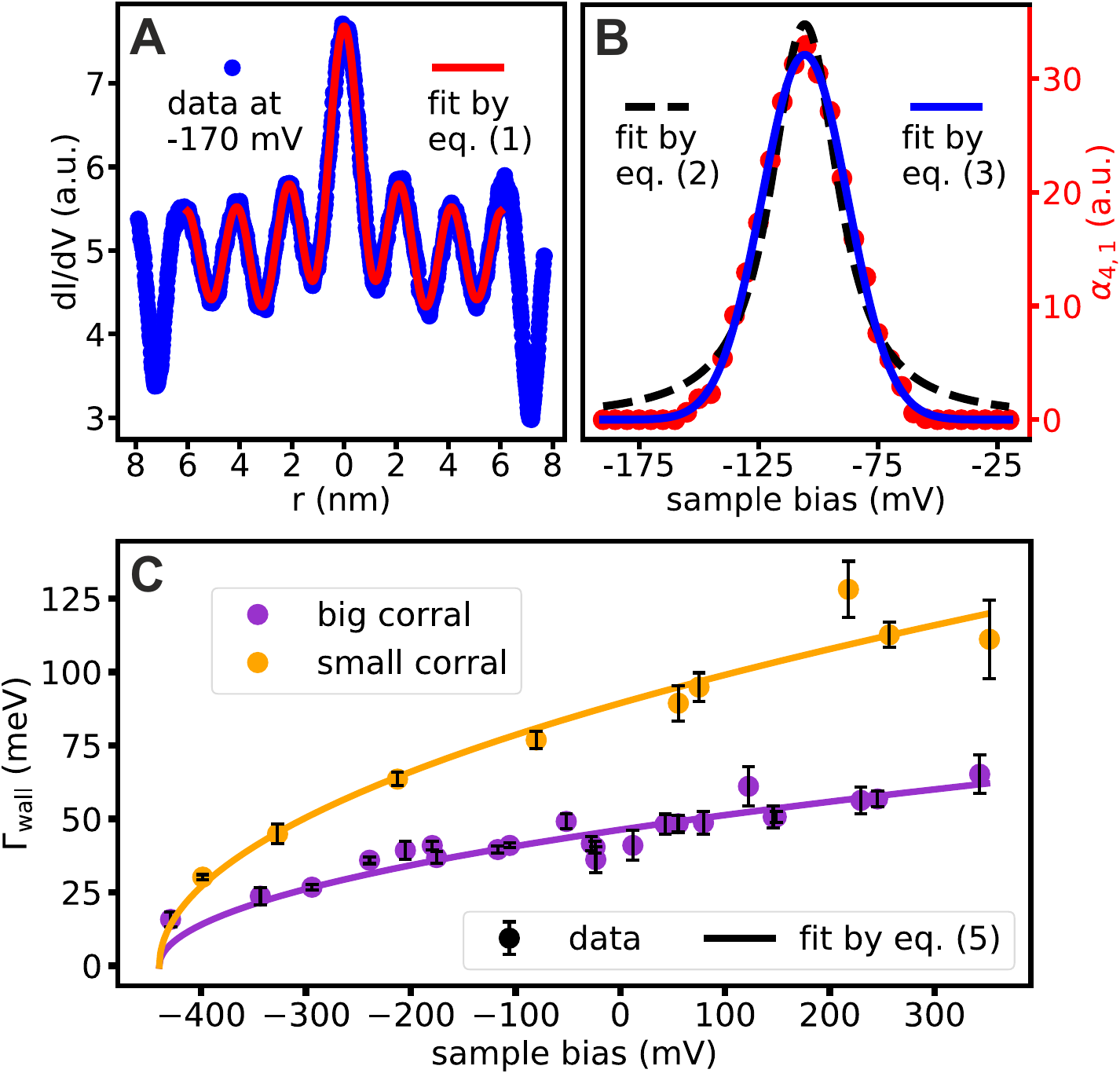}%
	\caption{\label{Fig2}
		\textbf{A:} Blue: Measured d$I$/d$V$ line scan at a sample bias of $V_\mathrm{B} = - 170\,$meV across the big corral. Red: Fit of equation (\ref{eq:2}) to the measured curve.
		\textbf{B:} Red dots: Energy distribution $\alpha_{4,1}$. The fit by equation (\ref{eq:firstfitting}) (Lorentzian \& broadening terms), depicted as the black dashed line, exhibits a relatively poor agreement. The blue full line represents the fit by equation (\ref{eq:fitting}), which includes a Lorentzian (associated width $\Gamma_\mathrm{L}$), a Gaussian (associated width $\Gamma_\mathrm{wall}$) and broadening terms, provides a better agreement.
		\textbf{C:} Purple markers: $\Gamma_\mathrm{wall}$ for the big corral ($R_{48\mathrm{,big}} = 7.13\,$nm). Orange markers:  $\Gamma_\mathrm{wall}$ for the small corral ($R_{24\mathrm{,small}} =3.57\,$nm). Full lines: Fit of the average path length model, described by equation (\ref{eq:sqrt_width}), to the data points with the results: $x_{48\mathrm{,big}} = (9.1 \pm 0.4)\,$nm and $x_{24\mathrm{,small}} = (4.7\pm0.3)\,$nm.
	}
\end{figure}

As an example a line scan obtained with $- 170\,$mV across the big corral and the corresponding fit from eq. (\ref{eq:2}) is depicted in FIG. \ref{Fig2}A. 
The local density of states maps obtained from the right-hand side of equation (\ref{eq:2}) for the full bias range are presented in the supplemental material (SM5) \cite{SupplMat}, confirming that the corral states are well described by the linear combination of states with the weighting function $\alpha_{n,l}(eV_\mathrm{B})$.

Fitting eq. (\ref{eq:2}) to the data allowed us to determine the energy distributions of several corral states, including $l\neq0$ states. 
As a representative example, the energy distribution $\alpha_{4,1}$ ($n = 4$ and $l=1$) of the big corral is shown in FIG. \ref{Fig2}B. 
All energy distributions $\alpha_{n,l}$ are given in the supplemental material (SM9) \cite{SupplMat}.

During d$I$/d$V$ measurements there will always be three broadening mechanisms which distort the shape of spectral peaks. First there is the temperature dependent widening of the Fermi-Dirac distribution \cite{TERNES}, which we will call Fermi-broadening. A second source of spectral broadening is radio-frequency-broadening (RF-broadening) \cite{Peronio2019, Assig2013, Bladh2003, LeSueur2006}. Both of these mechanisms cause a Gaussian shaped broadening of spectral features. Mathematically the combined broadening can be described as a convolution of two Gaussian terms: $G_\mathrm{Fermi, RF} = G_\mathrm{Fermi}*G_\mathrm{RF}$. For the RF-broadening we estimate a FWHM (full width at half maximum) of $\Gamma_\mathrm{RF} \approx 9\,$meV \cite{Peronio2019, footnote2} and for the Fermi-broadening at $\approx 5.7\,$K we calculate $\Gamma_\mathrm{Fermi} \approx 2\,$meV \cite{TERNES}. In total these two mechanism cause a Gaussian shaped broadening of spectroscopic peaks of $\Gamma_\mathrm{Fermi, RF}  \approx 9.2\,$meV. The third mechanism of spectral broadening is modulation broadening which causes a broadening with the shape of a semicircle $\chi$ with a diameter of $2eV_\mathrm{mod}$ \cite{TERNES,Klein1973}.

In order to characterize the energetic behavior of the corral states we analyzed the energy distributions $\alpha_{n,l}$ in more detail. 
We first attempted to fit the spectral line shapes of the energy distributions by accounting for a natural Lorentzian line shape $L$ with associated spectral width $\Gamma_L$ as well as the broadening mechanisms $G_\mathrm{Fermi, RF}$ and $\chi$ discussed above:
\begin{equation}\label{eq:firstfitting}
\alpha_{n,l} \propto L * G_\mathrm{Fermi, RF} * \chi.
\end{equation}
However, the fit was rather poor. This can be seen by the black dashed line for the energy distribution $\alpha_{4,1}$ in FIG. \ref{Fig2}B. We therefore incorporated a Gaussian line shape $G_\mathrm{wall}$ with associated width $\Gamma_\mathrm{wall}$ into the fitting function:
\begin{equation}\label{eq:fitting}
\alpha_{n,l} \propto L* G_\mathrm{wall} * G_\mathrm{Fermi, RF} * \chi.
\end{equation}
In this equation the widths $\Gamma_\mathrm{L}$ and $\Gamma_\mathrm{wall}$ are parameters which account for the Lorentzian and Gaussian widths.
The introduction of  $G_\mathrm{wall}$ notably enhanced the fit quality, as evidenced by the improved agreement shown by the full blue line in FIG. \ref{Fig2}B. Interestingly, even with this refinement, accurately characterizing all $\alpha_{n,l}$ curves remained challenging. This challenge has been shown to be particularly prominent for energy distributions of states with the same quantum number $n$ and a large angular momentum quantum number $l$. Such states show a highly similar spatial behavior. As our approach involves extracting energy distributions from line scans, the fitting algorithm, employing eq. (\ref{eq:2}), encounters complications in distinguishing between these closely related states. To characterize the accuracy of the energy distributions, a quality factor was introduced, as described in the supplemental material (SM6)  \cite{SupplMat}. We constrain the remainder of the discussion to those energy distributions with a sufficient quality factor.

A representative example of a good fit can be found in the $\alpha_{4,1}$ curve, depicted in FIG. \ref{Fig2}B.
The following FWHM for the Lorentzian and Gaussian wall components were obtained: $\Gamma_\mathrm{L} = ( 0 \varpm 2)\,$meV \& $\Gamma_\mathrm{wall} = (41 \pm 1)\,$meV. Throughout this article, the error bars represent two times the standard deviation obtained from fitting the data. These fitting parameters show, surprisingly, that the energy distribution $\alpha_{4,1}$ is purely described by a Gaussian peak shape. 
All energy distributions and the complementary fits with equation (\ref{eq:fitting}) are shown in SM9 \cite{SupplMat}. 

We find the FWHM of the Gaussian component to be significantly larger than the Lorentzian component for all energy distributions.
Moreover, the Lorentzian component for most fits is negligible (equal to zero within uncertainty).
The magnitudes (widths) of the Lorentzian and Gaussian wall components of the big and small  corral are shown in the supplemental material (SM7) \cite{SupplMat}.

To further validate the results of our line scan fitting method, we compared the widths of the energy distributions and peak widths obtained from a static d$I$/d$V$ measurement at the center of the big corral. 
This static measurement is particularly sensitive to the $l=0$ states. 
This analysis is shown in the supplemental material  (SM8) \cite{SupplMat} and yields the same values as the line scan fitting method.

By analyzing the properties of the energy distributions $\alpha_{n,l}(eV_\mathrm{B})$, we make several important conclusions.
First, the energy distributions of corral states are poorly fit with a Lorentzian.
Second, all distributions are fit well by a Gaussian curve (see SM7 \& SM9 \cite{SupplMat}).
The width of the Gaussian curve, $\Gamma_\mathrm{wall}$, shows a monotonic energy dependence (see markers in FIG. \ref{Fig2}C).
Finally, the overall magnitude of $\Gamma_\mathrm{wall}$ depends on the size of the corral.

\section{Characterizing the spectral widths of a corral with a single paramater}\label{section:IV}

As discussed in the introduction, there are several mechanisms that limit the lifetime of surface electrons: electron-phonon scattering, electron-electron scattering and electron-defect scattering.
As presented in the previous section, the spectral widths show a monotonic energy dependence and a dependence on the corral size (see FIG. \ref{Fig2}C), in agreement with previous theoretical calculations by Crampin \textit{et al.} \cite{Crampin2005}.
Similar to their results, we argue that the predominant lifetime limitation comes from the interaction of the surface electrons with the corral wall. 
This includes transmission (e.g. tunneling) through the potential barrier given by the CO-molecules and coupling to bulk states.

Due to the spatial homogeneity of electron-phonon scattering this decay mechanism is not sensitive to the size of quantum structures \cite{Kliewer2000, Braun2002}.  Additionally this mechanism results in a Lorentzian line shape. 
Therefore it would not correlate with the observation that the spectral width is strongly correlated with the corral size and is better described by a Gaussian.

Calculations by Stilp \textit{et al.} \cite{Stilp2021a} showed that the total electron density for the small and big corral are the same ($\approx 0.64\;$e$^-$/nm$^2$). 
Since electron-electron scattering relates with the total electron density \cite{Quinn1958, Echenique2004, Hofmann2009} one can conclude that electron-electron scattering is also independent on the size of the quantum structure and it would also result in a Lorentzian line shape.

The negligible impact of electron-electron and electron-phonon scattering on the lifetime (inverse spectral width) of confined electron states in quantum structures was also observed in vacancy islands on Ag(111) which had a similar size to our corrals \cite{Jensen}.

Since both electron-electron and electron-phonon scattering cannot explain the observed Gaussian line shape and the behavior of the size-dependent widths $\Gamma_\mathrm{wall}$, we conclude that in our analyzed corrals the dominant lifetime limitation is mainly dictated by the interaction of the surface electrons with the corral walls. 
This involves lateral transmission through the corral wall (e.g. tunneling) and scattering into bulk states (absorption).
A detailed discussion about this will be presented in chapter \ref{chapter:VI}.

Crampin \textit{et al.}  \cite{Crampin2005} used a similar interaction mechanism to the one originally proposed by Heller \textit{et al.} \cite{Heller1994} to theoretically determine the lifetime of corral states including lossy scattering (that is, scattering including a loss term). Furthermore, the model by Crampin \textit{et al.} adeptly captures the trends we introduced earlier, demonstrating that smaller corrals host broader spectral peaks (see our data in Fig. \ref{Fig2}C). However, their theoretically derived energy dependence of the spectral peak widths shows a steeper slope than we observe in our experiments. Consequently, energetically lower lying peaks are underestimated in width, while those at higher energies are overestimated. This discrepancy most likely stems from the fact that the discussed theoretical calculations are not founded on CO-based quantum structures but rather on the ``black-dot'' limit ($\delta$-peak like scattering potentials with a strongly absorbing channel), as introduced by Heller \textit{et al.} for Fe-based corrals \cite{Heller1994}.

Here, we present a classical model to understand and characterize the dependence of the widths $\Gamma_\mathrm{wall}$ on energy. At its core, this classical description establishes a connection between $\Gamma_\mathrm{wall}$ and the average path length of a surface state electron. The model also yields an excellent fit to our data.

The average path length $x$ that an electron with a velocity of $v_\mathrm{e}$ can travel during its lifetime $T_\mathrm{life}$ is $x = v_\mathrm{e}\cdot T_\mathrm{life}$. 
Assuming that the velocity $v_\mathrm{e}$ can be deduced classically from the energy $E = e V_\mathrm{B}+440\;$meV at which the electron is situated above the onset of the surface state band \cite{Kevan1983, Crommie_Nature}, the following equation is derived:

\begin{equation}\label{eq:distance}
	x = \sqrt{\frac{2(e V_\mathrm{B}+440\;\mathrm{meV})}{m_\mathrm{e}^*}} \cdot T_\mathrm{life}.
\end{equation}

Here $m_\mathrm{e}^*$ is the effective mass of surface state electrons on a Cu(111) surface which is $0.38$ times the mass of a free electron \cite{Kevan1983, Crommie_Nature}. 

By relating the spectral width to the lifetime via Planck's reduced constant, $T_\mathrm{life} = \hbar / \Gamma$, the spectral width can be written:

\begin{equation}\label{eq:sqrt_width}
	\Gamma = \frac{\hbar \sqrt{\frac{2(e V_\mathrm{B}+440\;\mathrm{meV})}{m_e^*}}}{x}.
\end{equation}
The average path length $x$ is then a parameter that is energy independent and can be determined for each corral by fitting equation (\ref{eq:sqrt_width}) to the data points of $\Gamma_\mathrm{wall}$ from FIG. \ref{Fig2}C. 
The resulting fits yield values of $x_{48\mathrm{,big}} = (9.1 \pm 0.4)\,$nm and $x_{24\mathrm{,small}} = (4.7\pm0.3)\,$nm and are shown in FIG. \ref{Fig2}C.

The advantage of this model is threefold and yields excellent agreement with the data.
First, it is a good description of the energy dependence of $\Gamma_\mathrm{wall}$. This again indicates that the dominant lifetime limitation of the corral states is the interaction of confined electrons with the corral wall.
Second, we can compare the values of $x_{48\mathrm{,big}}$ and $x_{24\mathrm{,small}}$ and notice that they differ by a factor of $1.9\pm 0.3$ which is the same factor of the difference in radius between the corrals ($R_{48\mathrm{,large}} / R_{24\mathrm{,small}} = 2$).
Third, it allows us to characterize a description of the lifetimes of the corral states (which are a function of energy) with a single parameter, the average path length $x$.

\section{Variation of the wall density}\label{chapter:V}

So far, we have presented data that shows that the lifetime of the corral states is directly related to the interaction of the surface state electrons with the corral walls.
It is clear that lossy scattering is essential in describing the spectral widths, and therefore the lifetimes, of corral states. Recently scattering at the corral walls has been shown to couple corral states to bulk states of a superconducting substrate \cite{Schneider2023}.
To better characterize the interaction with the walls, we determined the spectral widths of corrals states for a range of wall densities.
The wall density $\rho_\mathrm{wall}$ is defined as the number of CO molecules in the wall divided by the circumference of the corral.

\begin{figure}
	\includegraphics[width=0.7\columnwidth]{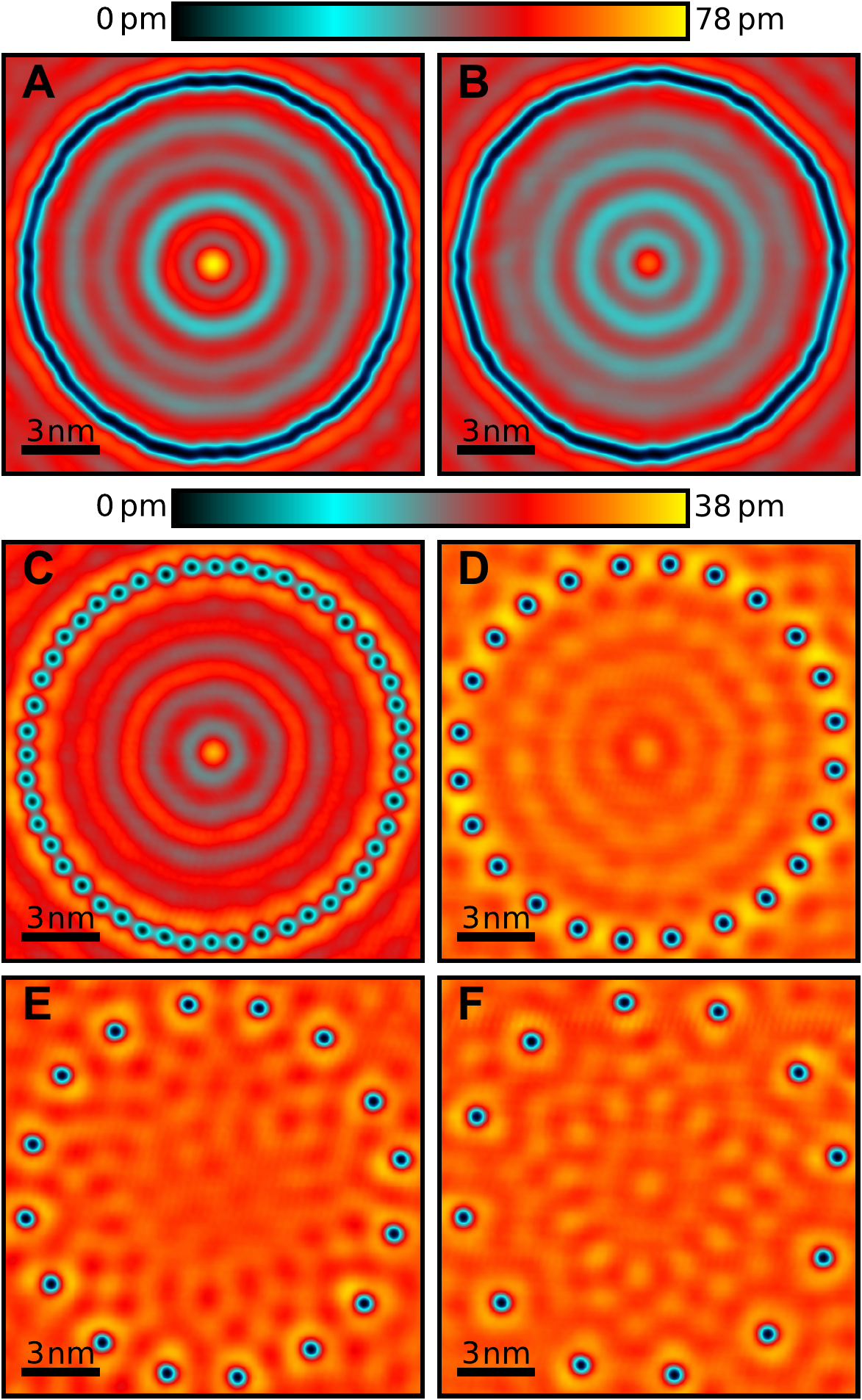}%
	\caption{\label{Fig_versch_dichten} Constant current STM images ($-10\,$mV/$100\,$pA) of the two 96-CO (\textbf{A} \& \textbf{B}), 48-CO (\textbf{C}),  24-CO (\textbf{D}), 16-CO (\textbf{E}) and 12-CO (\textbf{F}) quantum corrals. All depicted corrals, characterized by their wall densities, are constructed based on the CO adsorption sites of the 48-CO corral (C). Consequently, they share a common radius of $7.13\,$nm.}
\end{figure}

The starting point for this investigation was the larger 48-CO corral, shown in FIG. \ref{Fig1}A and FIG. \ref{Fig_versch_dichten}C.
We built corrals of lower and higher wall densities, keeping the radius the same ($R_\mathrm{wall\;density} = 7.13\;$nm), but incorporating 96, 24, 16 and 12 CO molecules in the walls.
These corrals are shown in FIG.~\ref{Fig_versch_dichten}.
Notably, in the case of the 96-CO corral, there are two possible configurations ($96.1 \; \& \; 96.2$), with one atomic adsorption site different for twenty four CO molecules. 
The construction details, including the exact adsorption sites for all corrals, can be found in the supplemental material SM2 \cite{SupplMat}. 
The average inter molecular distances (inverse wall density) of the 12-, 16-, 24-, 48-, 96.1- and 96.2-CO-corrals are as follows: $3719\,$pm, $2792\,$pm, $1876\,$pm, $938\,$pm and $477\,$pm (for both 96-CO-corrals). 
The 96-CO corrals have the maximum density, given that we require stable corral walls. In a previous study by Heinrich \textit{et al.} \cite{Heinrich2002}, the stability of densely packed CO structures on a Cu(111) surface was investigated. The findings revealed that higher density configurations, those necessary for corrals with a radius of $7.13\;$nm and more than 96 CO molecules, undergo significant changes in their initial geometry within 1 minute to 1 second. Therefore, the construction of a denser corral is not feasible.

\begin{figure}
	\includegraphics[width=0.7\columnwidth]{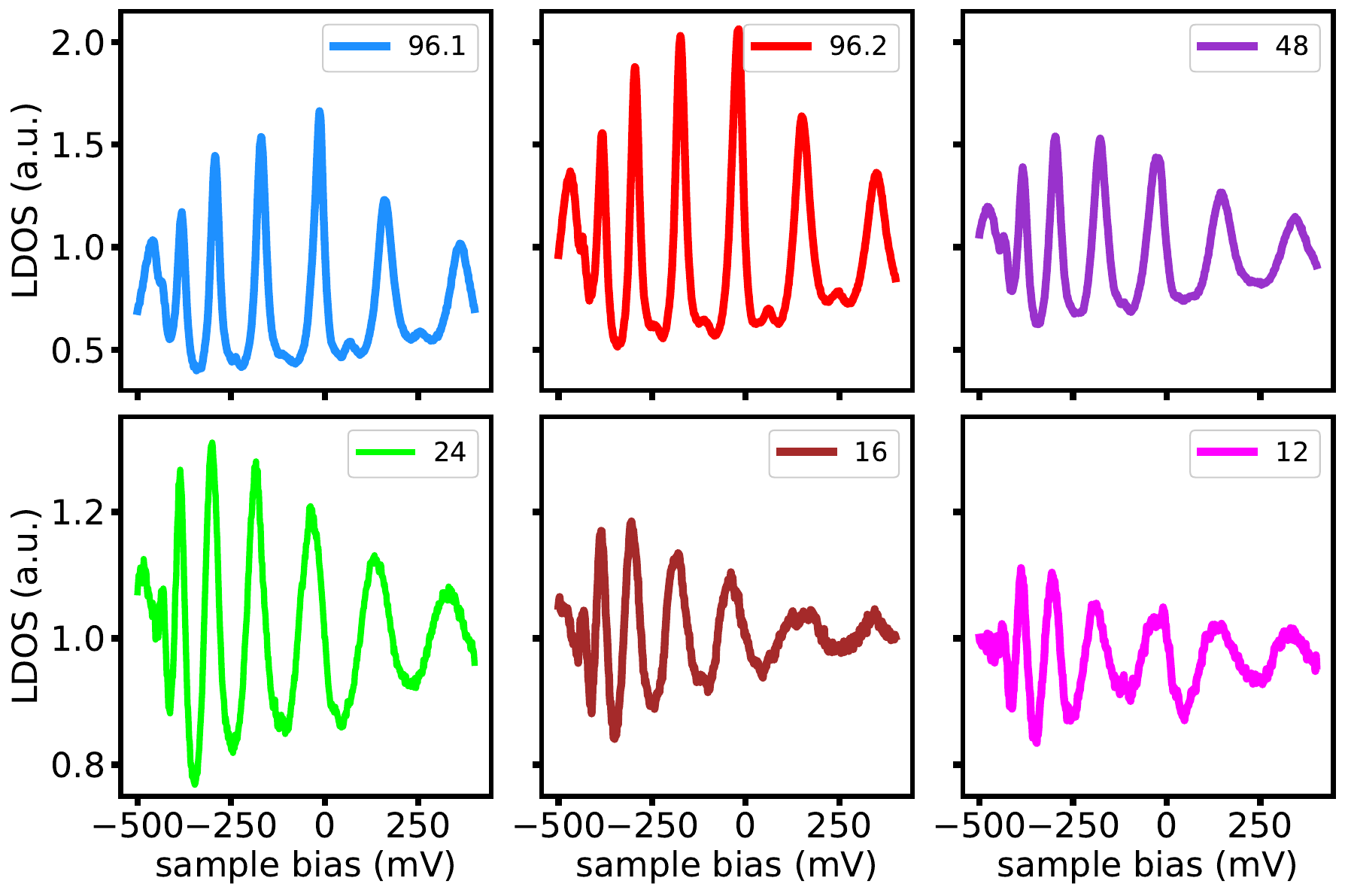}%
	\caption{\label{Fig_didv_versch_dichten} Background removed differential conductance measurements \cite{Wahl2008} over the center of CO corrals with various wall densities and identical radius of $7.13\,$nm. Each spectrum is normalized at $0\;$mV sample bias. Each corral had the same radius but a varying amount of CO molecules in the wall. The number in the legend resembles the number of CO molecules in the corral wall. A large number of COs means a high wall density and vice versa.}
\end{figure}

To characterize the effect of wall density on the lifetime (inverse spectral width) of corral states, we conducted static d$I$/d$V$ measurements over the center of each corral. 
The background removed spectra~\cite{Wahl2008} are shown in FIG \ref{Fig_didv_versch_dichten}. 
Every spectrum in this plot is normalized at $0\;$mV sample bias.

The most apparent change when comparing the spectra are the widths of the spectral peaks, which decrease with increasing wall density. 

In both 96-CO spectra, small additional peaks emerge between the main $l=0$ peaks. 
These smaller peaks can be attributed to $l\neq0$ states. 
We believe this additional influence comes from the non-infinitesimal shape of the tip apex.
For these measurements, we constructed a tip apex that ends in three metal atoms (verified with the COFI method \cite{Welker2012}). The increased sensitivity for $l\neq0$ states in d$I$/d$V$ measurements over the center of a 96-CO corral with a non-infinitesimal tip apex is further amplified by the generally narrower widths of spectral peaks hosted in these denser corrals.

\begin{figure}
	\includegraphics[width=0.7\columnwidth]{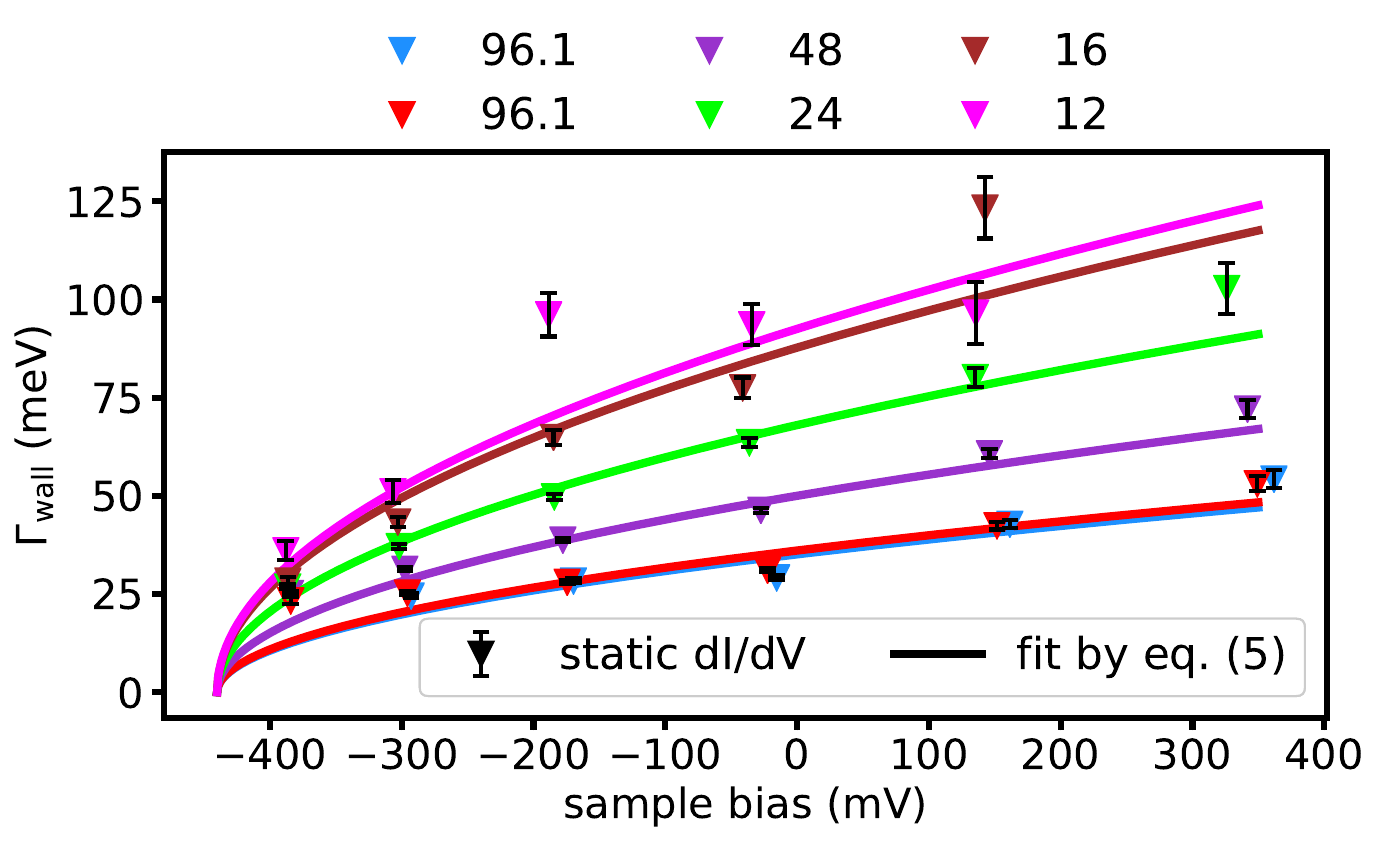}%
	\caption{\label{Fig_versch_wurzelplot} Plot of $\Gamma_\mathrm{wall}$ for $l=0$ states measured in the center of CO corrals with various wall densities and a common radius of $7.13\,$nm. The full lines are the fit of equation (\ref{eq:sqrt_width}) to the datapoints to determine the average path length $x$ of electrons in the quantum corral. The results: $x_{96.1} = (11.3 \pm 1.8)\,$nm, $x_{96.2} = (11.5 \pm 2.0)\,$nm, $x_{48} = (8.0 \pm 0.6)\,$nm, $x_{24} = (6.1 \pm 0.3)\,$nm, $x_{16} = (4.8 \pm 0.5)\,$nm and $x_{12} = (4.3 \pm 0.5)\,$nm.}
\end{figure}

The resulting widths $\Gamma_\mathrm{wall}$ are depicted in FIG. \ref{Fig_versch_wurzelplot}.
Again there is a strong energy dependence, as previously discussed.
These data are also well described by eq. (\ref{eq:sqrt_width}), as shown by the solid lines in FIG. \ref{Fig_versch_wurzelplot}.

\begin{figure}
	\includegraphics[width=0.7\columnwidth]{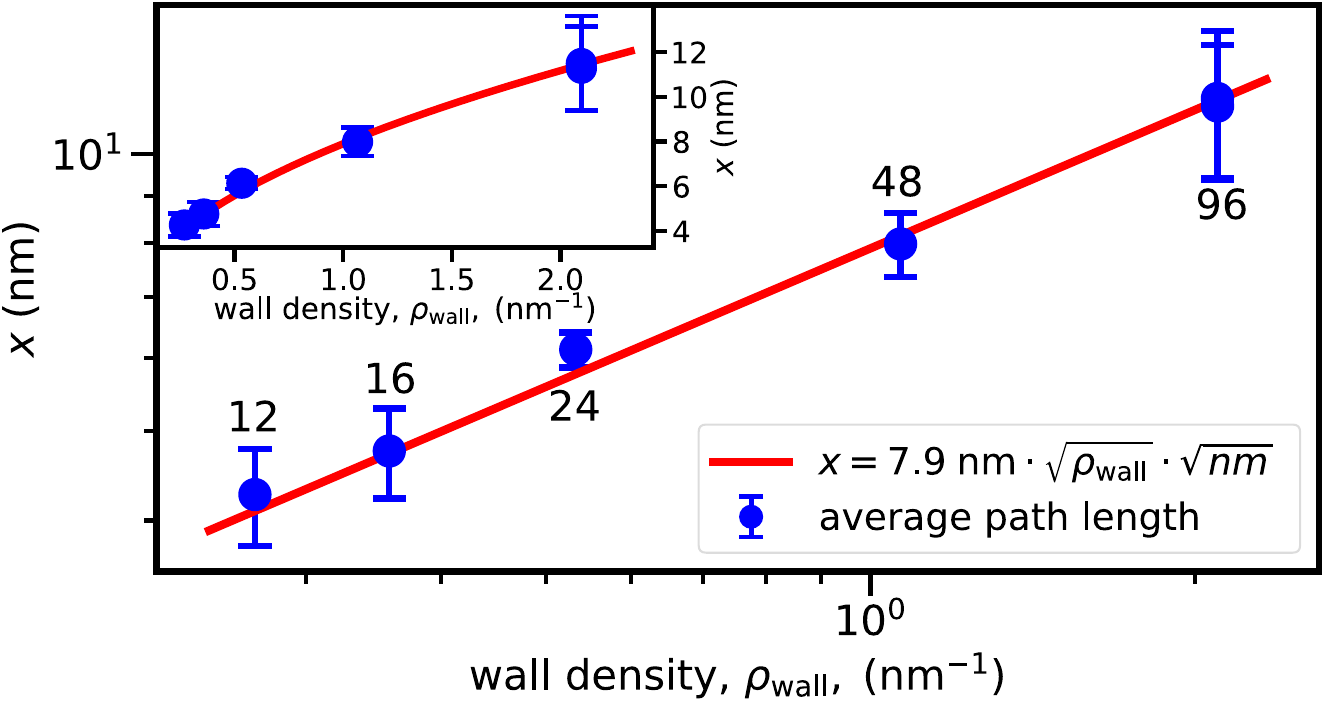}
	\caption{\label{Fig_APL_plot} Average path length $x$ plotted with respect to the inter molecular distance $imd$ in a logarithmic coordinates. Data points are depicted as blue markers. The numbers correspond to the number of CO-molecules in the corral wall. In the inset, a non-logarithmic presentation of the main plot is provided. The solid red line represents the empirically found behavior.}
\end{figure}

FIG. \ref{Fig_APL_plot} is a plot of the average path lengths of the surface electrons as a function of the wall density.
As the CO molecules in the corral wall become more densely packed, the path length of surface state electrons increases. 
This is again in agreement with the theoretical consideration of Crampin \textit{et al.} \cite{Crampin2005}.
An empirical fit reveals that the average path length $x$ exhibits a $\sqrt{\rho_\mathrm{wall}}$ dependency (see full red line in FIG.  \ref{Fig_APL_plot}).

\section{Discussion}\label{chapter:VI}
\subsection{Relating spectral widths and corral wall density}
As highlighted in section \ref{section:IV} , the primary factor limiting the lifetime of electrons in a quantum corral is their interaction with the wall. Generally, three mechanisms can occur when an electron interacts with the corral wall: reflection, absorption into bulk states and lateral transmission. The latter two represent lossy channels. If we consider these three possibilities, the total probability of an electron interacting with the wall is expressed as $ P_\mathrm{R} + P_\mathrm{A} + P_\mathrm{T} = 100 \%$. However, the respective probabilities ($P_\mathrm{R}$ for reflection, $P_\mathrm{A}$ for absorption, and $P_\mathrm{T}$ for transmission) change as a function of the wall density, $\rho_\mathrm{wall}$.

\begin{figure}
	\includegraphics[width=0.7\columnwidth]{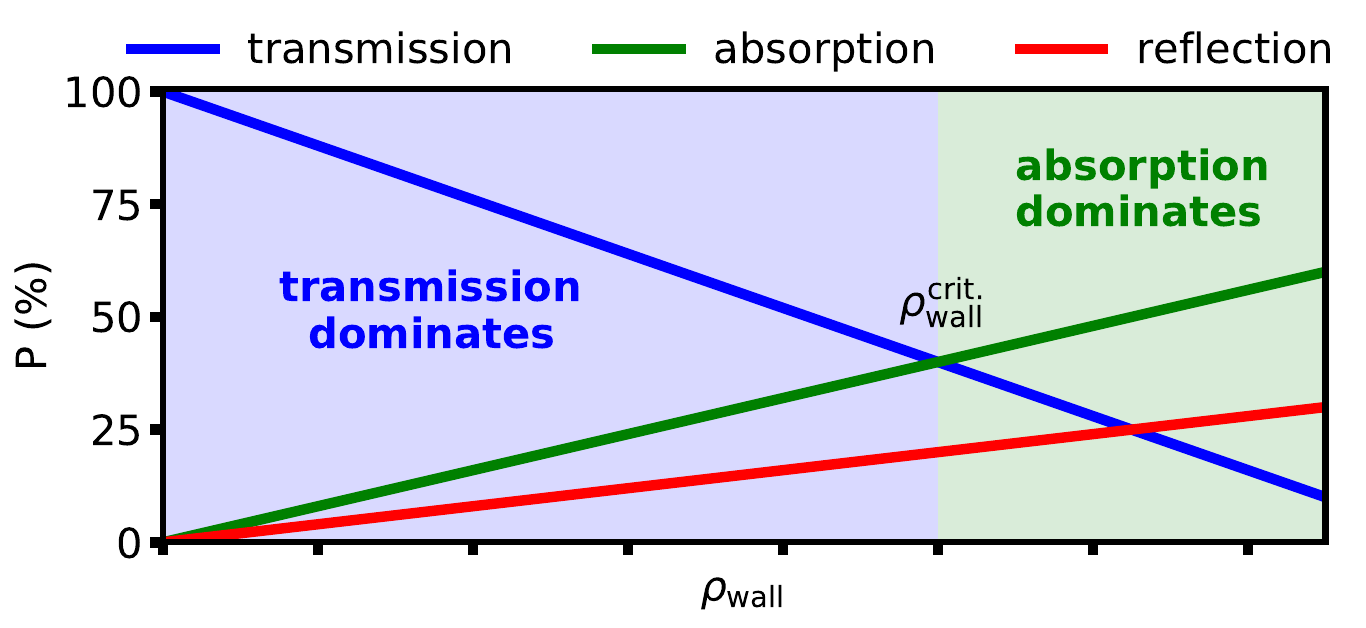}%
	\caption{\label{FIG:dominant} Consideration about the three possibilities for electron-wall interaction: reflection, transmission (lossy) and absorption (lossy). The sum of all probabilities has to be equal to $100\%$. Our analysis revealed an increasing lifetime with a higher wall density, $\rho_\mathrm{wall}$ . At the critical wall density,  $\rho_\mathrm{wall}^\mathrm{crit.}$, the probability for transmission and absorption are equal. The linear representation of these probabilities in relation to $\rho_\mathrm{wall}$ is employed for illustrative purposes.}
\end{figure}

For $\rho_\mathrm{wall} = 0\,\mathrm{nm}^{-1}$, the transmission probability $P_\mathrm{T}$ is $100\%$ as absorption and reflection at the boundary are not possible. As $\rho_\mathrm{wall}$ increases, $P_\mathrm{T}$ decreases and approaches 0. Our experimental observations clearly show that corral lifetimes increase monotonically with wall density (see FIG. \ref{Fig2}C \& \ref{Fig_versch_wurzelplot}), meaning $P_\mathrm{R}$ monotonically increases with $\rho_\mathrm{wall}$.
If we accept that interaction of the surface state electrons with the corral walls includes a lossy channel, then as $\rho_\mathrm{wall}$ increases, the number of lossy scatterers increases and therefore $P_\mathrm{R}$ monotonically increases.
These relationships: the monotonic decrease of $P_\mathrm{T}$ from $100 \%$ with increasing $\rho_\mathrm{wall}$, the monotonic increase of $P_\mathrm{A}$ and $P_\mathrm{R}$, and the requirement that the probabilities sum to $100\%$, are shown in FIG. \ref{FIG:dominant}.
An interesting observation can be made: beyond a critical point ($\rho_\mathrm{wall}^\mathrm{crit.}$), the dominant lossy channel changes from transmission to scattering to the bulk.
Further this analysis shows that transmission is more sensitive to $\rho_\mathrm{wall}$ than absorption.

Several theoretical models propose different perspectives on the dominant lifetime limiting process in quantum corrals. Some point towards bulk coupling (absorption) as the primary process \cite{Heller1994, Crampin1994, Crampin1996}, while García-Calderón and Chaos-Cador \cite{Garcia-Calderon2006}, for instance, suggest that coherent processes, such as tunneling (transmission), play a more significant role. On the other hand, some theoretical considerations rely entirely on elastic mechanisms without coupling to the bulk \cite{Harbury1996,Rahachou2004}. Respecting all these theoretical considerations, we thus propose the estimate that corrals with a low wall density lie in the regime where transmission dominates, while for high wall density corrals, absorption is the dominant lifetime limiting process. However, within this work we cannot determine the exact value of  $\rho_\mathrm{wall}^\mathrm{crit.}$.

\subsection{Gaussian spectral shape}
Up to now, literature uses Lorentzian-shaped curves to fit spectral peaks of quantum corrals \citep[e.g.][]{Crommie1993, Crampin2000, Kliewer2001, Jensen}. This implies an exponentially decaying survival probability of electrons within the corral. To satisfy this condition, electrons must exhibit a uniform decay probability for all times. However, our findings highlight that the primary mechanism limiting the lifetime of electrons in a quantum corral is their interaction with the wall. A single-particle approach, considering lossy interactions solely at the corral wall, is enough to yield a survival probability that does not exponentially decrease with time, as explicitly shown in (SM10) \cite{SupplMat}. This challenges the exclusive description of spectral corral peaks with Lorentzian curves. 
We propose that a more complete theoretical picture will yield line shapes that are better described by a Gaussian rather than a Lorentzian function.

Gaussian-shaped spectral peaks in d$I$/d$V$ measurements have been documented in literature, as observed in defects in salt layers \cite{Repp2005a} and quantum dots \cite{Jdira2008}. The Gaussian shapes were attributed to a strong coupling of a phonon mode with electrons. In the context of a corral, a plausible mode could be the coherent radial motion of CO molecules, also called breathing mode. The possibility of a direct, inter molecular interaction among individual CO molecules can be dismissed due to the substantial distance between them (e.g., for the 48-corral, $938\;$pm). Nevertheless, it is established that adsorbates can interact over extended distances through the surface state \citep[e.g.][]{Lau1978, Gumhalter1995, Hyldgaard2000, Hyldgaard2003}. Thus, the radially symmetric corral states could induce coherent motion of the COs. In the scenario of strong electron-phonon coupling, this coherent motion might offer an explanation for the Gaussian shape observed in the energy distributions presented in this paper. However, more comprehensive calculations are required that are out of the scope of this work.

The breathing mode could also directly affect of the energy of the corral states.
The modulation of the corral radius induced by the coherent radial motion of the CO molecules would also influence the energetic position of the corral states.
A reduction in radius elevates the states in energy, whereas an expansion in radius causes a downward shift. This variation in energies results in a widening of the measured spectral peak. 
To account for the experimentally determined spectral widths (see FIG. \ref{Fig2}C and \ref{Fig_versch_wurzelplot}) via the effect of the breathing mode on corral states, radial amplitudes of $200\;$pm would be necessary. Details about this calculation are given in the supplemental material (SM11) \cite{SupplMat}. Given that this value exceed the radius of a copper atom ($127\;$pm) and no shift in CO molecule positions occurred during the whole measurement period, it is inferred that the energy shift of corral states due to the breathing mode does not represent a predominant mechanism for broadening.

Since the required amplitude is too large to fully account for the spectral width, we ask if the breathing mode could reasonably contribute to the measured spectral widths assuming oscillation amplitudes as previously determined for single CO molecules.
Viewing the breathing mode as a quantum mechanical oscillator (QMO) allows us to estimate its radial amplitude. The oscillation governs the corral's radius, and like any QMO, its ground state is characterized by a Gaussian curve. Consequently, the corral radius, modulated by the breathing mode, also adheres to a Gaussian curve.
An approximation of the uncertainty in the position of a CO results in a Gaussian FWHM of about $33\;$pm \cite{Okabayashi2018}. Therefore if the motion of the CO adsorbates is coherent, this motion would, as a lowest estimate, contributes to a breathing mode induced Gaussian broadening of the energy distributions of approximately $4\;$meV at the Fermi energy and $\approx 8\;$meV for states $400\;$meV above the Fermi energy (for details see the supplemental material SM11 \cite{SupplMat}). 

In conclusion, we propose that the overall Gaussian shape of the energy distributions arises from (1) scattering of non-uniformly distributed scatterers, (2) strong coupling between electrons and a breathing mode, and (3) the effect of the breathing mode on the corral radii and the subsequent effect on the corral states.

\section{Summary and outlook}
\subsection{Summary}

To start, we built two differently sized, circular quantum corrals with the same wall density on Cu(111). 
By comparing the measured local density of states with a linear combination of the probability density of the wave functions obtained by the hard wall model we determined the energy distributions, including those of $l \neq 0$ states. 
Analyzing these distributions revealed that the natural line shape is best described by a Gaussian function for all states. 
It also showed a correlation between the size of the corral and the spectral width, leading to the conclusion that the dominant lifetime limitation is governed by the interaction of electrons with the corral wall.

We thus proposed a classical model to relate the spectral widths of the corral states with the average path length of a surface state electron. 
This model described the energy dependence of the spectral widths well.

To better characterize the electron-wall interaction, we constructed corrals with different wall densities. 
The widths of the spectral peaks decreased with increasing wall densities, clearly showing longer corral state lifetimes for denser walls. This behavior prompted us to conclude that lateral transmission through the corral wall is more sensitive to the wall density than coupling to bulk states.

Finally, we discussed three possibilities to explain the observed Gaussian spectral shape. Currently, literature employs Lorentzian functions to fit spectral peaks of quantum corrals, assuming an exponential decay in electron survival. However, our findings suggest that the primary limitation to electron lifetimes in the corral arises from their interaction with the wall.
This leads to a survival probability of the electrons in the corral that does not exponentially decrease with time, resulting in non Lorentzian shaped spectral peaks. 
Another potential explanation, previously discussed in literature for various sample systems, involves the coupling of electrons to specific phonon modes \cite{Repp2005a,Jdira2008}. In the limit of a strong electron-phonon coupling, Gaussian-shaped spectral peaks are observed. One plausible phonon mode for quantum corrals is the breathing mode, representing a coherent movement of CO molecules in the wall.
Moreover, not only can the coupling of electrons to this breathing mode induce Gaussian-shaped peaks in spectra, but the breathing mode itself can also influence the measured spectral line shape. While the estimated contribution of this variation in corral radius does not fully account for the measured widths, it may contribute to the overall Gaussian shape.

\subsection{Outlook}
By adjusting the wall density approximately one order of magnitude, we observe a change in the average path length of an equivalent order of magnitude. 
If we were able to make a corral wall with adsorbates that allow nearest-neighbour occupation on each Cu sites, we could hypothetically achieve a path length of $\approx 16\,$nm (by extrapolation of FIG.~\ref{Fig_APL_plot}).
While a denser corral wall is not feasible as discussed earlier, the radial thickness of the confinement can be increased by using a second or third row of CO molecules. 
It might be possible, with this approach, to reduce the lifetime limiting effects of tunneling (transmission) so drastically that the spectral widths are dominated by electron-electron, electron-phonon scattering and coupling to the bulk.
Future experiments such as these would allow for a deeper insight into the interaction between Shockley surface state electrons and impurities.

\begin{acknowledgements}
The authors thank A. Weindl and C. Setescak for carefully proofreading the manuscript and S. Tomsovic for fruitful discussions. 
\end{acknowledgements}

%


\pagebreak

\renewcommand\thetable{\arabic{figure}}
\renewcommand{\tablename}{Table}

\renewcommand\thefigure{S\arabic{figure}}
\renewcommand{\figurename}{Figure}

\renewcommand\thesection{SM \arabic{section}}
\renewcommand\thesubsection{SM \arabic{section}.\arabic{subsection}}

\setcounter{equation}{0}
\setcounter{figure}{0}
\setcounter{table}{0}
\setcounter{section}{0}

\pagestyle{plain}

\widetext
\begin{center}
	\textbf{\LARGE{Supplemental Material}}
\end{center}
\vspace{1mm}

\section{Experimental setup}
\noindent \textbf{Microscope} 
The experiments were performed on a home-built combined atomic force and scanning tunneling microscope with a base temperature of $\approx 5.7\,$K. The system operates with a qPlus sensor \cite{Giessibl1998} which was equipped with an electrochemically etched tungsten tip. The bias voltage is applied to the sample. The Cu(111) sample was prepared with standard sputter and anneal cycles. 

\begin{figure}[H]
	\begin{minipage}[c]{\textwidth}
		\centering
		\begin{minipage}[c]{0.49\linewidth}
			\includegraphics[width=\linewidth]{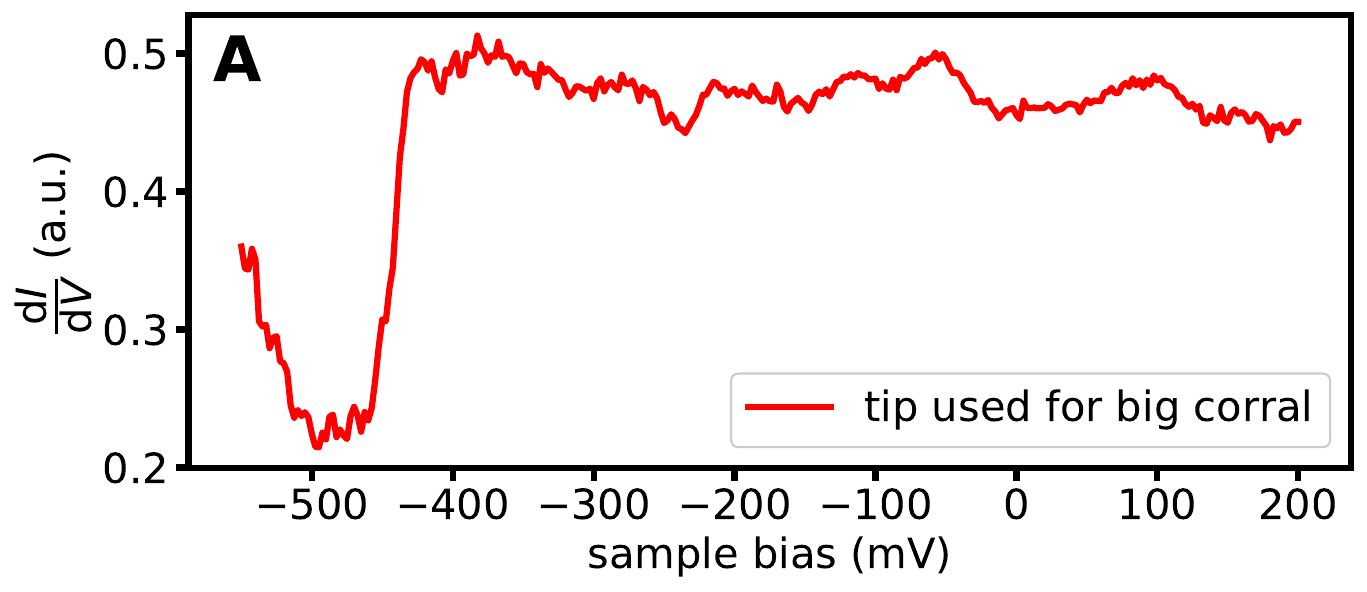}
		\end{minipage}
		\hfill
		\begin{minipage}[c]{0.49\linewidth}
			\includegraphics[width=\linewidth]{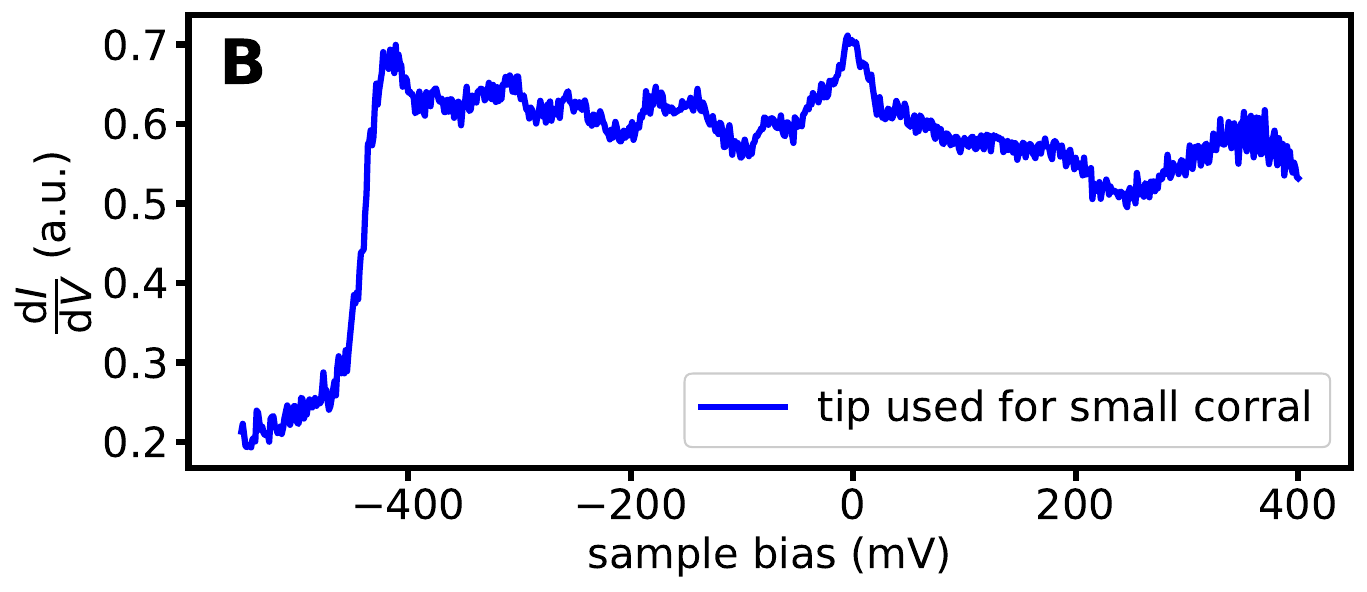}
		\end{minipage}
		\hfill
		\begin{minipage}[c]{0.49\linewidth}
			\includegraphics[width=\linewidth]{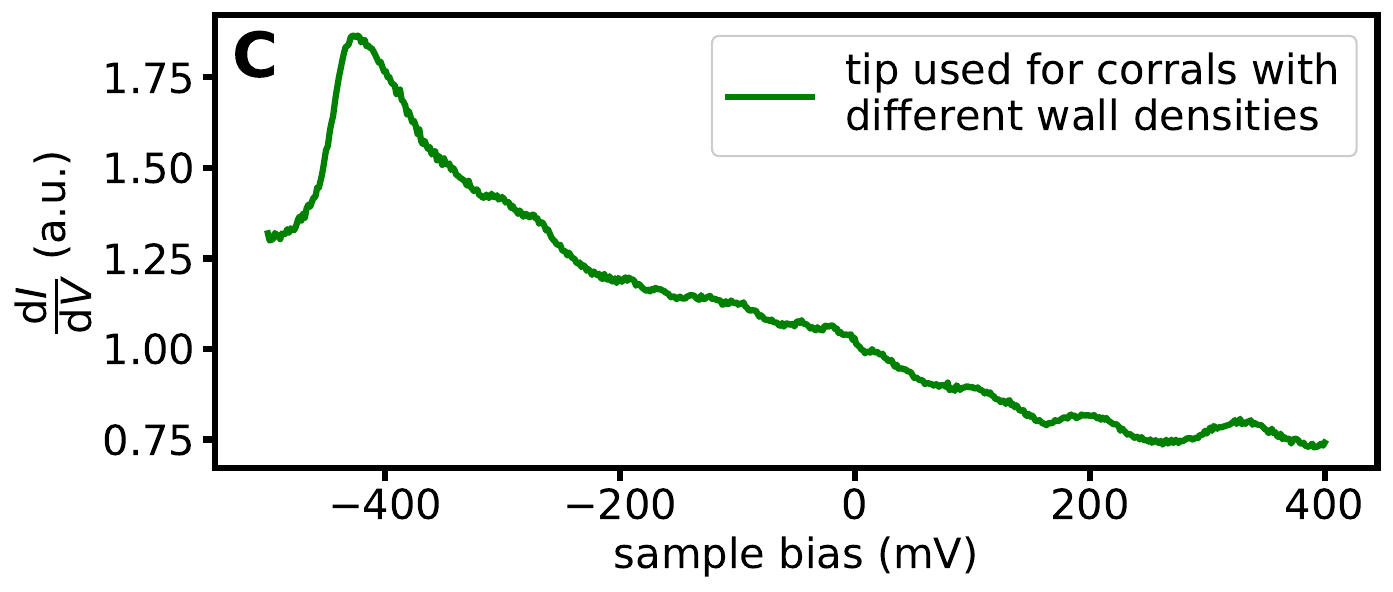}
		\end{minipage}
	\end{minipage}
	\caption{\label{tipDOS} Differential conductance measurement taken over the bare Cu(111) surface. For the big CO corral we used tip \textbf{A} and for the small CO corral we used tip \textbf{B}. For the static d$I$/d$V$ measurements in the corrals with various wall densities we employed tip \textbf{C}.}
\end{figure}

\noindent \textbf{Tip preparation} 
Differential conductance (d$I$/d$V$) measurements entail the convolution of the density of states (DOS) of both, the sample and the tip. To ensure that our measurements accurately represent the DOS of the corrals, we took special care to employ a tip with a flat DOS. To achieve this, we prepared the tip by carefully driving it into the sample multiple times until we obtained a featureless d$I$/d$V$ curve over bare copper. For the line scans presented in the main paper, we used one tip DOS configuration for the big 48-CO corral and another configuration for the small 24-CO corral. The d$I$/d$V$ spectra, where we assessed the flatness of the tip DOS, are displayed in Figure \ref{tipDOS} A \& B. The differential conductance measurement over the bare copper surface with the tip employed for characterizing the corrals with various wall densities, is displayed in Figure \ref{tipDOS} C.

\begin{figure}[H]
	\centering
	\includegraphics[width=0.8\linewidth]{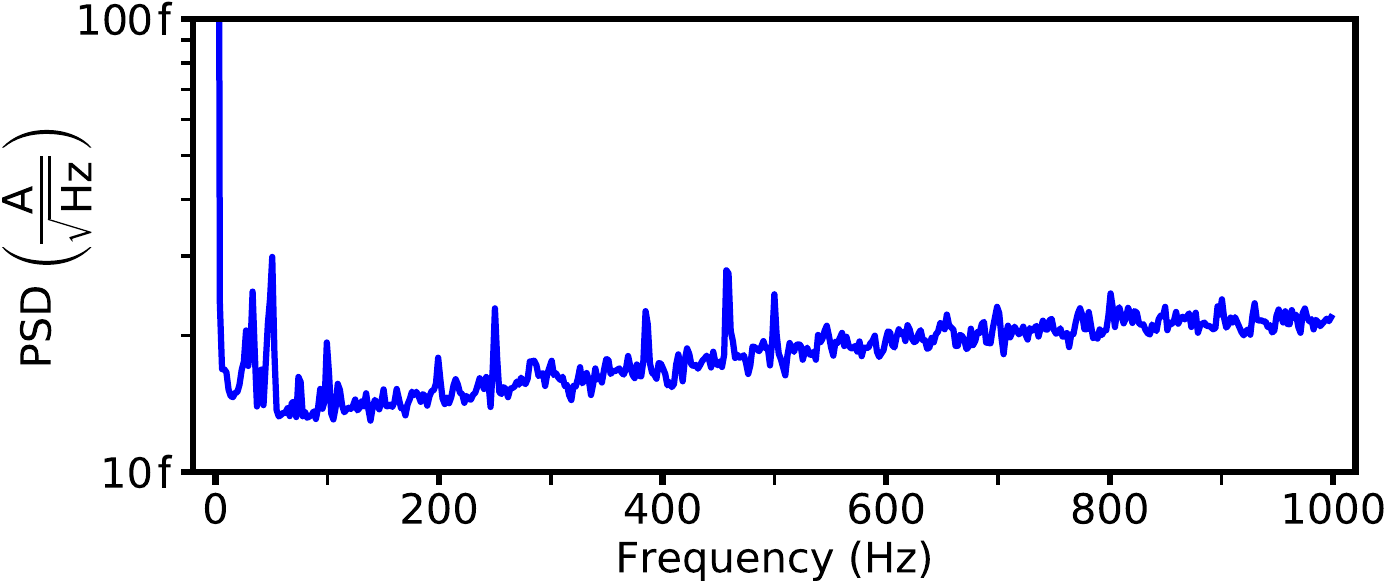}%
	\caption{\label{spectrum} Power spectral density (PSD) as a function of frequency in constant height over the center of the big 48-CO corral at a sample bias of $-10\,$mV and a tunneling current of $10\,$pA.} 
\end{figure}

\noindent \textbf{Measurement settings - line scans} 
The d$I$/d$V$ line scans across the corral diameter have a length of $18\,$nm for the big 48-CO corral and a length of $10\,$nm for the small 24-CO corral. Every line scan was performed in constant height. Each scan consists of an average over 7 line scans acquired with a scan speed of $2\,$nm/s and 1024 pixels per line. For the big 48-CO corral a setpoint of $-170\,$mV/$170\,$pA and for the small 24-CO corral a setpoint of $100\,$mV/$100\,$pA was used with the tip positioned at the center of each corral. For the acquisition of the differential conductance we modulated the sample bias by a sinusoidal AC voltage with an amplitude of $V_\mathrm{mod} = 5\,$mV and a frequency of $600\,$Hz. For the modulation frequency we chose a low noise region in the power spectral density spectrum of the tunneling current (see Figure \ref{spectrum}).

\newpage

\newpage

\section{Corral construction plans}
\begin{figure}[H]
	\begin{minipage}[c]{\textwidth}
		\centering
		\begin{minipage}[c]{0.49\linewidth}
			\includegraphics[width=\linewidth]{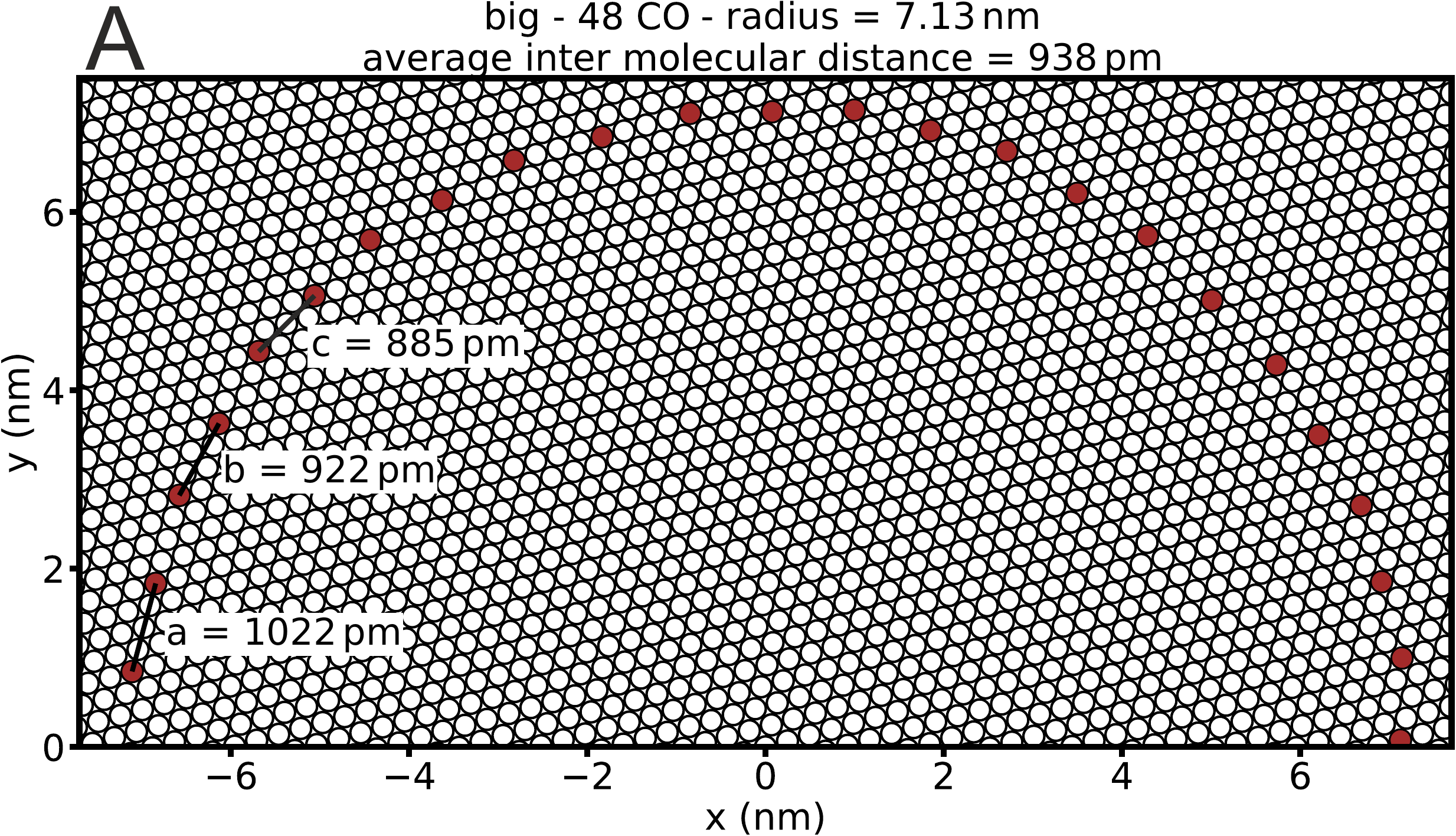}
		\end{minipage}
		\hfill
		\begin{minipage}[c]{0.49\linewidth}
			\includegraphics[width=\linewidth]{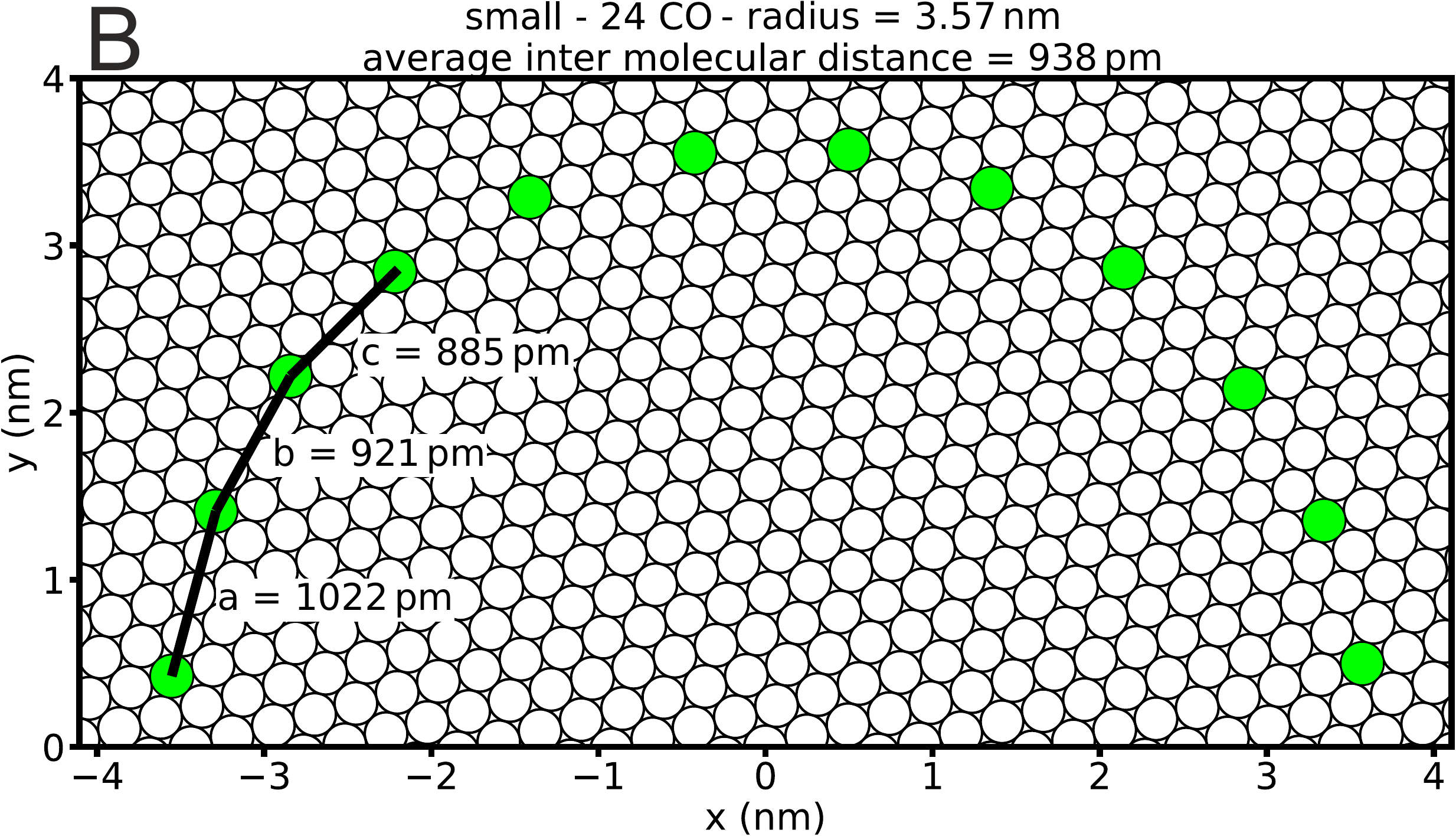}
		\end{minipage}
		\hfill
		\vspace{0,5cm}
		\begin{minipage}[c]{0.49\linewidth}
			\includegraphics[width=\linewidth]{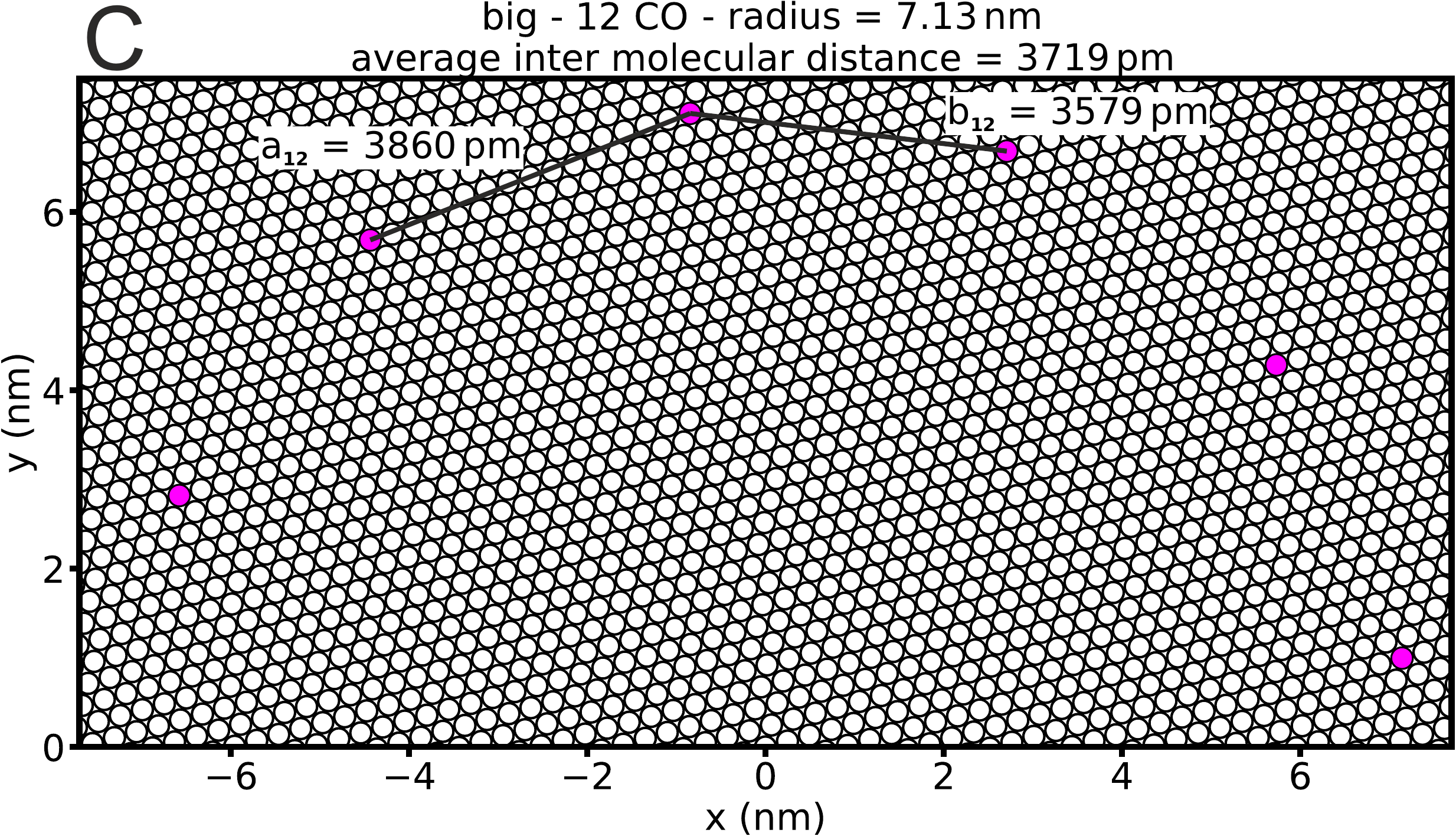}
		\end{minipage}
		\hfill
		\begin{minipage}[c]{0.49\linewidth}
			\includegraphics[width=\linewidth]{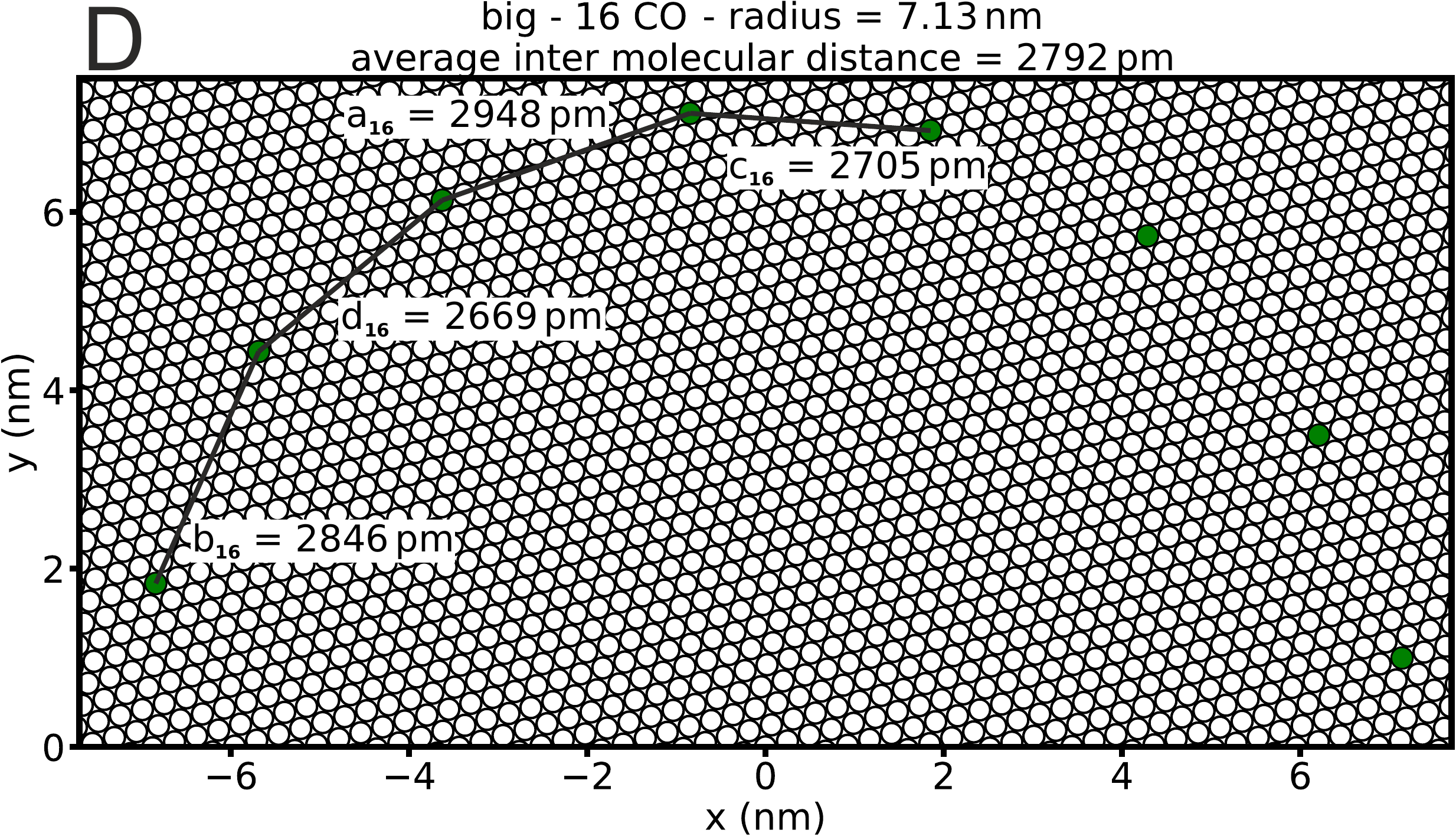}
		\end{minipage}
		\hfill
		\vspace{0,5cm}
		\begin{minipage}[c]{0.49\linewidth}
			\includegraphics[width=\linewidth]{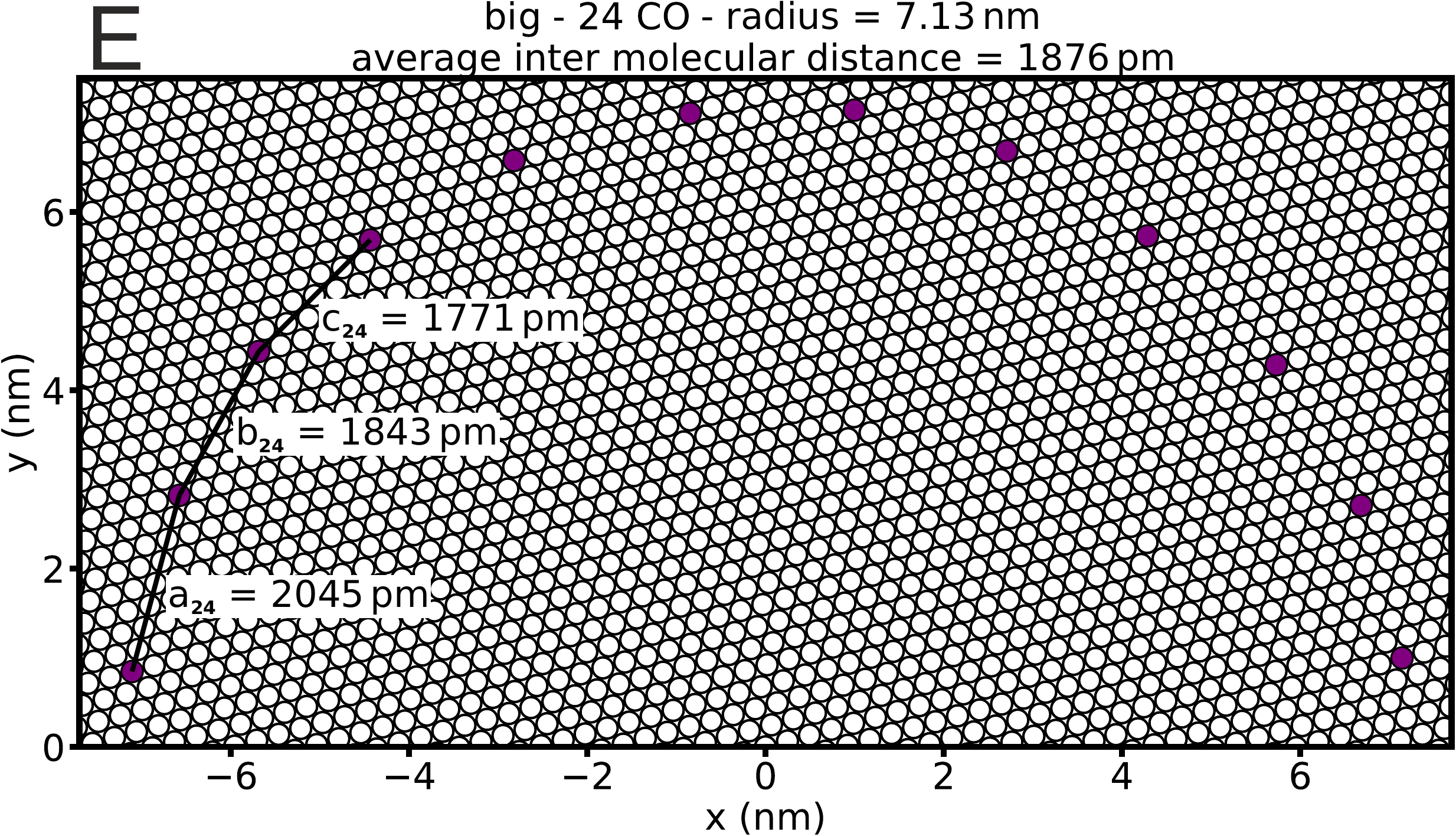}
		\end{minipage}
		\hfill
		\begin{minipage}[c]{0.49\linewidth}
			\includegraphics[width=\linewidth]{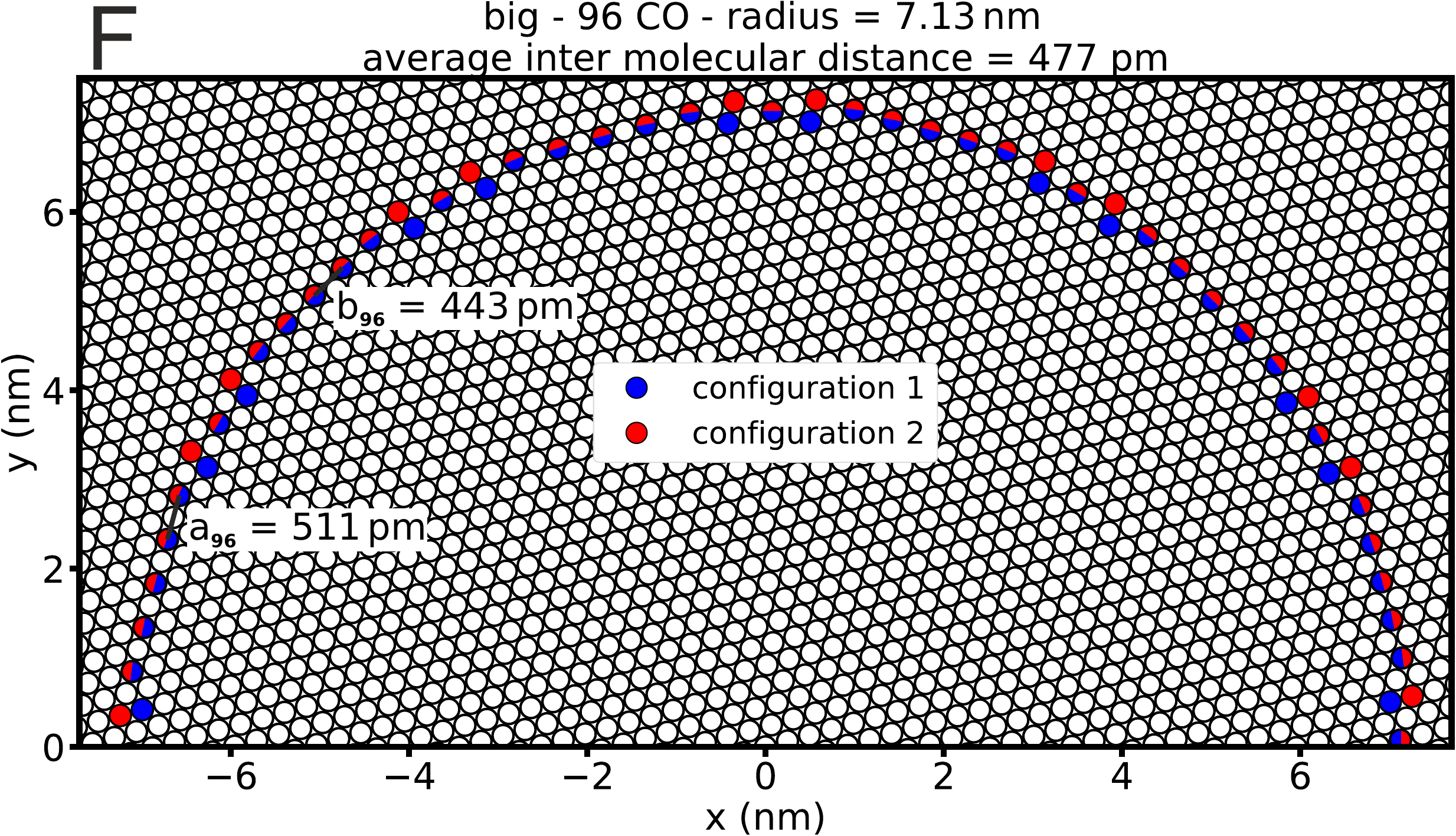}
		\end{minipage}	
		\hfill
		\vspace{0,5cm}
	\end{minipage}
	\caption{\label{bauplan} Building plans for the corrals. Black unfilled circles resemble the Cu(111) surface. Filled colored circles indicate the position of CO. The number in the title is the amount of CO molecules in the corral wall.}
\end{figure}

The construction plans for the different CO corrals are depicted in Figure \ref{bauplan}. CO adsorbs on the top site of the Cu(111) surface. Copper atoms are drawn as empty black circles and the CO molecules are drawn as colored circles. The atomic spacing of the copper substrate is $255\,$pm. 

In the first part of the main text, we analyzed a large corral (composed of 48 CO molecules and a radius of $7.13\,$nm) and a smaller one (composed of 24 CO molecules and a radius of $3.57\,$nm). Figures \ref{bauplan} A and B depict the arrangement of CO molecules in these two corrals.

For constructing corrals with a higher wall density we took the building plan of the big 48 CO corral (a radius of $7.13\,$nm) and inserted additional CO molecules. To decrease the wall density we excluded adsorption positions. The CO arrangement for the 12, 16, 24 and 96 CO corral are depicted in Figure \ref{bauplan} C-F.
The wall densities of the 12-, 16-, 24-, 48-, 96.1- and 96.2-CO-corrals are as follows: $0.269\,\mathrm{nm}^{-1}$, $0.358\,\mathrm{nm}^{-1}$, $0.533\,\mathrm{nm}^{-1}$, $1.066\,\mathrm{nm}^{-1}$ and $2.096\,\mathrm{nm}^{-1}$(for both 96-CO-corrals).

\newpage

\newpage

\section{Stationary dI/dV measurements in the big 48-CO corral}

As an example two stationary d$I$/d$V$ spectra are displayed in Fig \ref{Fig1}. The red curve is a measurement over the center of the $7.13\,$nm corral and is therefore only sensitive to the $l=0$ states. The second curve (green) was measured $900\,$pm off center. A d$I$/d$V$ spectrum at this position is sensitive to $l=0$ and $l=1$ states which results in additional spectral peaks. Because of an overlay of several peaks one can only see the top most part of the states energy distributions. Since a Gaussian and Lorentz distribution are very similar near their maxima, a proper line shape analysis is not possible. Therefore d$I$/d$V$ at single points above the corral for a detailed line shape study of $l\neq0$ states is not reliable.

\begin{figure}[h]
	\centering
	\includegraphics[width=0.8\linewidth]{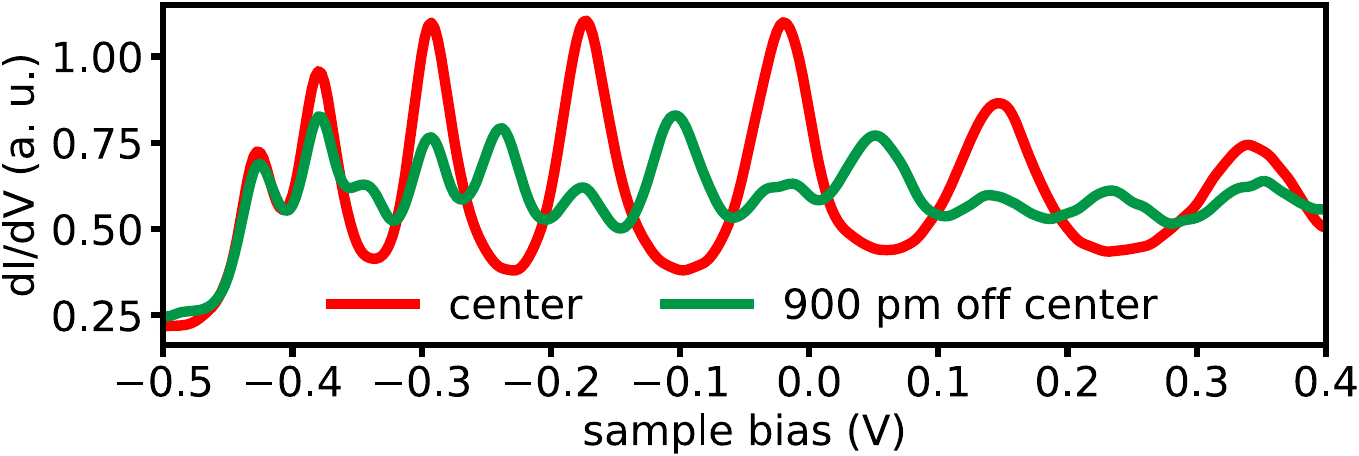}%
	\caption{\label{Fig1} Two d$I$/d$V$ spectra measured inside the big 48-CO corral at different positions. Red: center of the corral; Green: $900\,$mV off center.} 
\end{figure}

\newpage

\section{Fitting equation (1) of the main text to the line scans}	
Equation (1) of the main text involves a considerable number of free parameters (one $\alpha_{n,l}$ for every corral state) for the fitting process. Such a large number is neither effective nor reliable for fitting purposes. To achieve the most accurate fit for the linescans, a balance was sought between capturing essential states contributing to each d$I$/d$V$ linescan and minimizing the number of free parameters ($\alpha_{n,l}$). Consequently, a subset of states was chosen for each linescan, focusing on those most relevant to the specific bias voltage at which the scan was conducted. This approach significantly simplified the fits, ensuring that each fitting process involved a manageable number of parameters.

The decision on whether to consider a state for a line scan fit was not determined by a fixed bias voltage window. If the bias voltage of a line scan approached the expected energy of a state, a "test fit" was executed. If the newly considered state contributed to the fit function ($\alpha_{n,l} > 0$), it was retained; otherwise, if it did not contribute ($\alpha_{n,l} = 0$), it was excluded from the fit function, and the fit was performed again without the previously considered state. This procedure for including a state in the fit function was applied to all states and over the whole bias window. A similar strategy was employed for removing a state from the fitting function. When the $\alpha_{n,l}$-value of a state had went back to 0 over 2 to 3 linescans (10-15 mV), the state was excluded from the fitting function.
Using this procedure, the fitting uncertainties of $\alpha_{n,l}$ averaged to approximately single-digit percentages.

\newpage

\section{Assessing the unambiguousness of the line scan fits}
\begin{figure}[h]
	\centering
	\includegraphics[width=\linewidth]{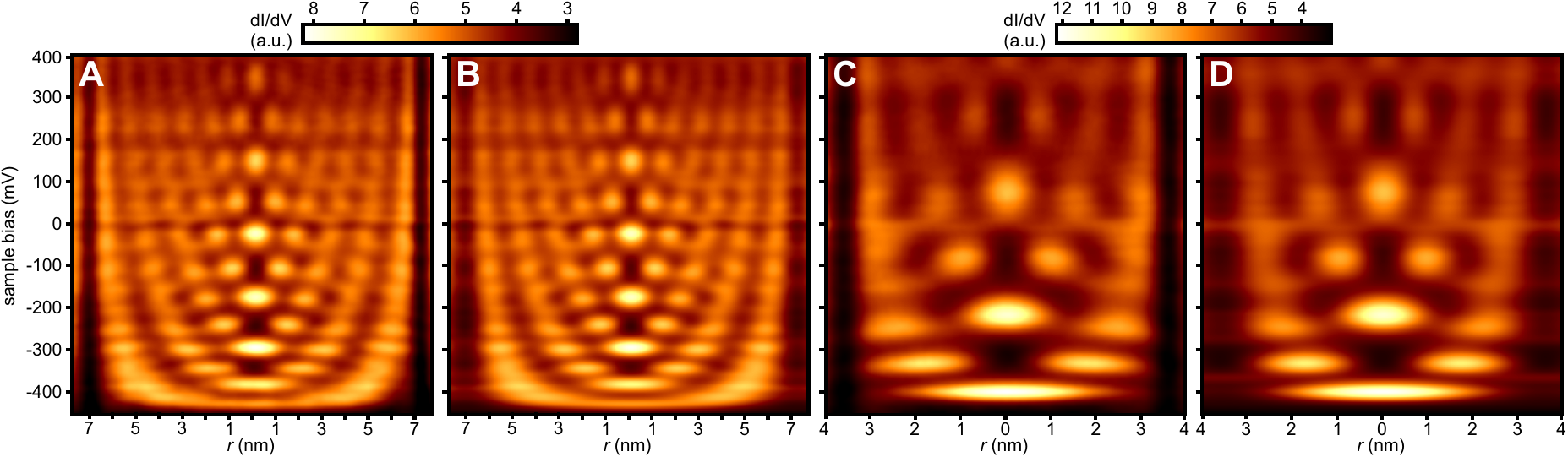}%
	\caption{\label{fig:comp_maps} Differential conductance (d$I$/d$V$) measurements are shown as a function of radial distance $r$ from the center and sample bias for the (\textbf{A}) large 48-CO corral ($R_{48\mathrm{,large}} = 7.13\,$nm) and the (\textbf{C}) small 24-CO corral ($R_{24\mathrm{,small}} =3.57\,$nm). The fits with equation (1) of the main text to the measurements are displayed in \textbf{B} and \textbf{D}. Notably, deviations are observed in the vicinity of CO molecules (at approximately $r \approx 7\,$nm in \textbf{A}\&\textbf{B} and $r \approx 3.5\,$nm in \textbf{C}\&\textbf{D}). These deviations result from the exclusion of LDOS behavior near the CO molecules in the model-based wavefunction fits. However, it's important to emphasize that the fits accurately reproduce the LDOS within the corral itself, which is the region of critical interest.} 
\end{figure}

To check the unambiguity of the line scan fits, several checks were done:

\noindent \textbf{Direct comparison between data and fit}
Comparing the local density of states (LDOS) maps obtained from measurements with those derived from the fitting procedure reveals a strong agreement between data and fit, as illustrated in Figure \ref{fig:comp_maps}.

While the agreement holds well within the interior of the corrals, distinctions emerge near the corral walls, where the CO molecules are positioned. Notably, for the large 48-CO corral depicted in Figure \ref{fig:comp_maps} A\&B, discrepancies become evident at a radial distance of $r \approx 7\;$nm. Similarly, in the case of the smaller 24-CO corral shown in Figure \ref{fig:comp_maps} C\&D, differences are notable around $r \approx 3.5\;$nm. These differences stem from the exclusion of LDOS behavior near CO molecules in the wavefunction fits based on the hard wall model. Nevertheless, it is crucial to underscore that our fits accurately replicate the LDOS within the corral itself, which constitutes the region of primary interest.

\noindent \textbf{Fit stability}
In order to assess the stability and reliability of our fitting procedure, we conducted a stability test. Our aim was to examine whether variations in the initial starting parameters for the fitting process would significantly impact the results.

For our main text analysis, we employed consistent starting parameters, setting the starting value of every $\alpha_{n,l}(eV_\mathrm{B})$ to $10$ for every fit in all line scans. To gauge the stability of our method, we studied different starting parameter values, ranging from 0 to 100 with increments of 1. This entailed conducting 100 iterations. Each iteration produced a set of energy distributions for the states within the quantum corrals.

Subsequently, we subjected the distributions resulting from each iteration to the same line shape analysis that we employed in our main text (equation (3)). Our key observation was that, regardless of the variations in the starting parameters, the line shapes of the energy distributions remained stable and the dominant Gaussian behavior was always reproduced (with an average error of $2$\% for the FWHM of the Gaussians). This consistency in the line shapes across all variations of starting parameters reaffirmed the robustness and reliability of our fitting method.

\begin{figure}[h]
	\begin{minipage}[c]{\textwidth}
		\centering
		\begin{minipage}[c]{0.49\linewidth}
			\includegraphics[width=\linewidth]{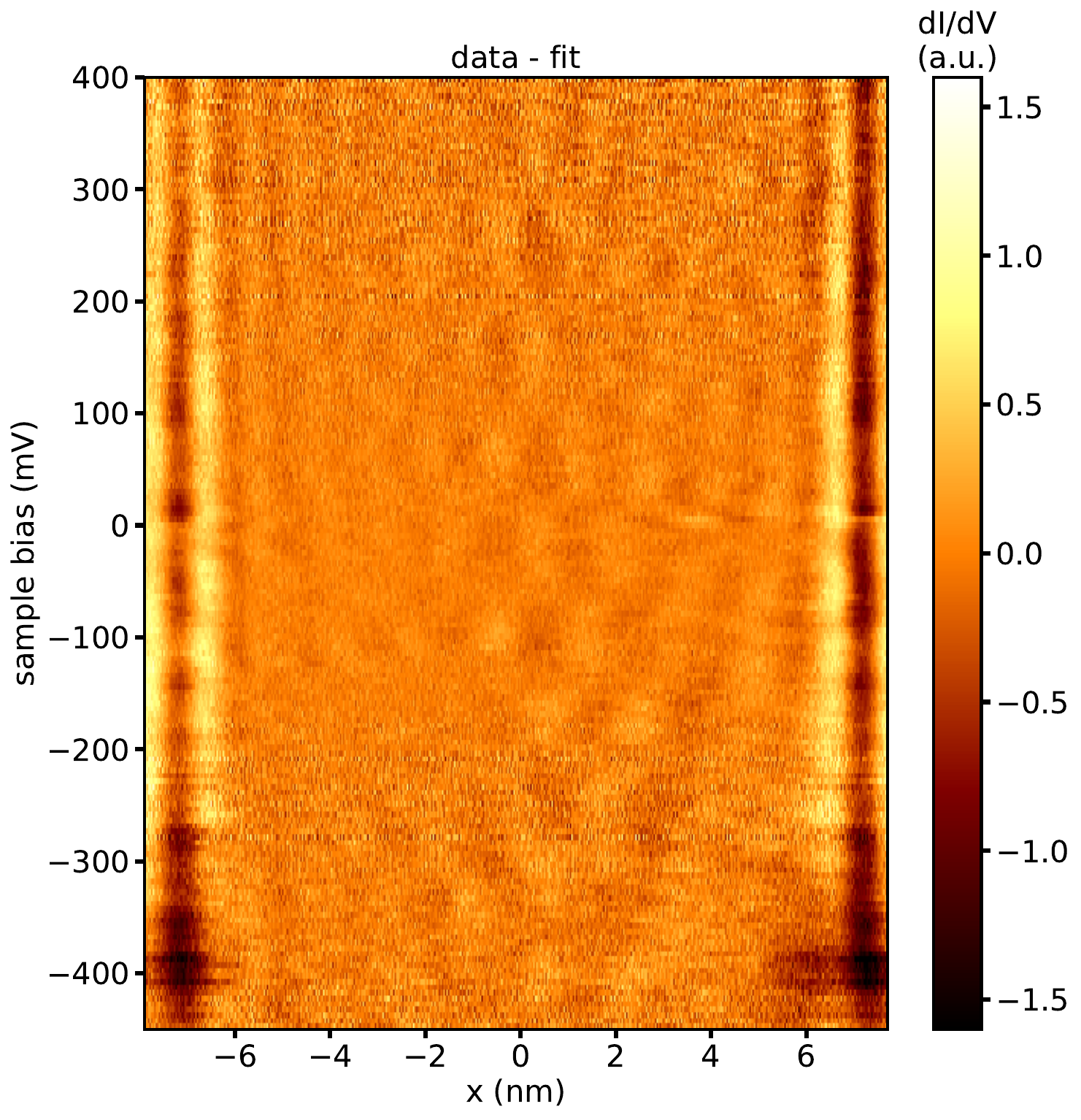}
		\end{minipage}
		\hfill
		\begin{minipage}[c]{0.49\linewidth}
			\includegraphics[width=\linewidth]{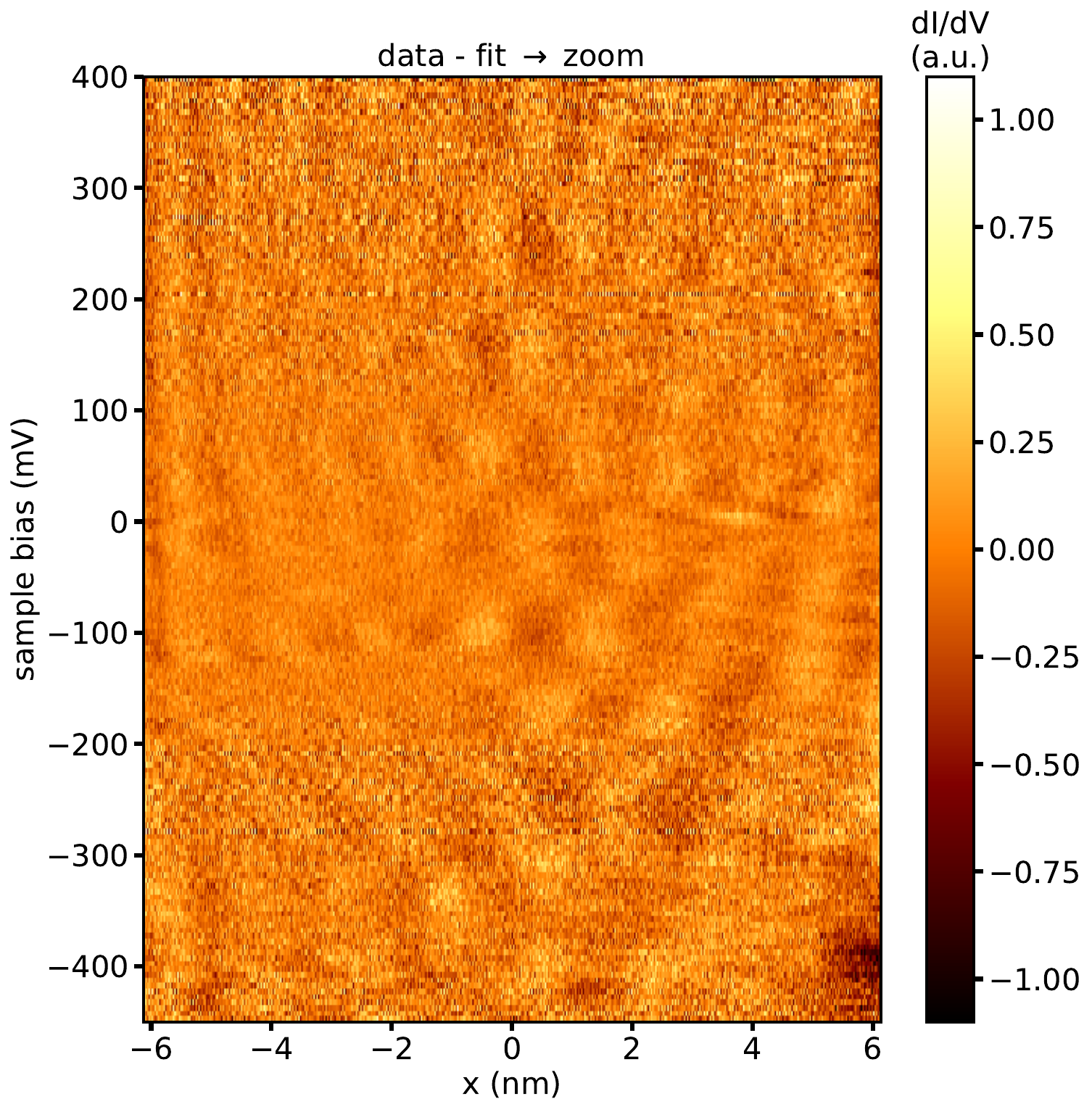}
		\end{minipage}
		\hfill
		\begin{minipage}[c]{0.3\linewidth}
			\includegraphics[width=\linewidth]{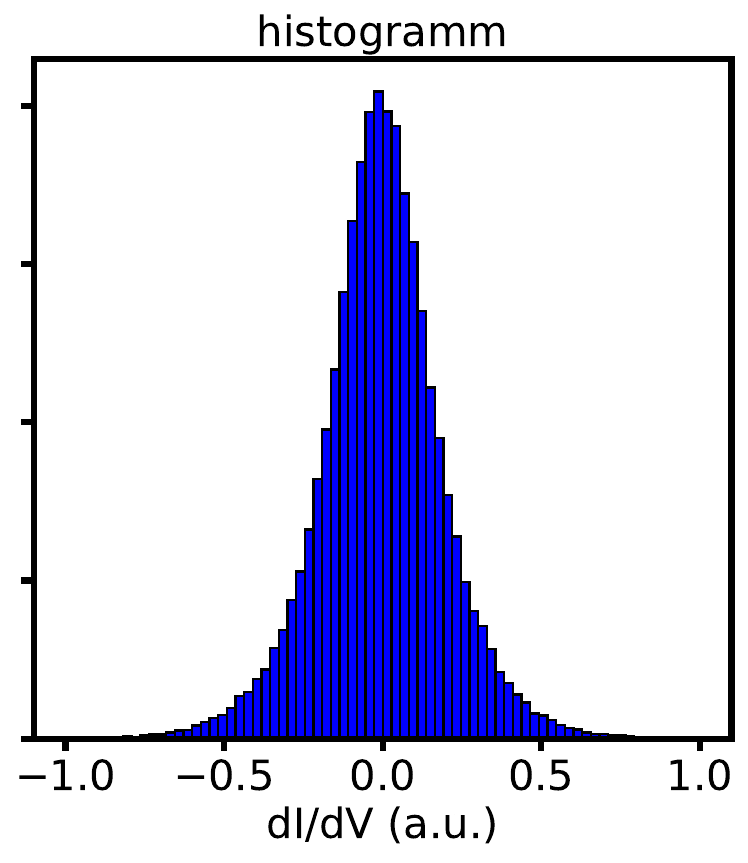}
		\end{minipage}
	\end{minipage}
	\caption{\label{residual_analysis} Residual analysis (data - fit) of the d$I$/d$V$ maps shown in the main text. Notably, deviations are visible in the vicinity of CO molecules (at approximately $x \approx \pm 7\,$nm). These deviations result from the exclusion of LDOS behavior near the CO molecules in the model-based wavefunction fits. However, it's important to emphasize that the fits accurately reproduce the LDOS within the corral itself, which is the region of critical interest. The upper right panel shows a zoom of the upper left plot to the inner of part of the corral. This image only shows minor deviations. The bottom image depicts a histogram of the zoomed in residual plot, only showing a noise distribution.}
\end{figure}

\noindent \textbf{Residual analysis}
To assess the quality of our fitted linescans, we conducted a residual analysis by comparing the difference between the measured data and the corresponding fit. The result for the big 48-CO corral, depicted in the upper row of FIG. \ref{residual_analysis}, reveals that the region spanning between $x \approx -6\,$nm to $x \approx 6\,$nm is predominantly influenced by noise, indicating that our fits accurately replicate all measured features. A histogram of the residual plot from  $x \approx -6\,$nm and $x \approx 6\,$nm is displayed at the bottom row, showing a noise distribution. 

Like already mentioned above, deviations are more pronounced in the vicinity of CO molecules, as shown in FIG. \ref{residual_analysis} (upper row, left plot) around $x \approx \pm 7 \,$nm. This is due to the exclusion of LDOS behavior near CO molecules in the hard wall model based wavefunction fits. However, our fits accurately reproduce the LDOS within the corral itself, which is the region of critical interest.

\newpage

\newpage

\section{Quality of the reconstructed energy distributions}\label{SM:roughness}
To quantify the quality of the reconstructed energy distributions $\alpha_{n,l}$ we used the relative quadratic deviation between every distribution and the fit with equation (3) of the main text.

\begin{equation}\label{SM:1}
\mathrm{RQD} = \dfrac{\mathlarger{\sum\limits_{i = 0}^{n}}\biggl[\alpha_{n,l} \Bigl(V_{B}^{(i)}\Bigr) - \mathrm{F}\Bigl(V_{B}^{(i)}\Bigr)\biggr]^2 }{\mathlarger{\sum\limits_{i = 0}^{n}}\biggl[\mathrm{F}\Bigl(V_{B}^{(i)}\Bigr)\biggl]^2}
\end{equation}

In this formula $\alpha_{n,l} \Bigl(V_{B}^{(i)}\Bigr)$ describes the value of the energy distribution at the sample voltage $V_{B}^{(i)}$ and $\mathrm{F}\Bigl(V_{B}^{(i)}\Bigr)$ is given by the value of the fit of equation (3) in the main text at the same bias voltage. As a final step the quality factor $\xi$ of the energy distribution  can be defined as

\begin{equation}\label{SM:2}
\xi = 1-\mathrm{RQD}
\end{equation}

\begin{figure}[h]
	\centering
	\includegraphics[width=0.8\linewidth]{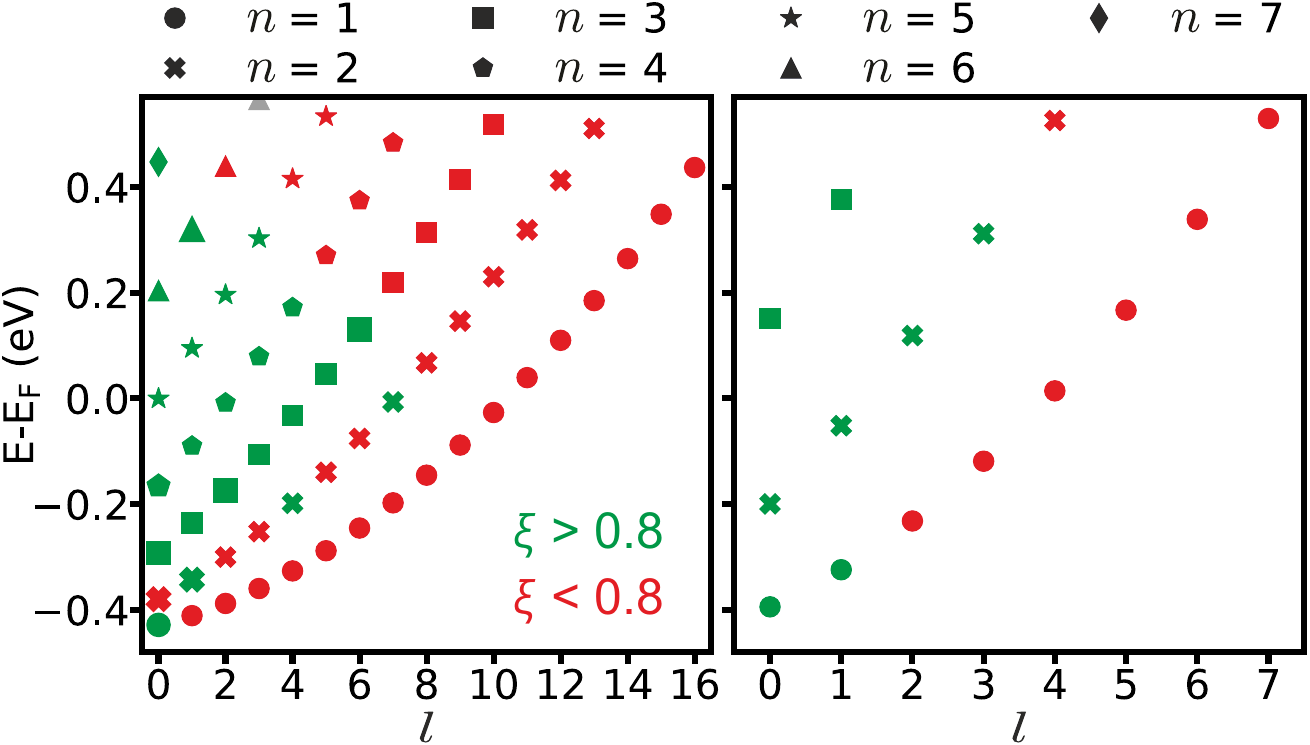}%
	\caption{\label{fig:goodornot} Energy
		spectra of the big (left) and small (right) corral. Red markers: Due to an insufficient quality these energy distributions were excluded form the line shape analysis.
		Green markers: Energy distributions
		with a sufficient quality factor $\xi$.} 
\end{figure}

In order to maximize the over all accuracy of the line shape analysis presented in the main text only distributions with a $\xi$-value between $1$ and $0.80$ were analyzed further.
Distributions with $\xi > 0.80$ are depicted in figure \ref{fig:goodornot} as green markers and low quality energy distributions ($\xi <0.80$) are shown as red.

\newpage

\section{Gaussian and Lorentzian components of the reconstructed energy distributions}
\begin{figure}[h]
	\centering
	\includegraphics[width=0.6\linewidth]{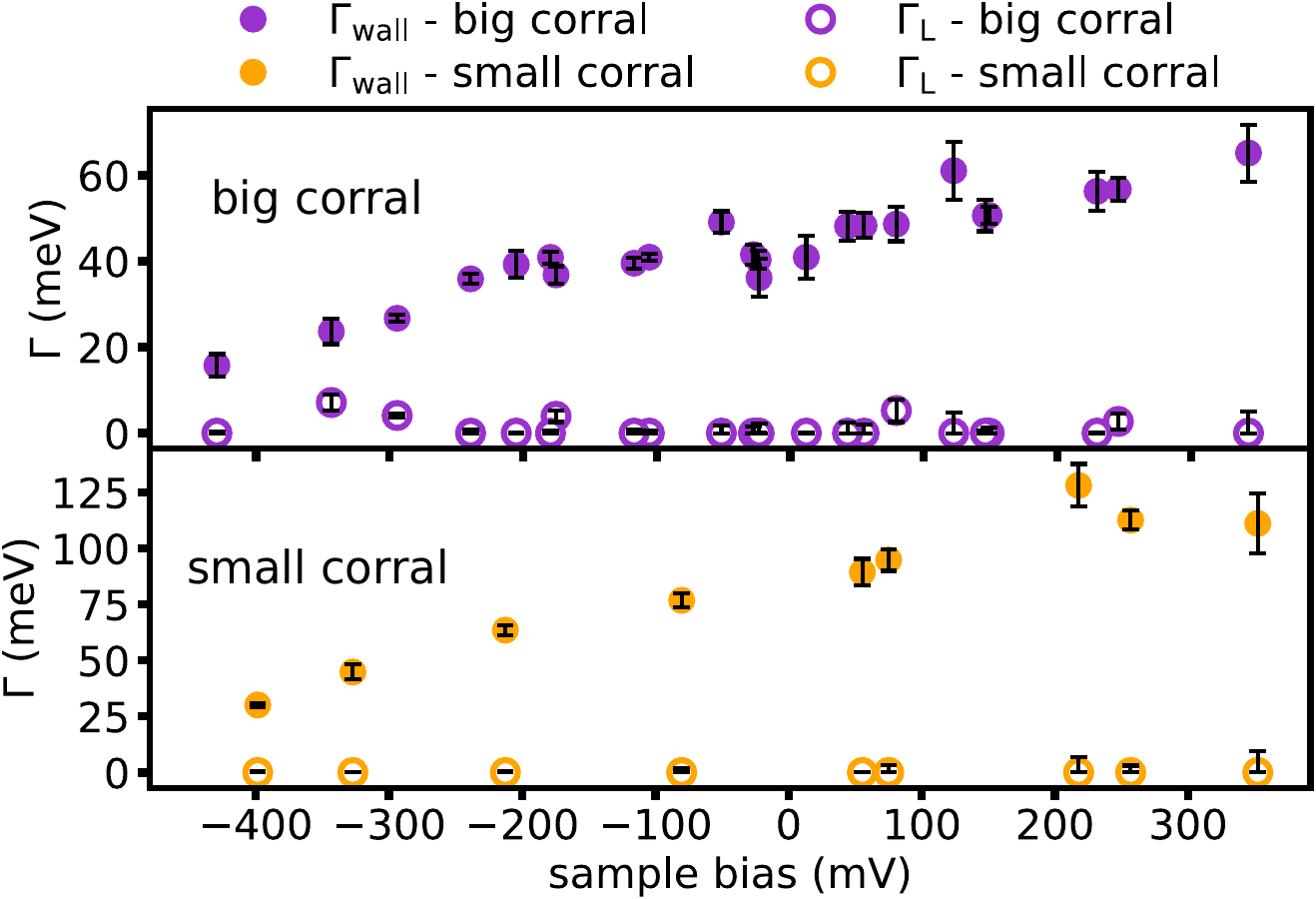}%
	\caption{\label{fig:comparison_Gauss_Lorentz}  Comparison of the width parameters $\Gamma_\mathrm{wall}$ \& $\Gamma_\mathrm{L}$ of the large ($R_{48\mathrm{,large}} = 7.13\,$nm) and small corral  ($R_{24\mathrm{,small}} =3.57\,$nm). This highlights that $\Gamma_\mathrm{wall}$ is the dominant component of the energy distributions and it depends on the size of the corral.} 
\end{figure}

The magnitudes of the Lorentzian component $\Gamma_\mathrm{L}$ and the Gaussian wall component $\Gamma_\mathrm{wall}$ are depicted in Figure \ref{fig:comparison_Gauss_Lorentz}. This visual representation illustrates that the line shapes of the reconstructed energy distributions for both the small and large corrals are primarily dictated by the Gaussian-shaped $\Gamma_\mathrm{wall}$.

\newpage

\section{Comparing line shapes determined by line scan fitting and static $\boldsymbol{\mathrm{d}} \boldsymbol{I} \boldsymbol{/} \boldsymbol{\mathrm{d}} \boldsymbol{V}$}

\begin{figure}[h]
	\centering
	\includegraphics[width=0.5\linewidth]{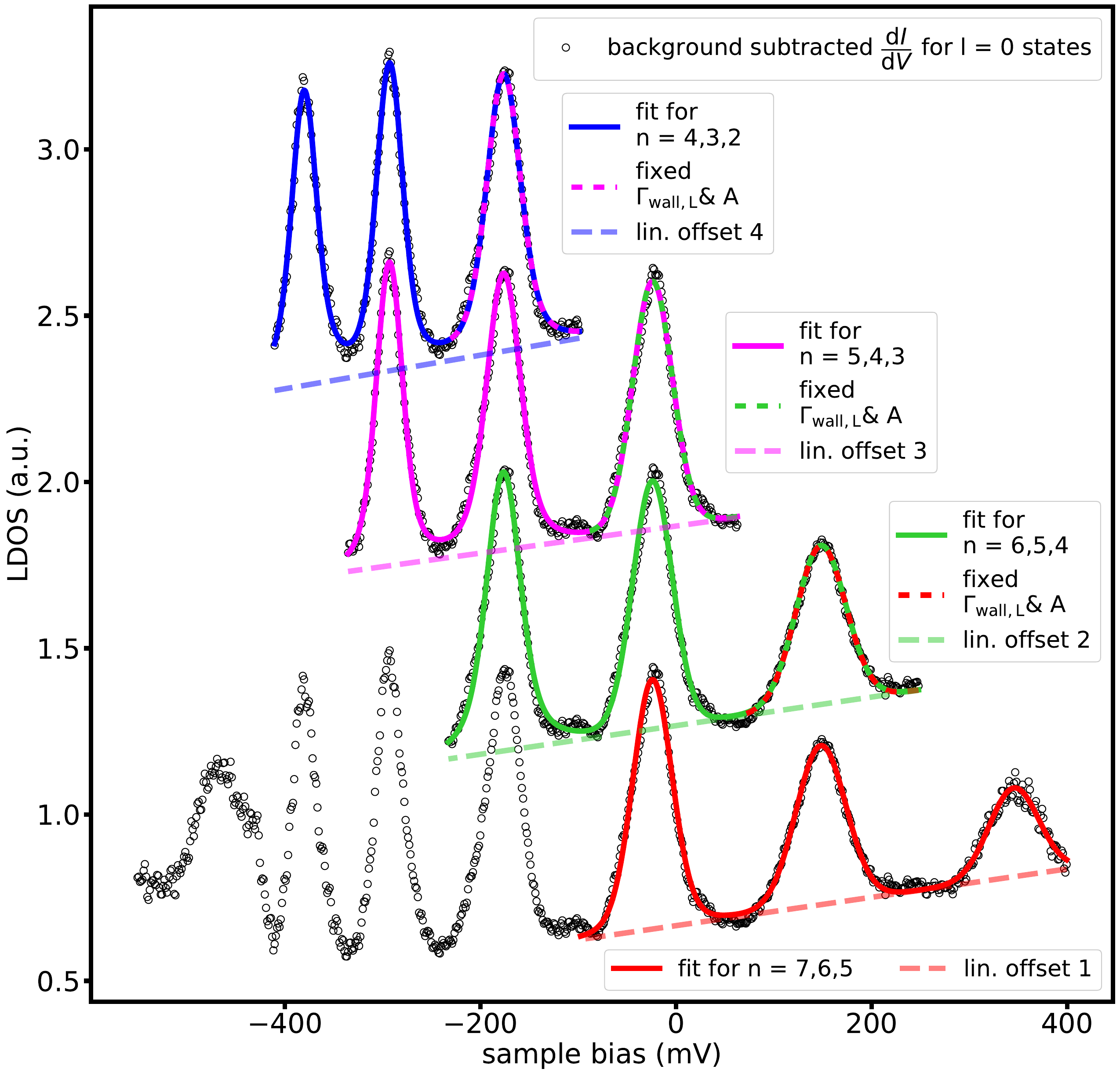}%
	\caption{\label{fig:sequential} Tunneling spectrum over the center of the big 48-corral after the background removal \cite{Wahl2008} (open black circles). The spectrum is normalized to $1$ at $0\,$mV sample bias. Peaks at higher energies are well separated (n=7,6,5). An offset which is increasing with energy is visible. The segmented fitting procedure is also depicted and explained in the text.}
\end{figure}

To further validate the results of our line scan fitting method, we conducted a comparison between the widths of the reconstructed energy distributions and peak widths obtained from a static d$I$/d$V$ measurement at the center of the big corral. This static measurement is particularly sensitive to the $l=0$ states. To eliminate the influence of the tip-DOS in the static measurement, we employed the full deconvolution method described by Wahl \textit{et al.} \cite{Wahl2008}, resulting in the background-removed spectrum shown in figure \ref{fig:sequential} as black unfilled markers. The background removed spectra are further normalized to 1 at $0\,$mV sample bias.

Upon close examination of this background-removed spectrum, we observed distinct separation of peaks at higher energies, particularly for the $n=5,6 \; \mathrm{\&}\;7$ states, with flat regions between the peaks. It appears that the individual peaks emerge from a smooth background that grows with energy.

To address this growing offset, a specialized fitting approach is required. Given our lack of knowledge regarding the energy dependence of this smooth offset, we analyzed the spectrum incrementally. Fitting a single peak is insufficient due to the potential influence of neighboring peaks. Therefore, we analyzed the first three peaks ($n=7, 6, 5$) using a fitting function comprising a linear offset plus three peaks as described by equation (3) of the main text. The resulting fit of this first peak sequence is represented by the red solid line in figure \ref{fig:sequential}.

Subsequently, we proceeded to the next set of peaks ($n=6, 5, 4$). Utilizing the parameters ($\Gamma_\mathrm{wall}$ , $\Gamma_\mathrm{L}$ and Amplitude $A$) obtained for the n=6 state from the previous fitting run, we fixed these parameters for the $n=6, 5, 4$ fit. This resulted in fitting two peaks and a linear offset, as illustrated by the green solid line in figure \ref{fig:sequential}. The green-red dashed part of this curve symbolizes the fixation of the parameters of the $n=6$ state from the previous fit set.

The same approach was repeated for the $n=5, 4, 3$ states, fixing the parameters of the $n=5$ state. The outcome of this fitting step is depicted as the magenta-colored full line in figure \ref{fig:sequential} with the fixing of $n=5$ depicted as the green-magenta dashed part. We continued this fitting mechanism until we analyzed the set of states $n=4, 3, 2$.

Unfortunately, due to the merging of the $n=1$ state with the band edge of the Shockley surface state during our background removal treatment, we could not include the $n=1$ state in our analysis.

\begin{figure}[h]
	\begin{minipage}[c]{\textwidth}
		\centering
		\begin{minipage}[c]{0.51\linewidth}
			\includegraphics[width=\linewidth]{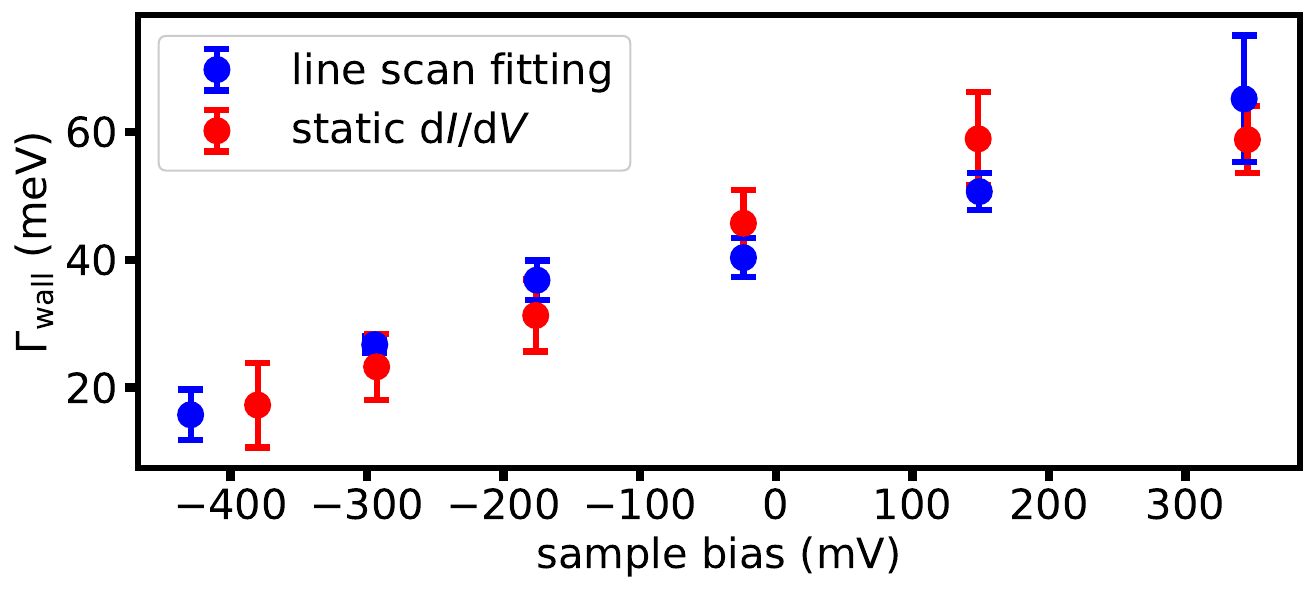}
		\end{minipage}
		\hfill
		\begin{minipage}[c]{0.51\linewidth}
			\includegraphics[width=\linewidth]{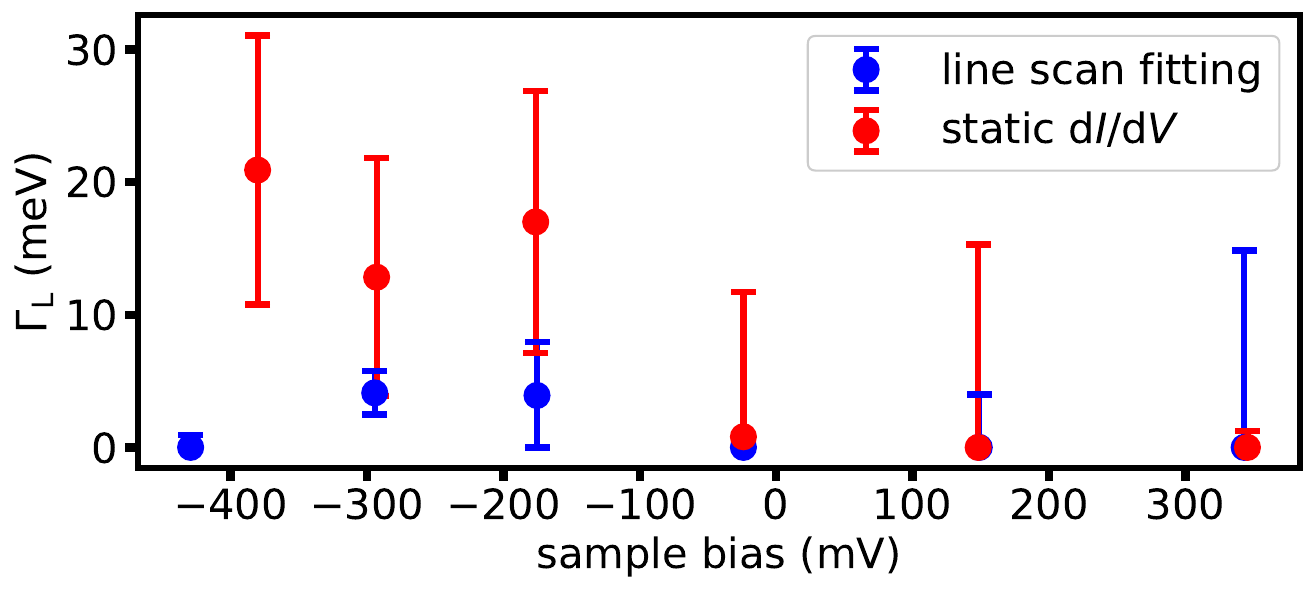}
		\end{minipage}
	\end{minipage}
	\caption{\label{fig:comparison} Comparison of Gaussian an Lorentzian width parameters ($\Gamma_\mathrm{wall}\,  \&\, \Gamma_\mathrm{L}$) obtained by line scan fitting (blue) and static dI/dV measurements (red) of the l=0 states in the big corral.}
\end{figure}

After completing this fitting procedure, we obtained width parameters for every peak in our static spectrum (except the $n=1$, $l=0$ state). 
Plotting the Gaussian $\Gamma_\mathrm{wall}$ of the static d$I$/d$V$ measurement results in the upper plot of figure \ref{fig:comparison} (red data points). Notably, these Gaussian components closely match the values obtained from the line scan fitting method (blue data points), within the margin of error. The same trend is also observed for the Lorentzian components, as shown in the lower pannel of figure \ref{fig:comparison}.

\newpage

\section{Energy distributions and fits for the big and small corral}\label{SM:distri_fit}

The fits of the distributions were done with equation (3) of the main text. Afterwards the quality value $\xi$ was determined as described in \ref{SM:roughness}. In this section all the reconstructed energy distributions with $\xi \geq 0.80$ are presented.

\subsection{$l=0$ states - big corral}

\begin{figure}[h]
	\begin{minipage}[c]{\textwidth}
		\centering
		\begin{minipage}[t][2cm][t]{0.325\linewidth}
			\includegraphics[height=5.6cm]{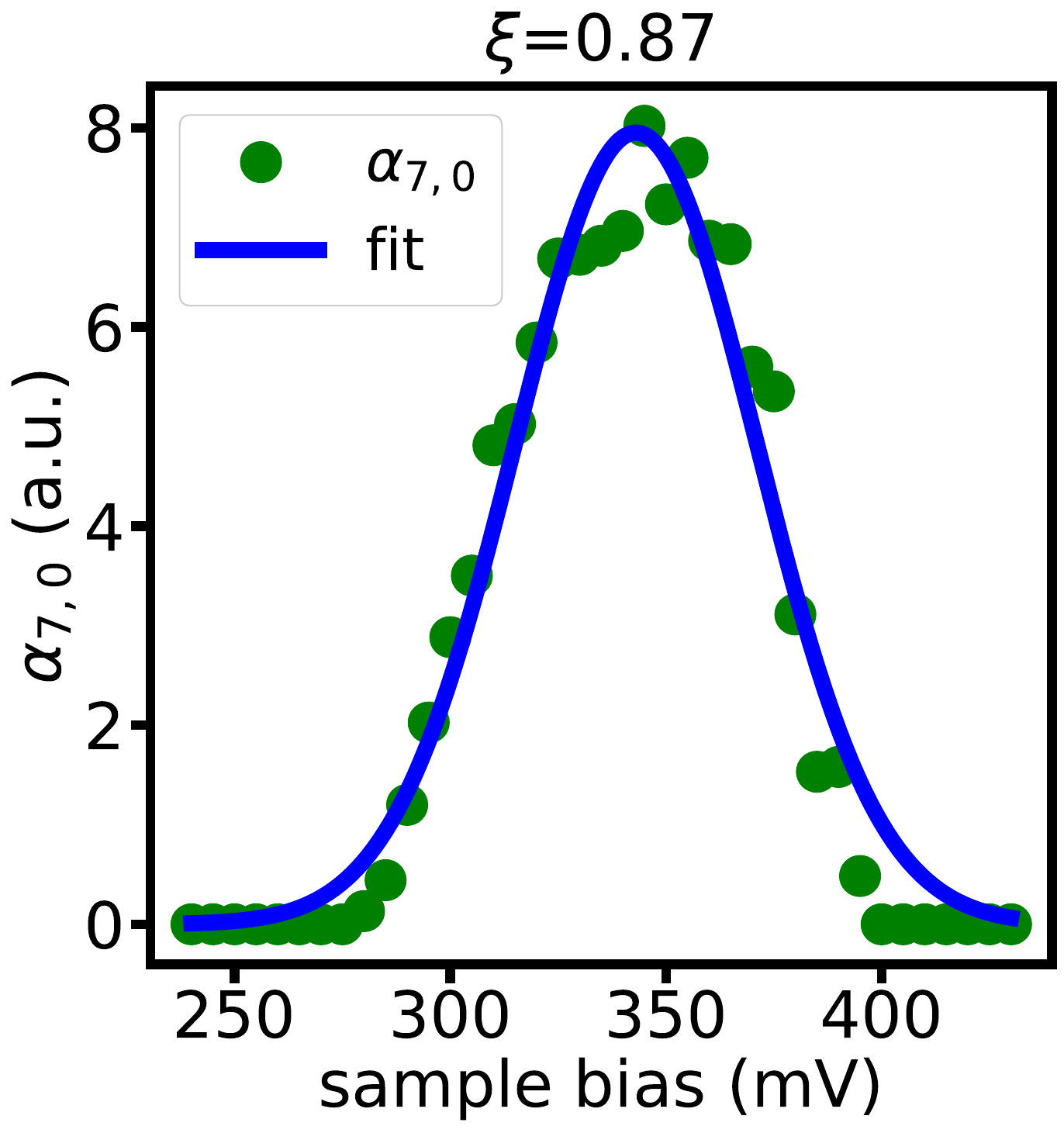}
		\end{minipage}
		\hfill
		\begin{minipage}[t][2cm][t]{0.325\linewidth}
			\includegraphics[height=5.6cm]{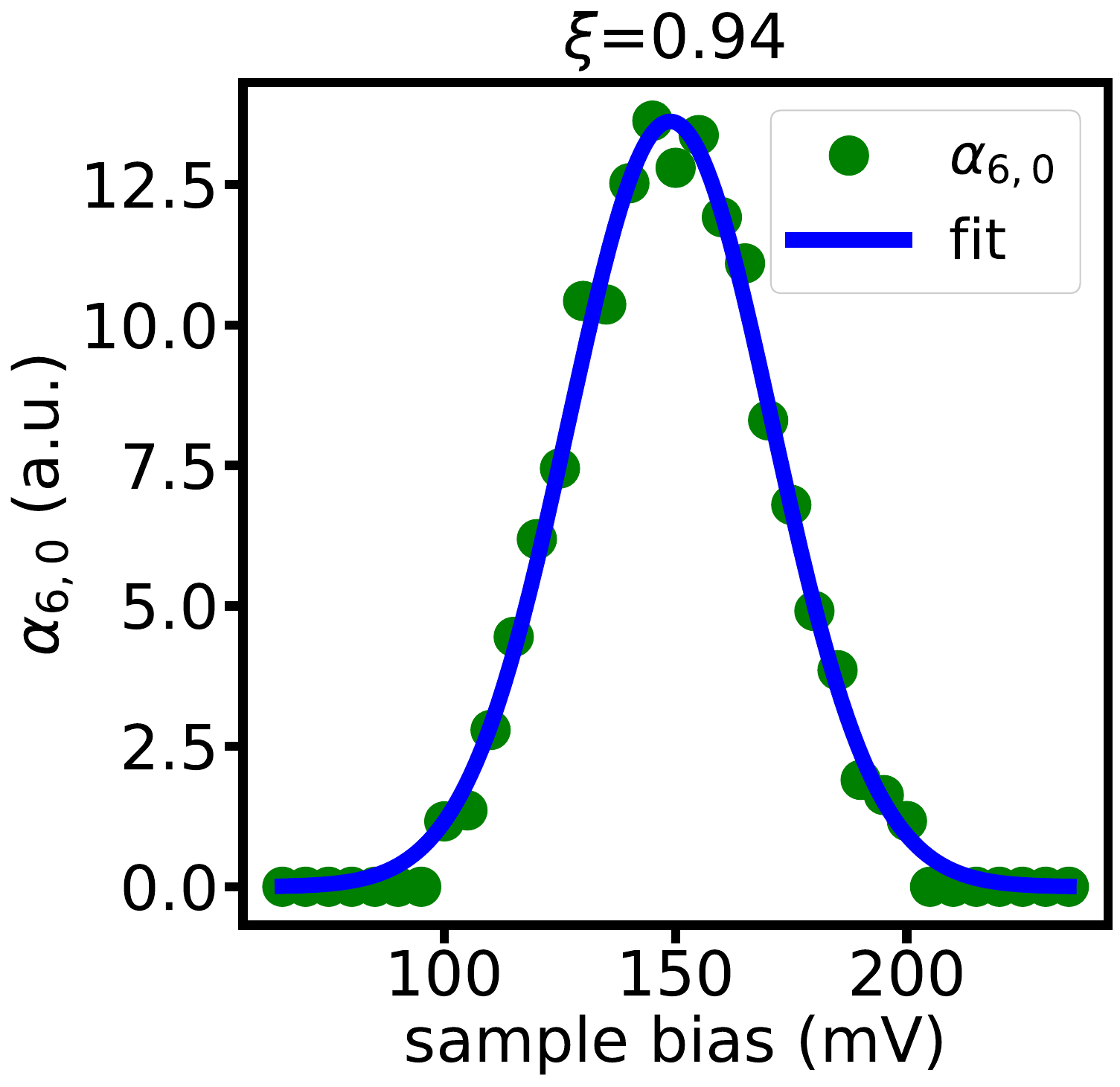}
		\end{minipage}
		\hfill
		\begin{minipage}[t][2cm][t]{0.325\linewidth}
			\includegraphics[height=5.6cm]{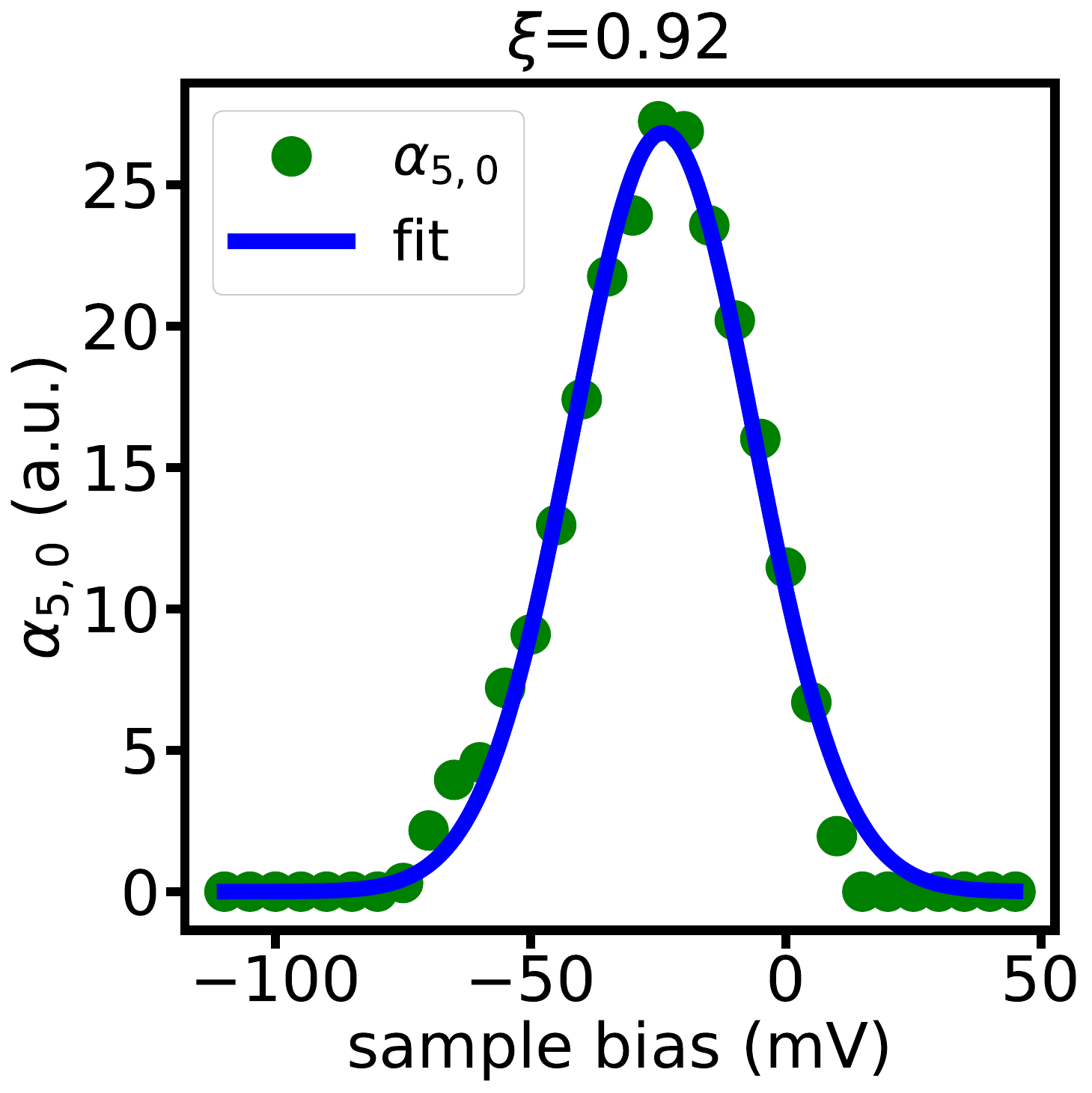}
		\end{minipage}
		\hfill
		\vspace{0,5cm}
		
		\begin{minipage}[c]{0.325\linewidth}
			\includegraphics[height=5.6cm]{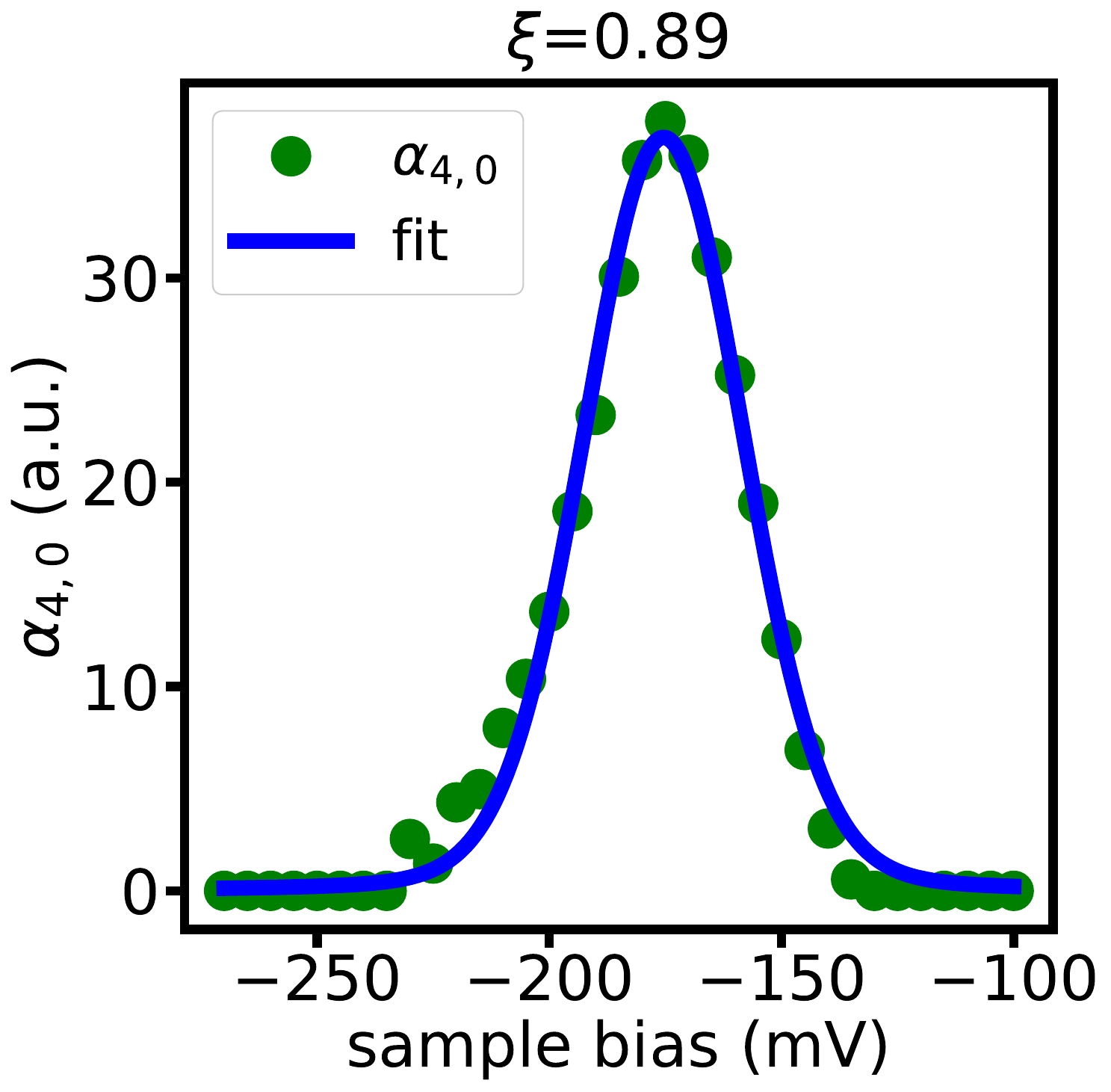}
		\end{minipage}
		\hfill
		\begin{minipage}[c]{0.325\linewidth}
			\includegraphics[height=5.6cm]{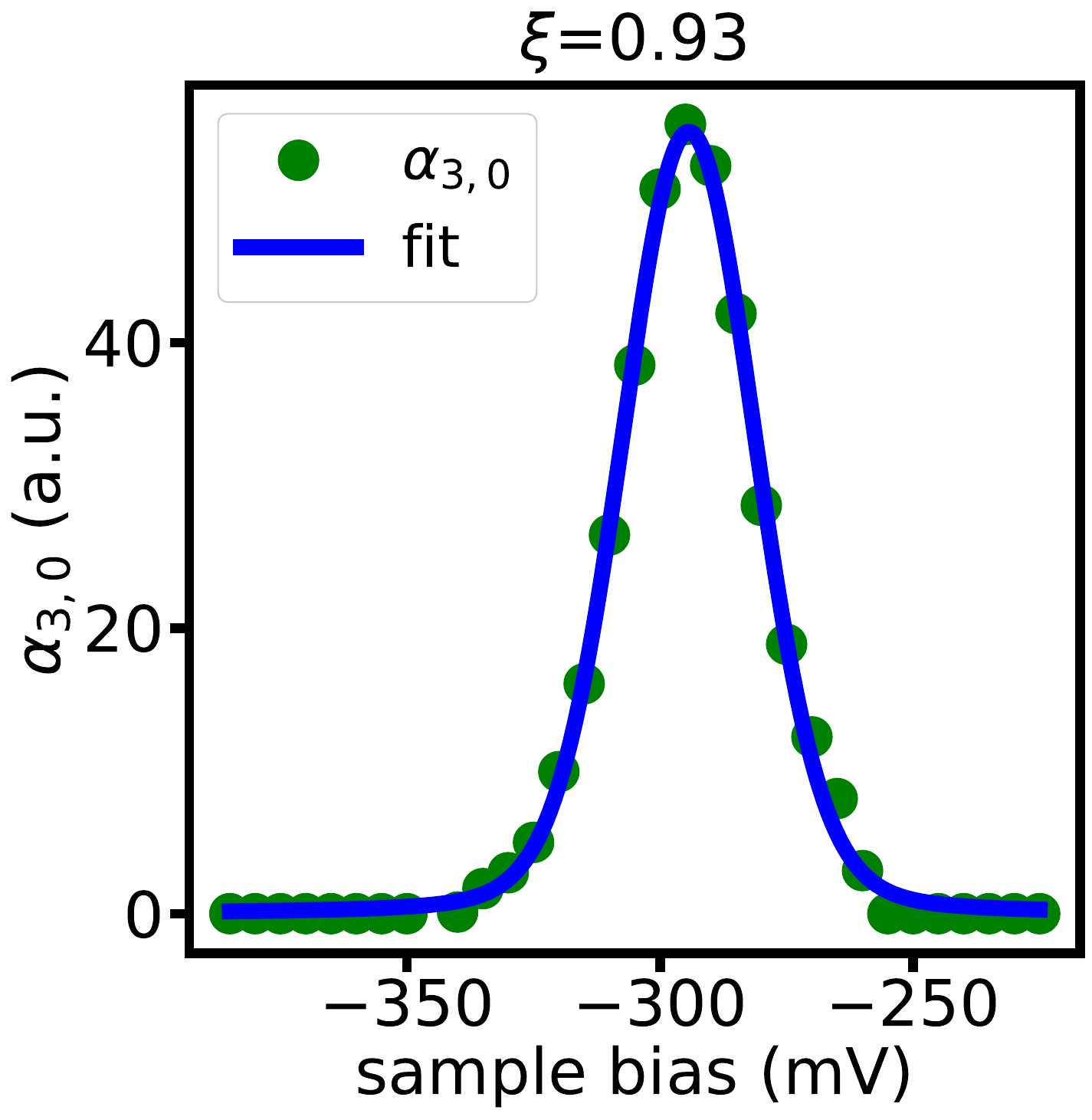}
		\end{minipage}
		\hfill
		\begin{minipage}[c]{0.325\linewidth}
			\includegraphics[height=5.6cm]{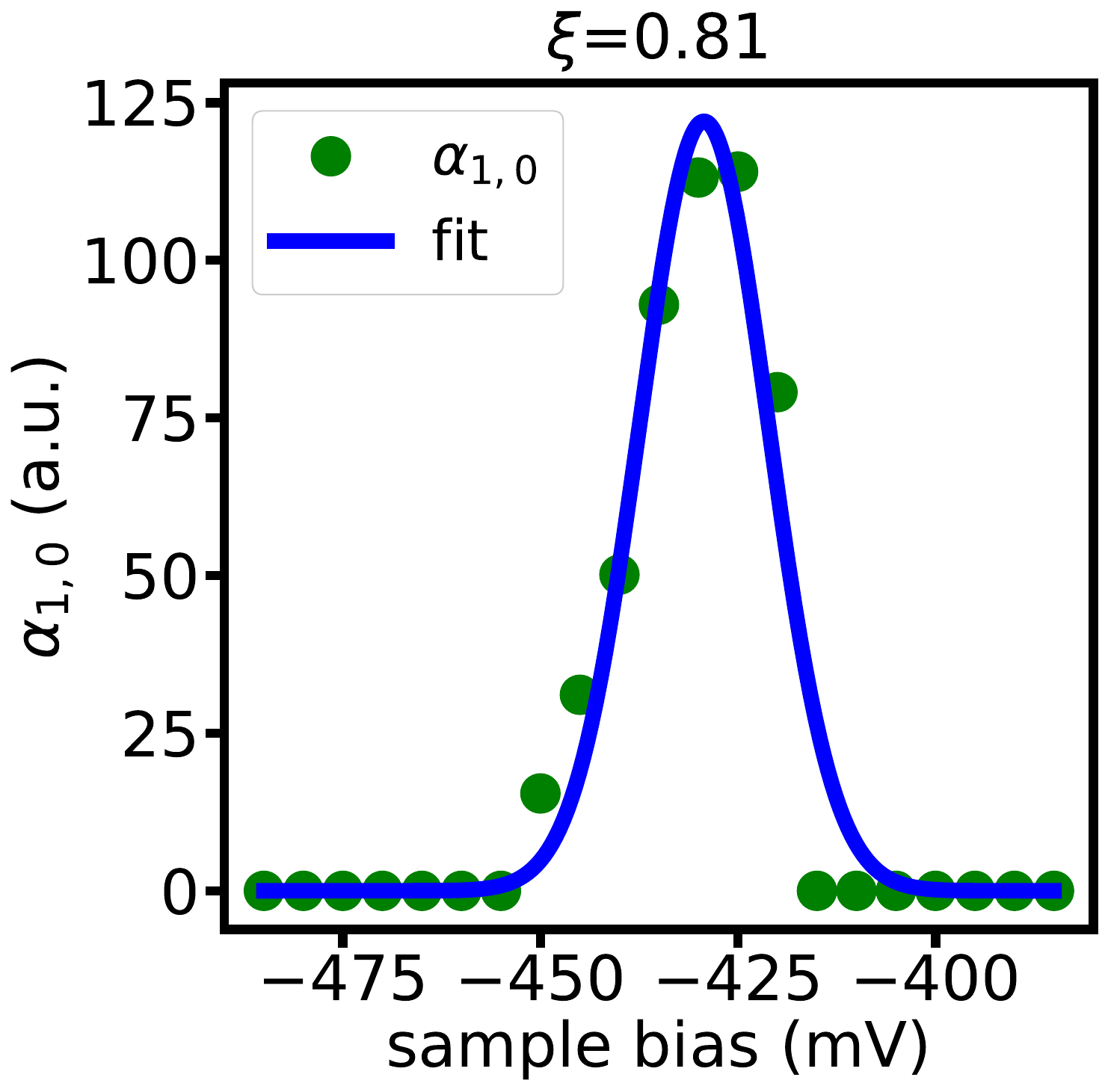}
		\end{minipage}	
		\hfill
		\vspace{0,5cm}
		
	\end{minipage}
	\caption{Reconstructed energy distributions for the $l = 0$ states (green dots). For the detailed line shape analysis equation (3) (main text) was fitted to the distributions. The fit is depicted with a full, blue line. The quality value $\xi$  can be found above each plot and was calculated as described in \ref{SM:roughness}.}
\end{figure}

\newpage

\begin{table}[H]
	\begin{minipage}[c]{\linewidth}
		\centering
		\begin{minipage}[c]{0.325\linewidth}
			\begin{tabular}{|c|c|}\hline
				
				$n=7,l=0$ & fit (meV)  \\ \hline
				
				$\Gamma_\mathrm{wall}$ & $65 \pm 7$ \\
				
				$\Gamma_\mathrm{L}$ & $0 \varpm  10$\\
				
				$E_\mathrm{7,0}$ & \textcolor{white}{$-$}$343.0 \pm 0.8$ \\ \hline
			\end{tabular}
		\end{minipage}
		\begin{minipage}[c]{0.325\linewidth}
			\begin{tabular}{|c|c|}\hline
				
				$n=6, l=0$ & fit (meV)  \\ \hline
				
				$\Gamma_\mathrm{wall}$ & $51 \pm 2$ \\
				
				$\Gamma_\mathrm{L}$ & $0 \varpm  3$ \\
				
				$E_\mathrm{6,0}$ &\textcolor{white}{$-$}$148.9 \pm 0.3$ \\ \hline
			\end{tabular}
		\end{minipage}
		\begin{minipage}[c]{0.325\linewidth}
			\begin{tabular}{|c|c|}\hline
				
				$n=5,l=0$ & fit (meV)  \\ \hline
				
				$\Gamma_\mathrm{wall}$ & $40 \pm 2$ \\
				
				$\Gamma_\mathrm{L}$ & $0 \varpm  1$ \\
				
				$E_\mathrm{5,0}$ &\textcolor{white}{$0$}$-24.0 \pm 0.3$ \\ \hline
			\end{tabular}
		\end{minipage}
		
		\begin{minipage}[c]{\linewidth}
			\vspace{0.3cm}
		\end{minipage}
		
		\begin{minipage}[c]{0.325\linewidth}
			\begin{tabular}{|c|c|}\hline
				
				$n=4, l=0$ & fit (meV)  \\ \hline
				
				$\Gamma_\mathrm{wall}$ & $37 \pm 2$ \\
				
				$\Gamma_\mathrm{L}$ & $4 \pm 3$ \\
				
				$E_\mathrm{4,0}$ & $-175.4 \pm 0.3$ \\ \hline
			\end{tabular}
		\end{minipage}
		\begin{minipage}[c]{0.325\linewidth}
			\begin{tabular}{|c|c|}\hline
				
				$n=3, l=0$ & fit (meV)  \\ \hline
				
				$\Gamma_\mathrm{wall}$ & $27 \pm 1$ \\
				
				$\Gamma_\mathrm{L}$ & $4 \pm 1$ \\
				
				$E_\mathrm{3,0}$ & $-294.3  \pm 0.2$ \\ \hline
			\end{tabular}
		\end{minipage}
		\begin{minipage}[c]{0.325\linewidth}
			\begin{tabular}{|c|c|}\hline
				
				$n=1, l=0$ & fit (meV)  \\ \hline
				
				$\Gamma_\mathrm{wall}$ & $16 \pm 3$ \\
				
				$\Gamma_\mathrm{L}$ & $0 \varpm  1$ \\
				
				$E_\mathrm{1,0}$ & $-429.3  \pm 0.4$ \\ \hline
			\end{tabular}
		\end{minipage}
		
		\begin{minipage}[c]{\linewidth}
			\vspace{0.3cm}
		\end{minipage}
		
		
		
		
		
	\end{minipage}
	\caption{Results for fitting equation (3) of the main text to the reconstructed  $l=0$ energy distributions of the big corral.}
\end{table}

\newpage

\subsection{$l=1$ states - big corral}
\begin{figure}[H]
	\begin{minipage}[c]{\textwidth}
		\centering
		\begin{minipage}[c]{0.325\linewidth}
			\includegraphics[height=5.6cm]{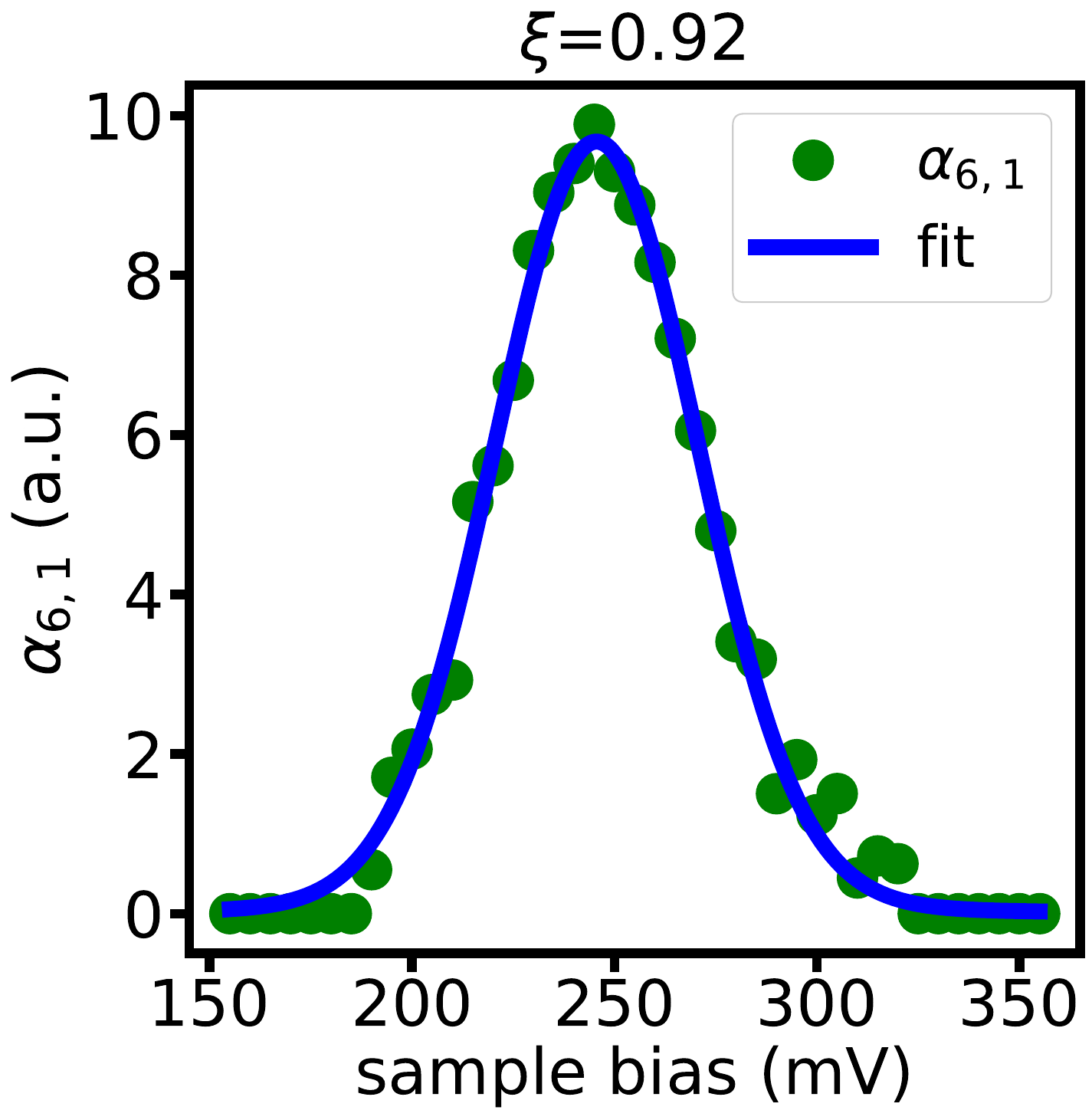}
		\end{minipage}
		\hfill
		\begin{minipage}[c]{0.325\linewidth}
			\includegraphics[height=5.6cm]{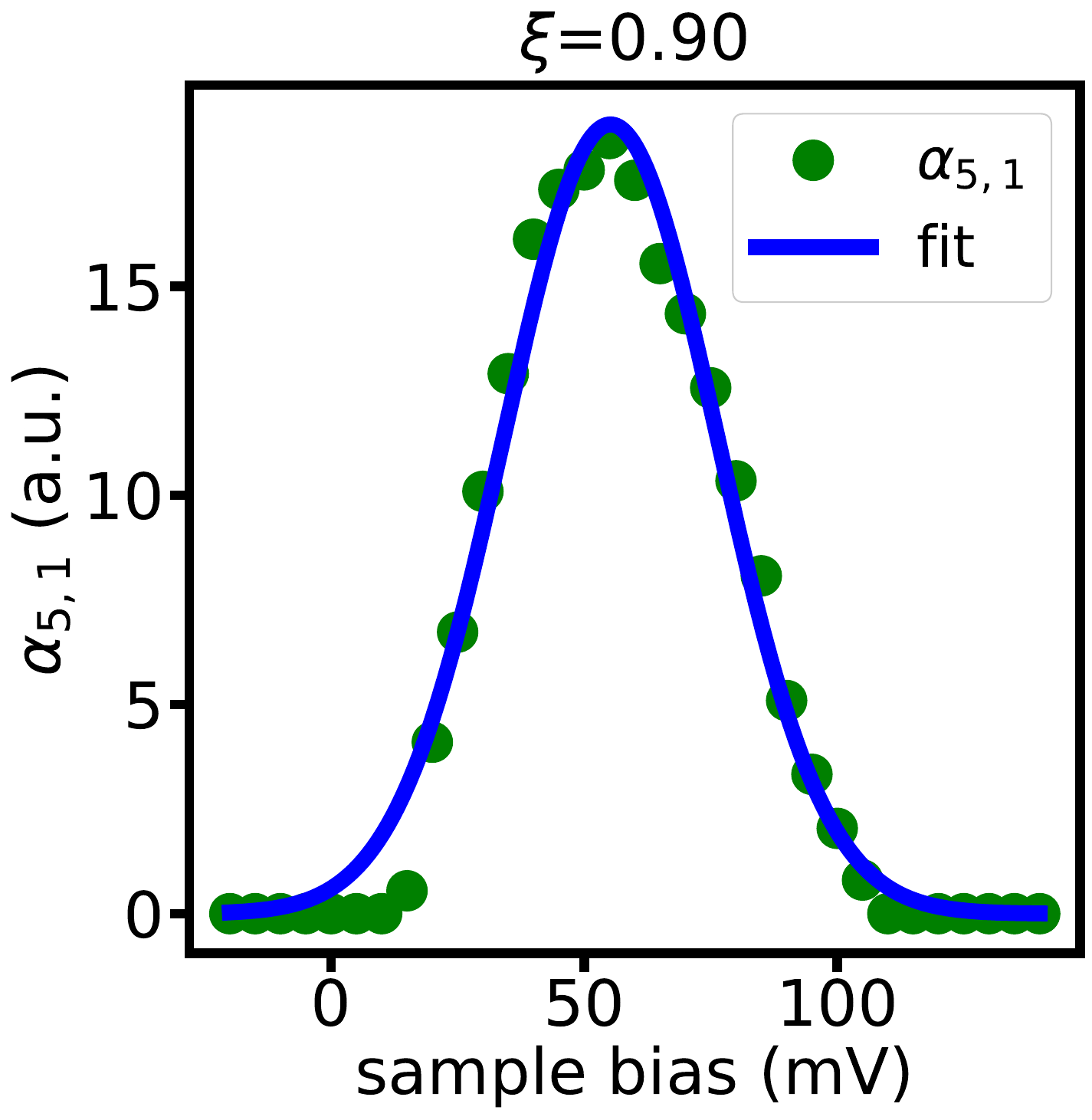}
		\end{minipage}
		\hfill
		\begin{minipage}[c]{0.325\linewidth}
			\includegraphics[height=5.6cm]{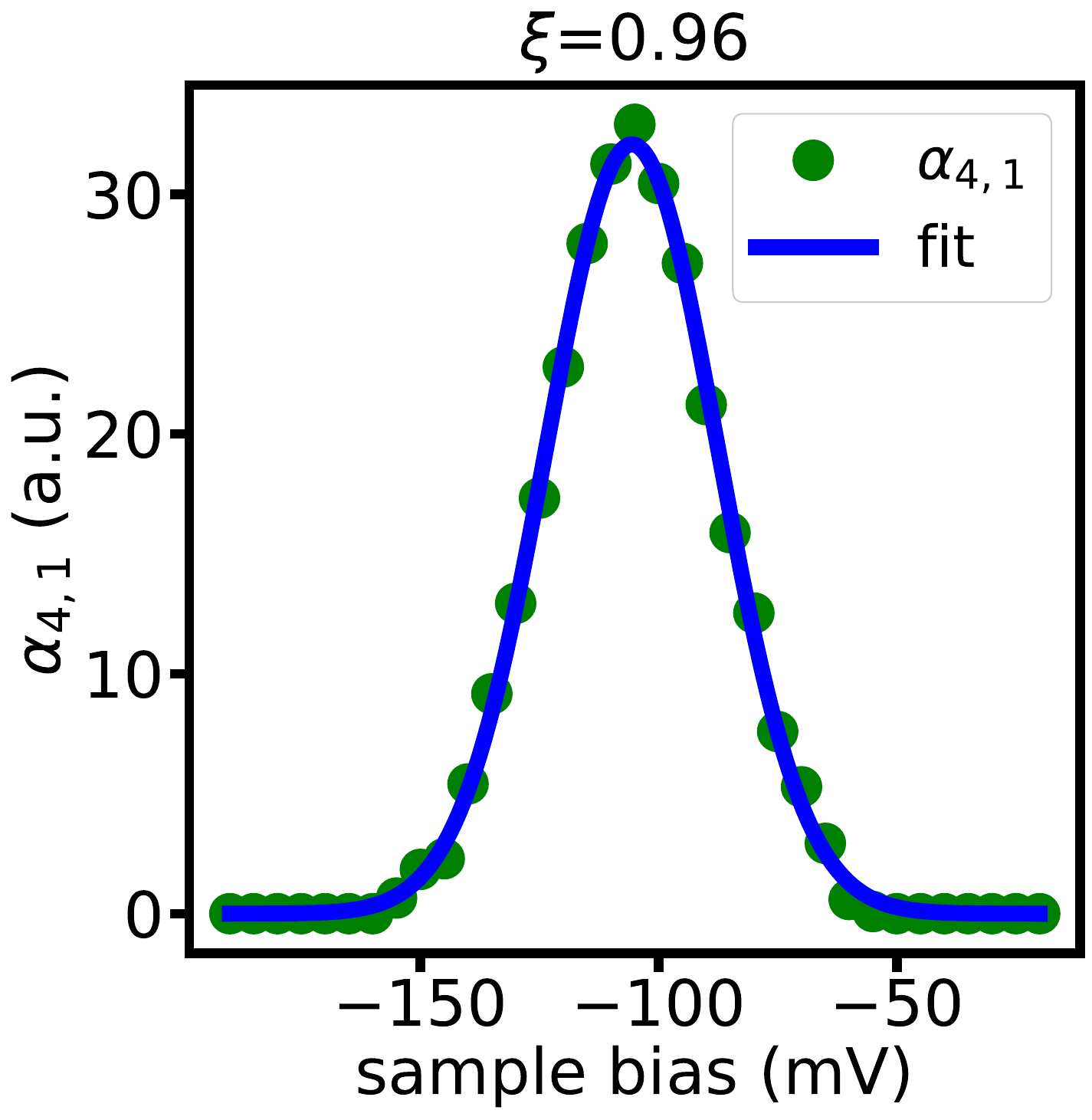}
		\end{minipage}
		\hfill		
		\vspace{0.5cm}
		
		\begin{minipage}[c]{0.325\linewidth}
			\includegraphics[height=5.6cm]{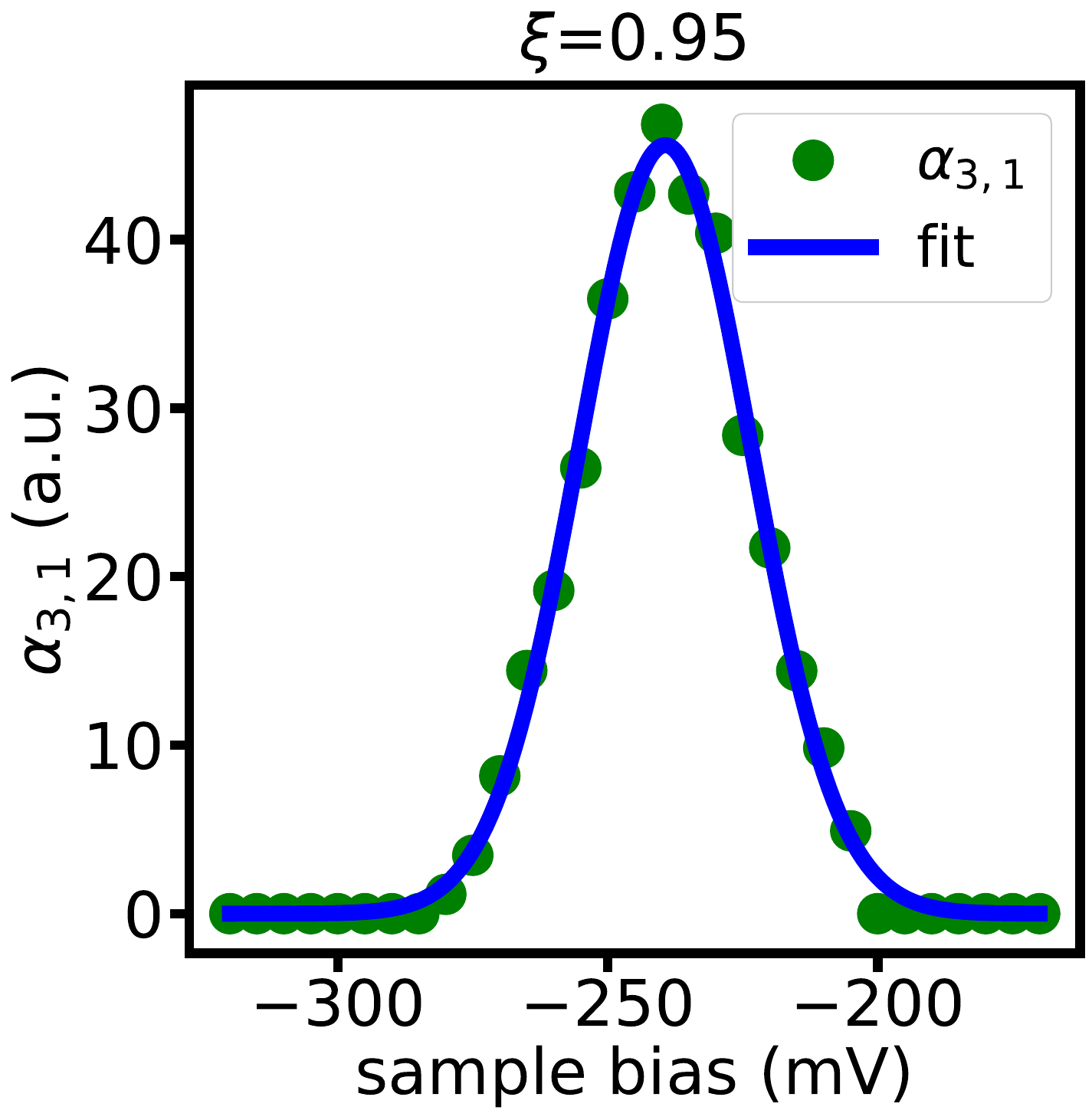}
		\end{minipage}
		\begin{minipage}[c]{0.325\linewidth}
			\includegraphics[height=5.6cm]{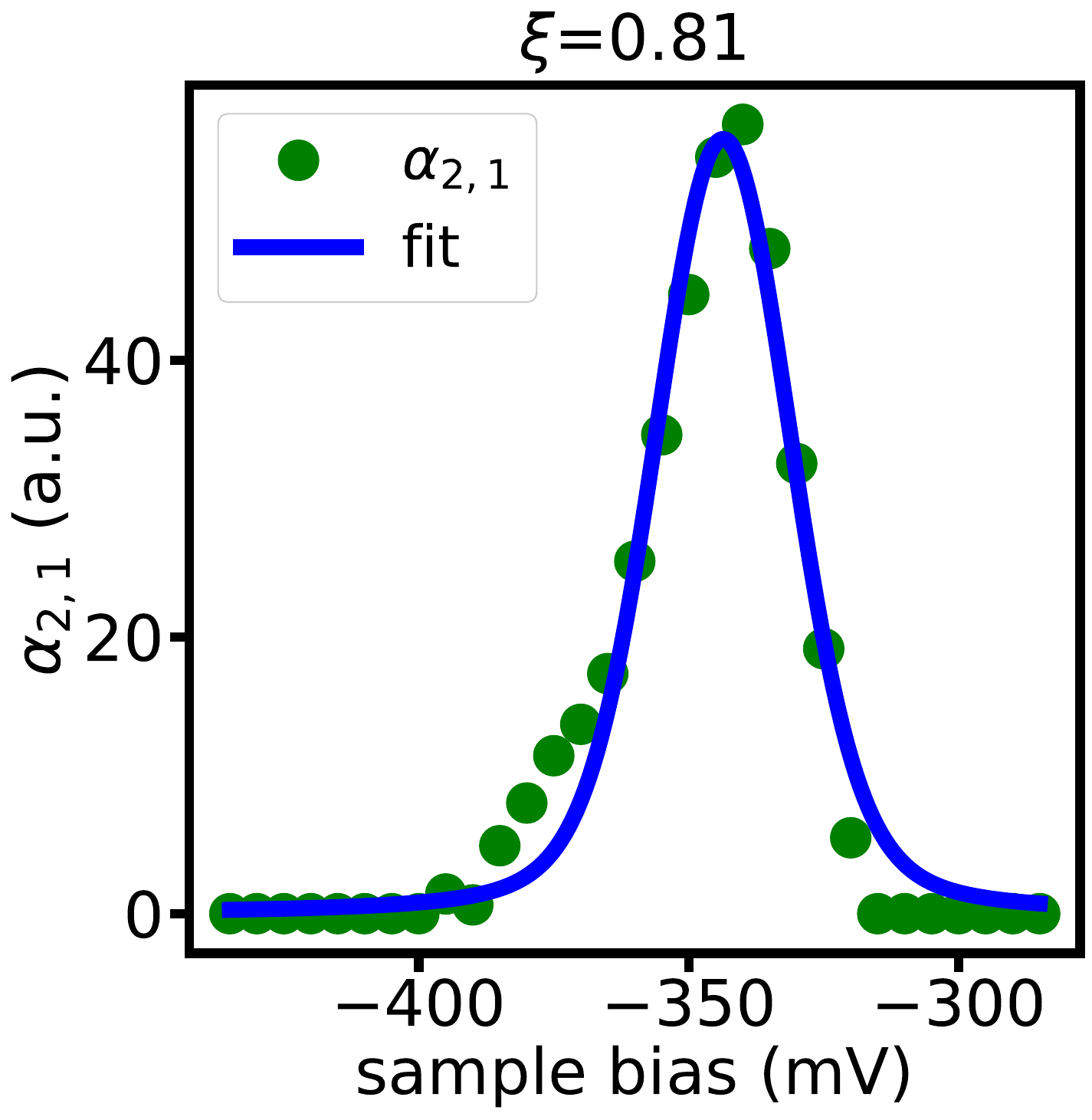}
		\end{minipage}
	\end{minipage}
	\caption{Reconstructed energy distributions for the $l = 1$ states (green dots). For the detailed line shape analysis equation (3) (main text) was fitted to the distributions. The fit is depicted with a full, blue line. The quality value $\xi$  can be found above each plot and was calculated as described in \ref{SM:roughness}.}
\end{figure}

\begin{table}[H]
	\begin{minipage}[c]{\linewidth}
		\centering
		\begin{minipage}[c]{0.325\linewidth}
			\begin{tabular}{|c|c|}\hline
				
				$n=6,l=1$ & fit (meV)  \\ \hline
				
				$\Gamma_\mathrm{wall}$ & $57 \pm 3$ \\
				
				$\Gamma_\mathrm{L}$ & $3 \pm  3$ \\
				
				$E_\mathrm{6,1}$ & \textcolor{white}{$-$}$245.6  \pm 0.4$ \\ \hline
			\end{tabular}
		\end{minipage}
		\begin{minipage}[c]{0.325\linewidth}
			\begin{tabular}{|c|c|}\hline
				
				$n=5, l=1$ & fit (meV)  \\ \hline
				
				$\Gamma_\mathrm{wall}$ & $48 \pm 3$ \\
				
				$\Gamma_\mathrm{L}$ & $0 \varpm  4$ \\
				
				$E_\mathrm{5,1}$ &\textcolor{white}{$-$}$55.2 \pm 0.5$ \\ \hline
			\end{tabular}
		\end{minipage}
		\begin{minipage}[c]{0.325\linewidth}
			\begin{tabular}{|c|c|}\hline
				
				$n=4, l=1$ & fit (meV)  \\ \hline
				
				$\Gamma_\mathrm{wall}$ & $41 \pm 1$ \\
				
				$\Gamma_\mathrm{L}$ & $0 \varpm  2$ \\
				
				$E_\mathrm{4,1}$ & $-105.6 \pm 0.2$ \\ \hline
			\end{tabular}
		\end{minipage}
		
		\begin{minipage}[c]{\linewidth}
			\vspace{0.3cm}
		\end{minipage}
		
		\begin{minipage}[c]{0.325\linewidth}
			\begin{tabular}{|c|c|}\hline
				
				$n=3, l=1$ & fit (meV)  \\ \hline
				
				$\Gamma_\mathrm{wall}$ & $36 \pm 2$ \\
				
				$\Gamma_\mathrm{L}$ & $0 \varpm  2$ \\
				
				$E_\mathrm{3,1}$ & $-239.4 \pm 0.2$ \\ \hline
			\end{tabular}
		\end{minipage}
		\begin{minipage}[c]{0.325\linewidth}
			\begin{tabular}{|c|c|}\hline
				
				$n=2, l=1$ & fit (meV)  \\ \hline
				
				$\Gamma_\mathrm{wall}$ & $24 \pm 3$ \\
				
				$\Gamma_\mathrm{L}$ & $7 \pm 4$ \\
				
				$E_\mathrm{2,1}$ & $-343.6  \pm 0.4$ \\ \hline
			\end{tabular}
		\end{minipage}
	\end{minipage}
	\caption{Results for fitting equation (3) of the main text to the reconstructed  $l=1$ energy distributions of the big corral.}
\end{table}

\newpage

\subsection{$l=2$ states - big corral}
\begin{figure}[H]
	\begin{minipage}[c]{\textwidth}
		\centering
		\begin{minipage}[c]{0.325\linewidth}
			\includegraphics[height=5.6cm]{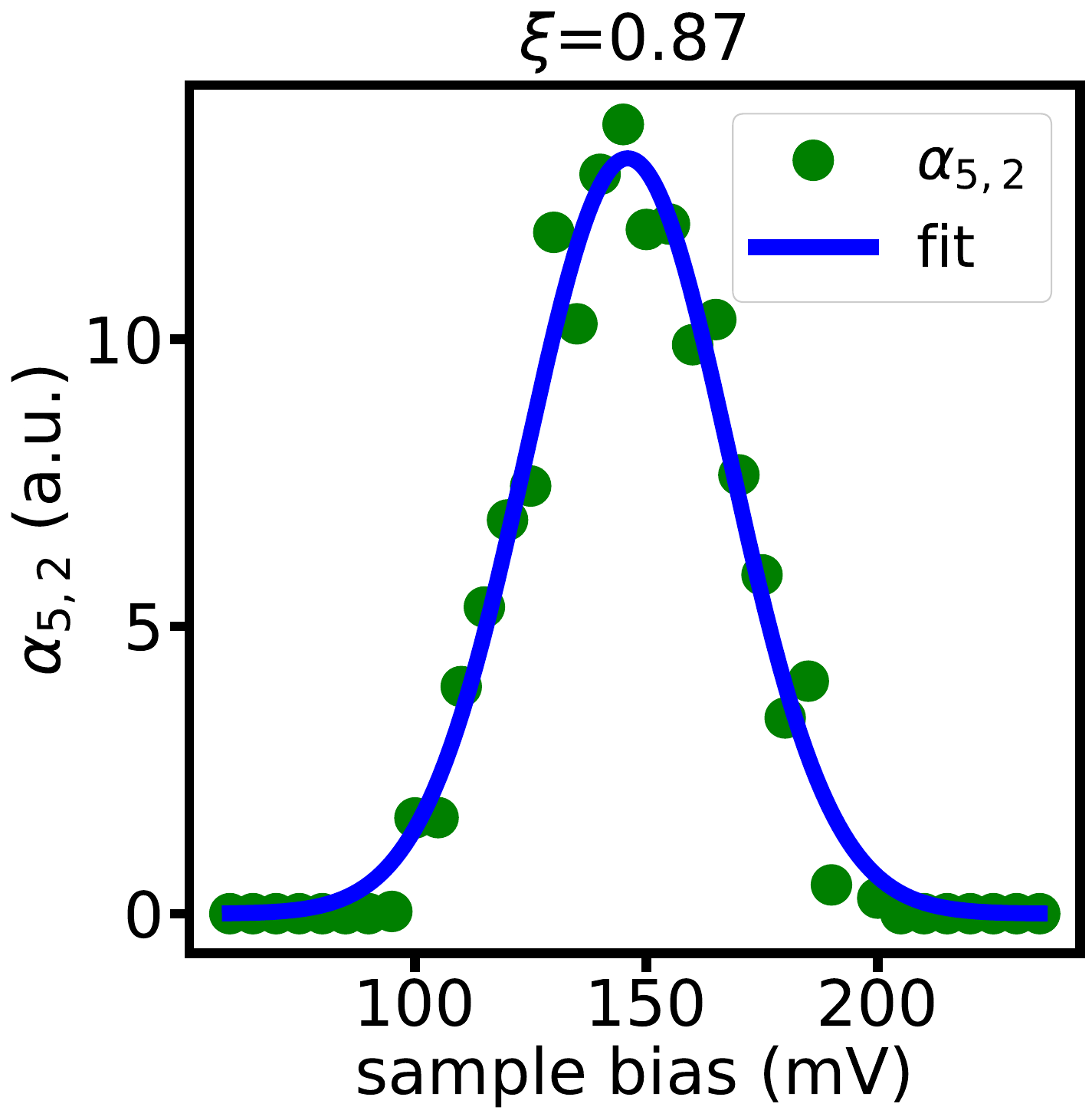}
		\end{minipage}
		\hfill
		\begin{minipage}[c]{0.325\linewidth}
			\includegraphics[height=5.6cm]{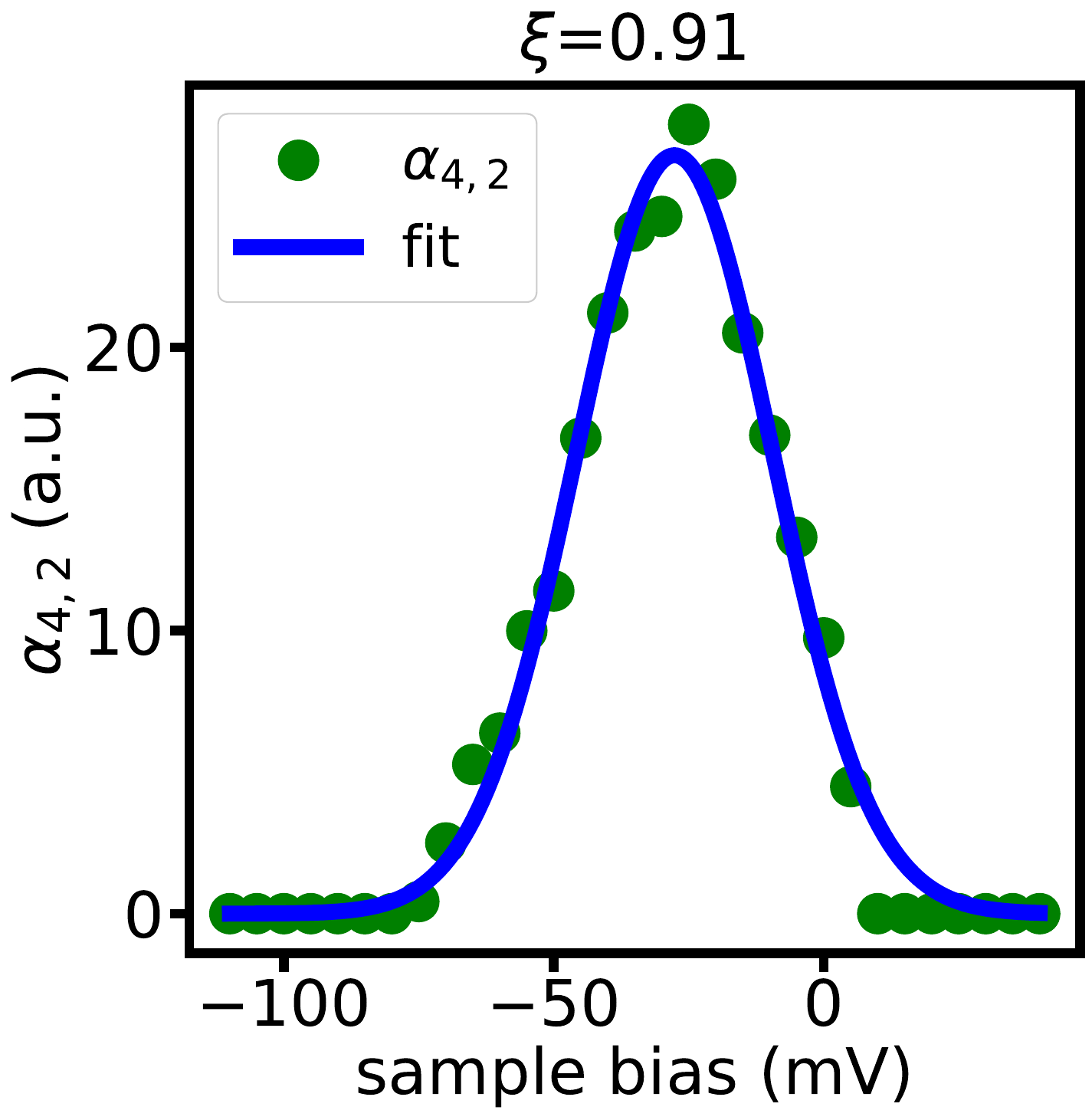}
		\end{minipage}
		\hfill
		\begin{minipage}[c]{0.325\linewidth}
			\includegraphics[height=5.6cm]{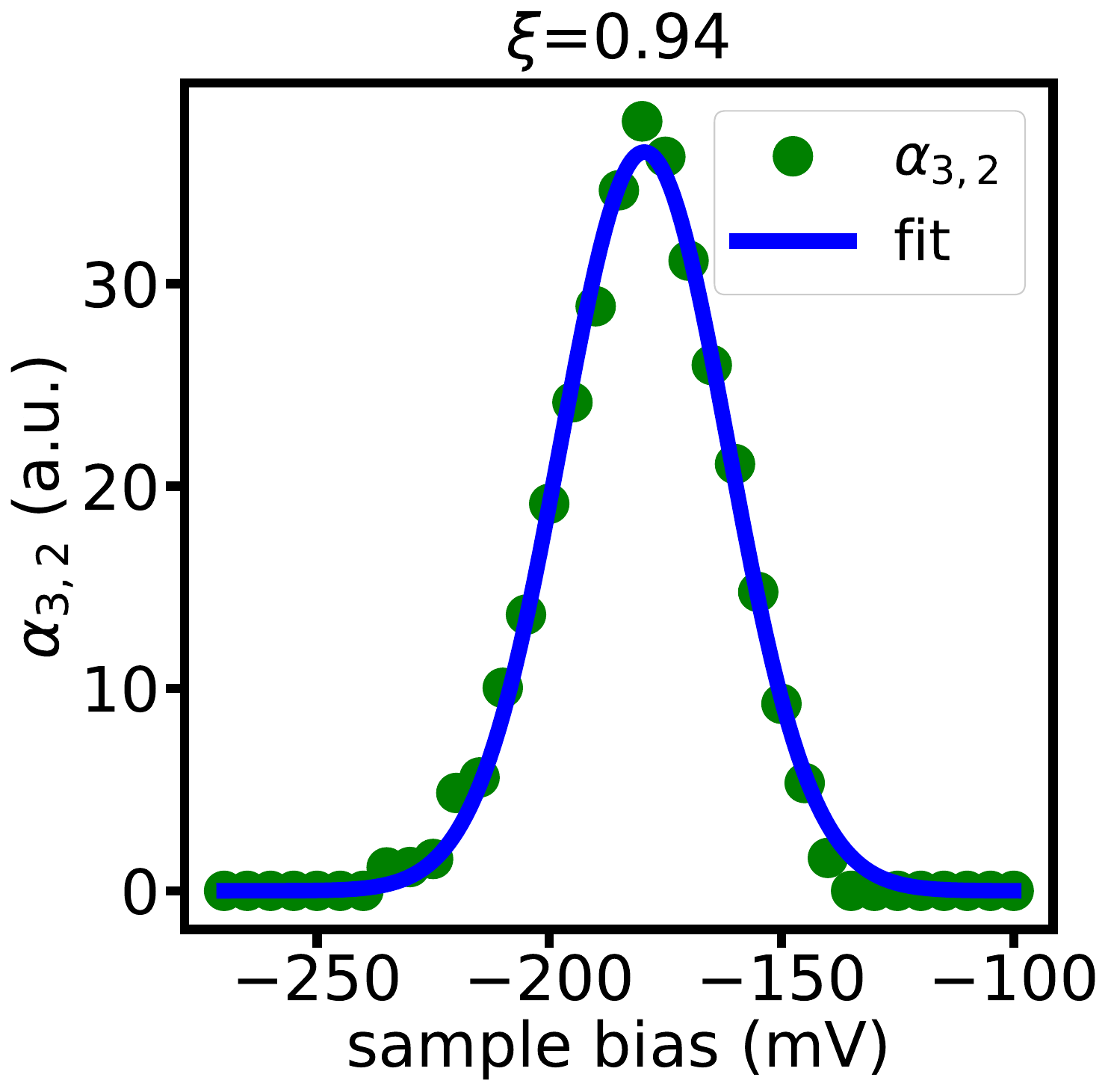}
		\end{minipage}
		\hfill
	\end{minipage}
	\caption{Reconstructed energy distributions for the $l = 2$ states (green dots). For the detailed line shape analysis equation (3) (main text) was fitted to the distributions. The fit is depicted with a full, blue line. The quality value $\xi$  can be found above each plot and was calculated as described in \ref{SM:roughness}.}
\end{figure}

\begin{table}[H]
	\begin{minipage}[c]{\linewidth}
		\centering
		\begin{minipage}[c]{0.325\linewidth}
			\begin{tabular}{|c|c|}\hline
				
				$n=5, l=2$ & fit (meV)  \\ \hline
				
				$\Gamma_\mathrm{wall}$ & $51 \pm 4$ \\
				
				$\Gamma_\mathrm{L}$ & $0 \varpm  2$ \\
				
				$E_\mathrm{5,2}$ &\textcolor{white}{$-$}$145.9 \pm 0.6$ \\ \hline
			\end{tabular}
		\end{minipage}
		\begin{minipage}[c]{0.325\linewidth}
			\begin{tabular}{|c|c|}\hline
				
				$n=4, l=2$ & fit (meV)  \\ \hline
				
				$\Gamma_\mathrm{wall}$ & $42 \pm 3$ \\
				
				$\Gamma_\mathrm{L}$ & $0 \varpm  4$ \\
				
				$E_\mathrm{4,1}$ & $-27.6 \pm 0.4$ \\ \hline
			\end{tabular}
		\end{minipage}
		\begin{minipage}[c]{0.325\linewidth}
			\begin{tabular}{|c|c|}\hline
				
				$n=3, l=2$ & fit (meV)  \\ \hline
				
				$\Gamma_\mathrm{wall}$ & $41 \pm 2$ \\
				
				$\Gamma_\mathrm{L}$ & $0\varpm 1 $ \\
				
				$E_\mathrm{3,2}$ & $-179.5  \pm 0.2$ \\ \hline
			\end{tabular}
		\end{minipage}
	\end{minipage}
	\caption{Results for fitting equation (3) of the main text to the reconstructed  $l=2$ energy distributions of the big corral.}
\end{table}

\newpage

\subsection{$l=3$ states - big corral}

\begin{figure}[H]
	\begin{minipage}[c]{\textwidth}
		\centering
		\begin{minipage}[c]{0.325\linewidth}
			\includegraphics[height=5.6cm]{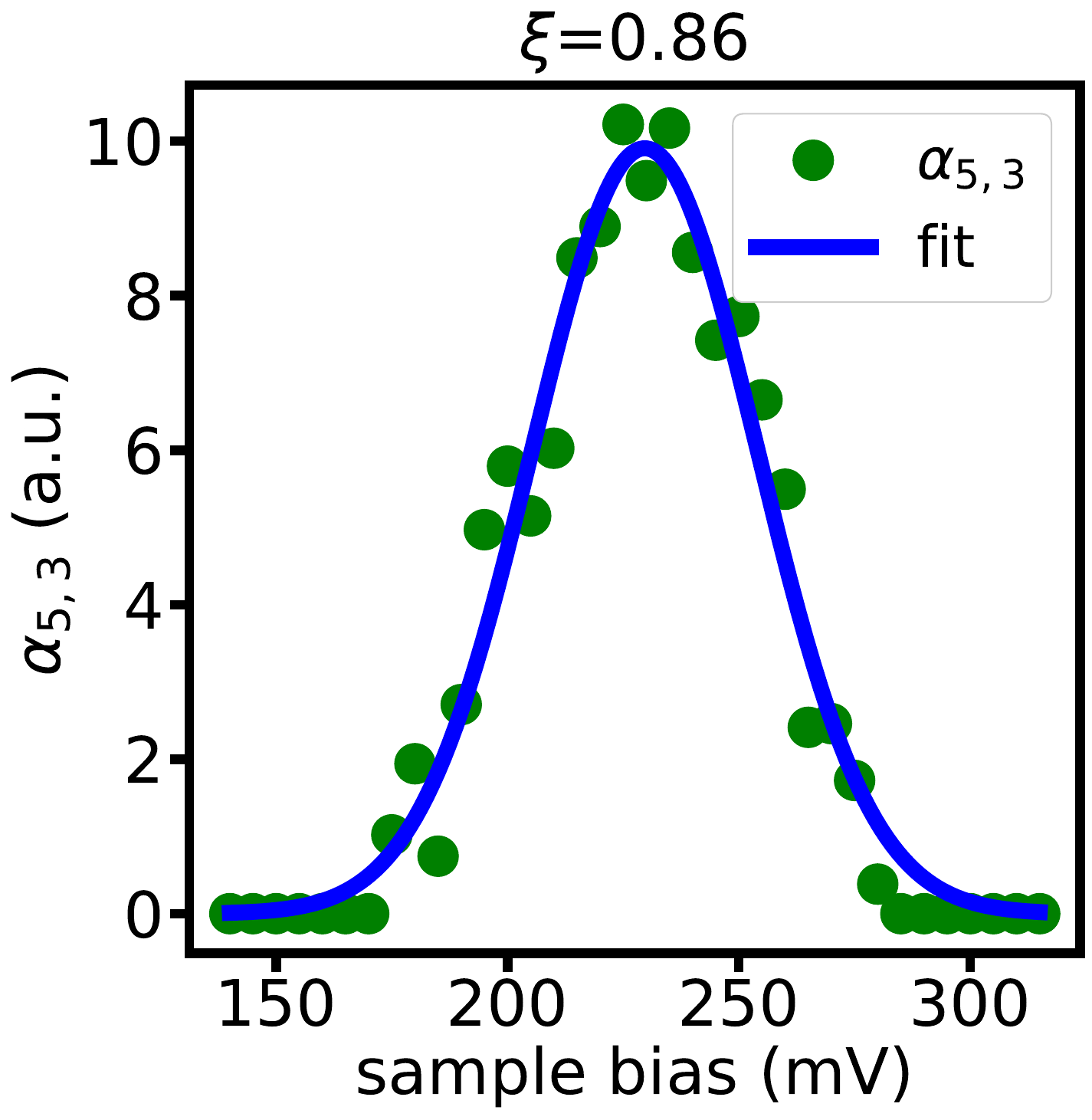}
		\end{minipage}
		\hfill
		\begin{minipage}[c]{0.325\linewidth}
			\includegraphics[height=5.6cm]{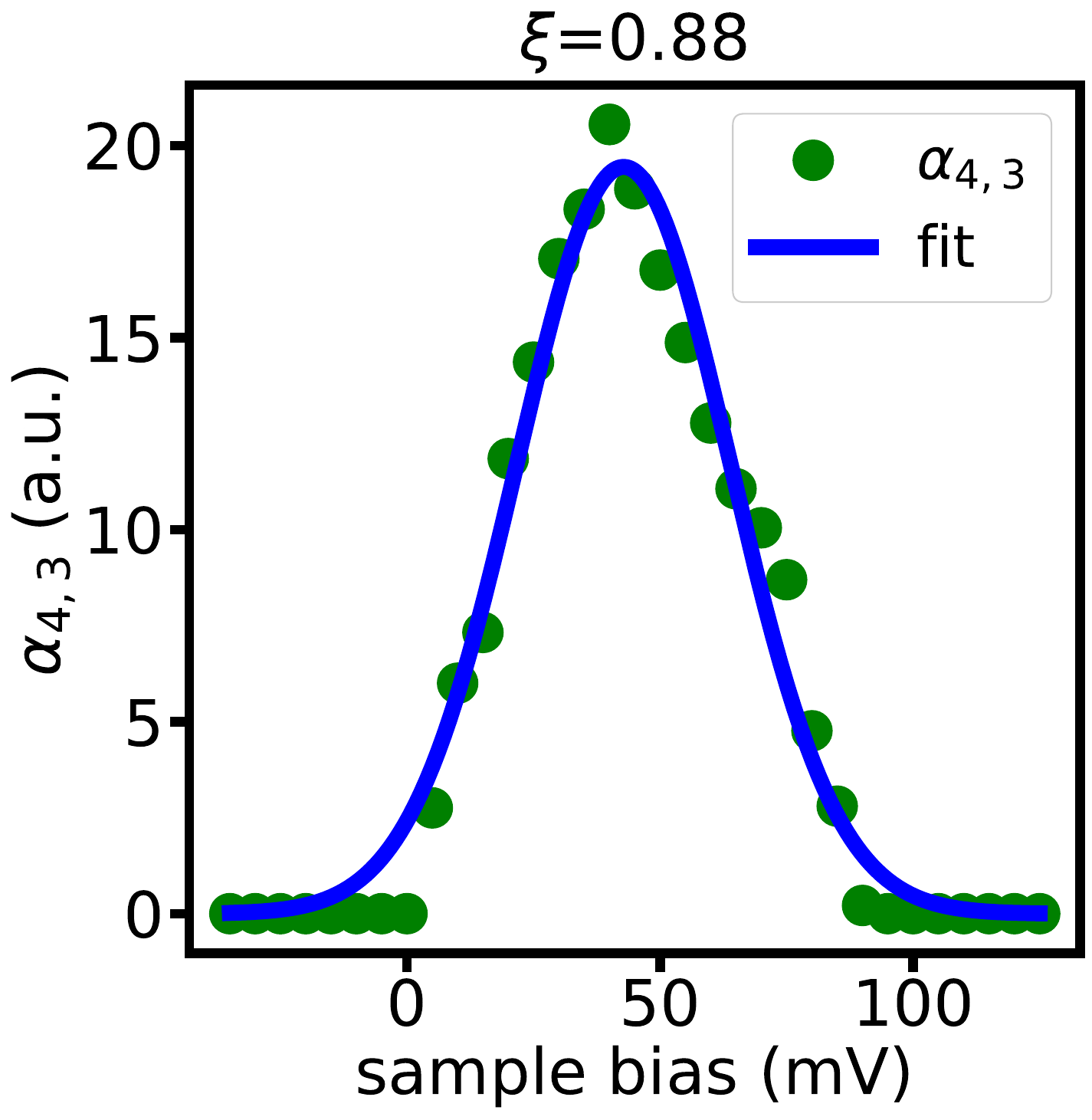}
		\end{minipage}
		\hfill
		\begin{minipage}[c]{0.325\linewidth}
			\includegraphics[height=5.6cm]{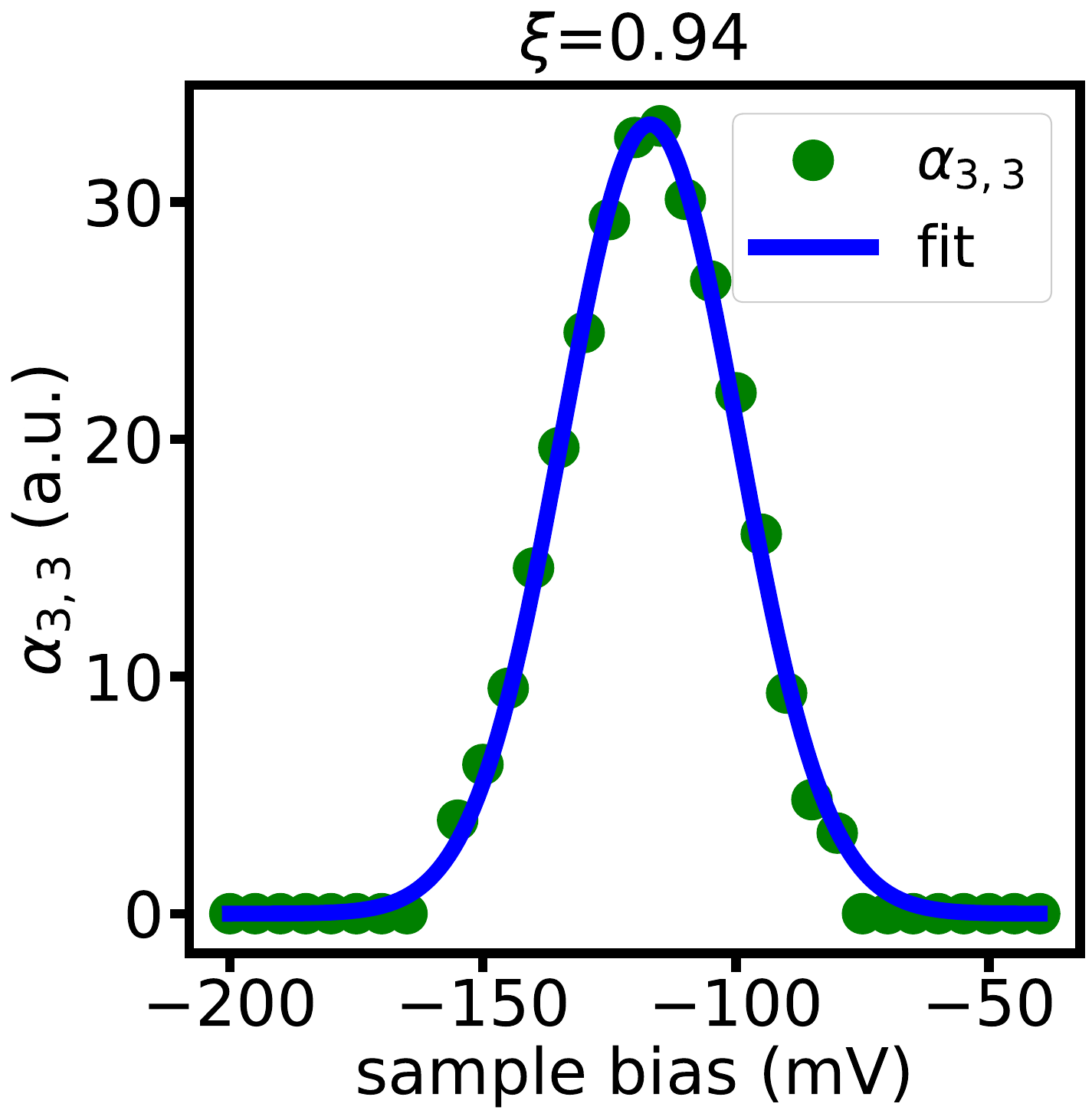}
		\end{minipage}
		\hfill
	\end{minipage}
	\caption{Reconstructed energy distributions for the $l = 3$ states (green dots). For the detailed line shape analysis equation (3) (main text) was fitted to the distributions. The fit is depicted with a full, blue line. The quality value $\xi$  can be found above each plot and was calculated as described in \ref{SM:roughness}.}
\end{figure}

\begin{table}[H]
	\begin{minipage}[c]{\linewidth}
		\centering
		\begin{minipage}[c]{0.325\linewidth}
			\begin{tabular}{|c|c|}\hline
				
				$n=5, l=3$ & fit (meV)  \\ \hline
				
				$\Gamma_\mathrm{wall}$ & $56 \pm 5$ \\
				
				$\Gamma_\mathrm{L}$ & $0 \varpm  1$ \\
				
				$E_\mathrm{5,3}$ &\textcolor{white}{$-$}$229.6 \pm 0.7$ \\ \hline
			\end{tabular}
		\end{minipage}
		\begin{minipage}[c]{0.325\linewidth}
			\begin{tabular}{|c|c|}\hline
				
				$n=4, l=3$ & fit (meV)  \\ \hline
				
				$\Gamma_\mathrm{wall}$ & $48 \pm 4$ \\
				
				$\Gamma_\mathrm{L}$ & $0 \varpm  5$ \\
				
				$E_\mathrm{4,3}$ & $42.8 \pm 0.6$ \\ \hline
			\end{tabular}
		\end{minipage}
		\begin{minipage}[c]{0.325\linewidth}
			\begin{tabular}{|c|c|}\hline
				
				$n=3, l=3$ & fit (meV)  \\ \hline
				
				$\Gamma_\mathrm{wall}$ & $40\pm 2$ \\
				
				$\Gamma_\mathrm{L}$ & $0 \varpm  2$ \\
				
				$E_\mathrm{3,3}$ & $-117.0  \pm 0.2$ \\ \hline
			\end{tabular}
		\end{minipage}
	\end{minipage}
	\caption{Results for fitting equation (3) of the main text to the reconstructed  $l=3$ energy distributions of the big corral.}
\end{table}

\newpage

\subsection{$l=4$ states - big corral}
\begin{figure}[H]
	\begin{minipage}[c]{\textwidth}
		\centering
		\begin{minipage}[c]{0.29\linewidth}
			\includegraphics[height=5.6cm]{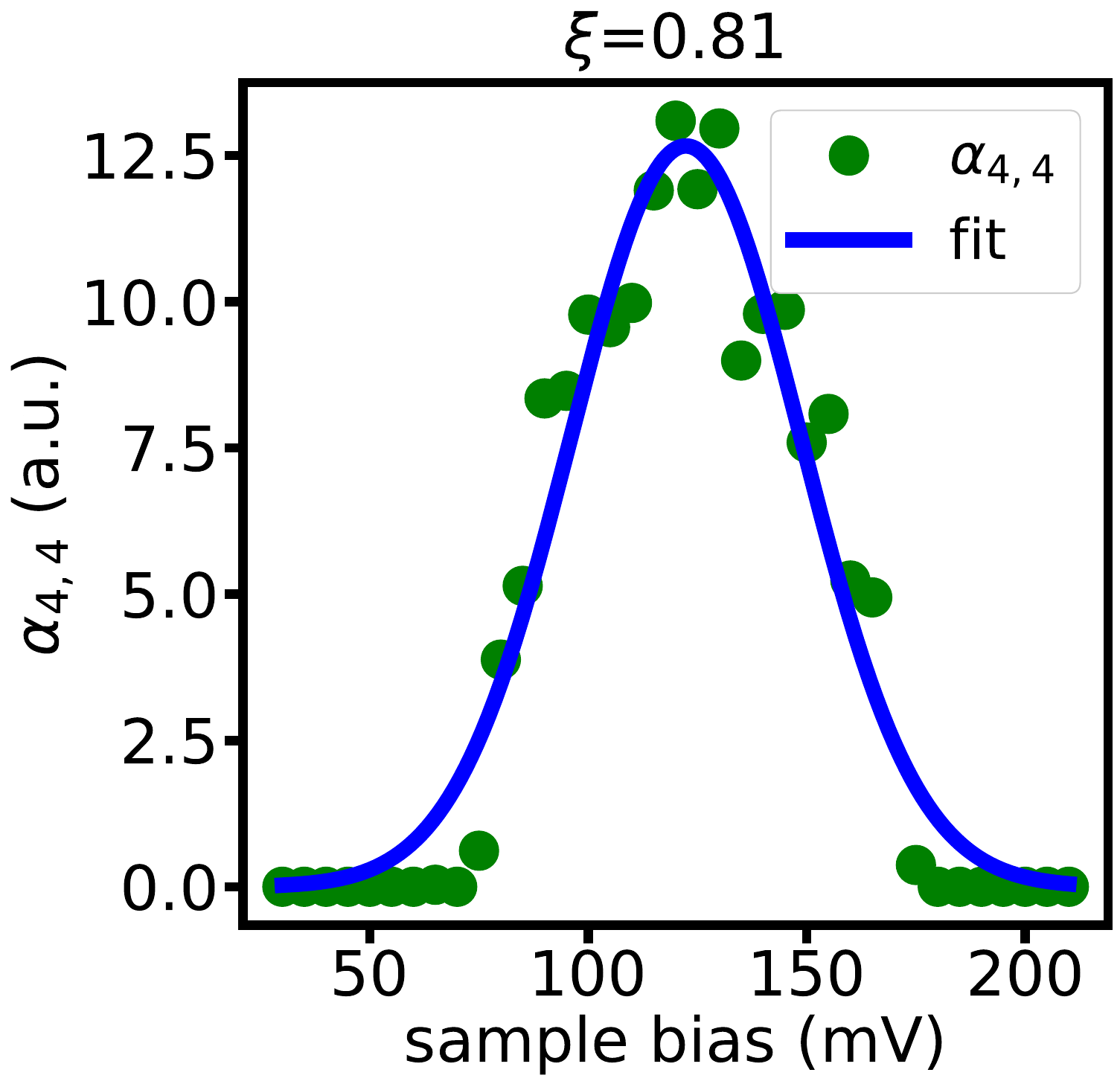}
		\end{minipage}
		\hfill
		\begin{minipage}[c]{0.29\linewidth}
			\includegraphics[height=5.6cm]{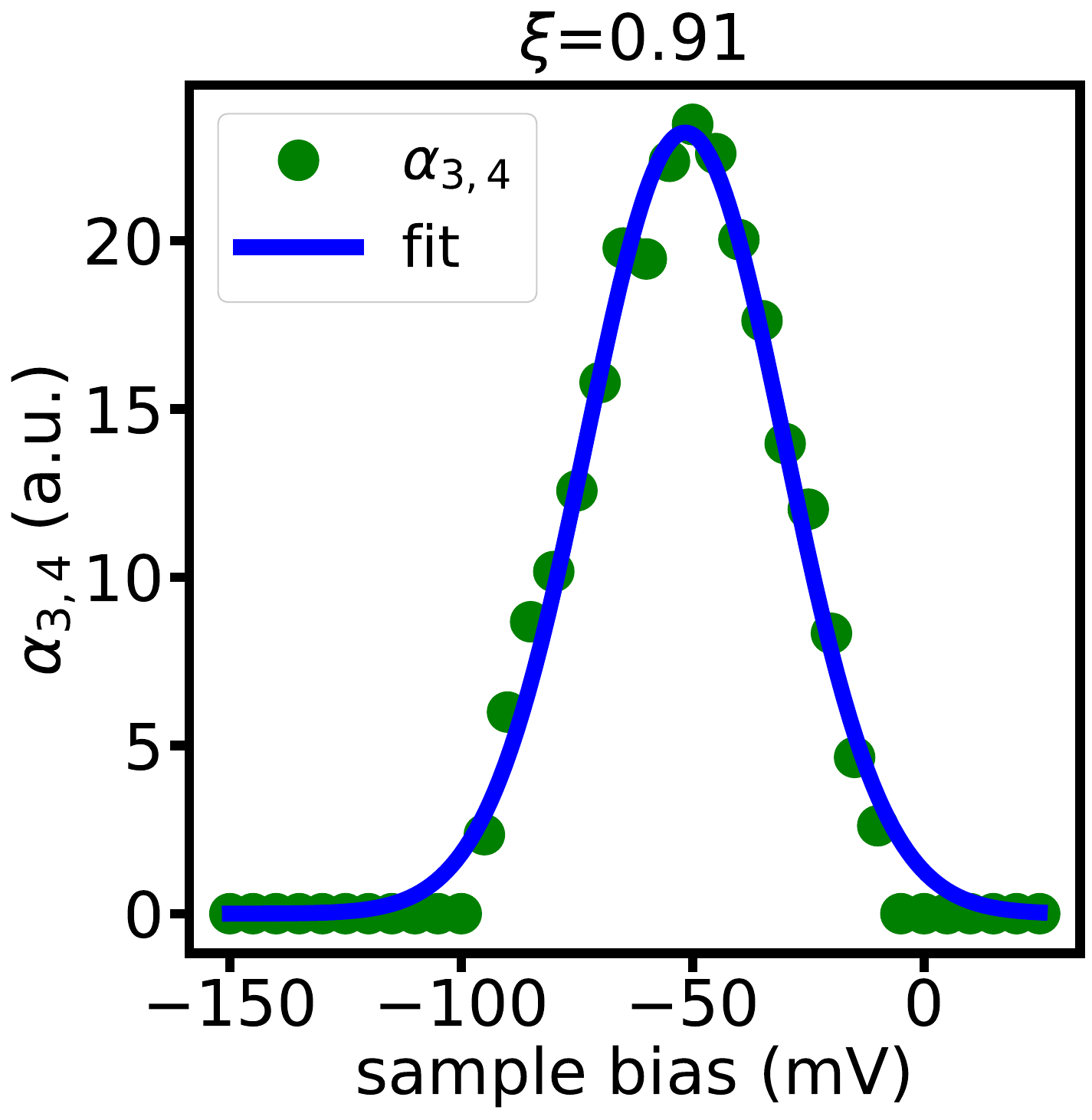}
		\end{minipage}
		\hfill
		\begin{minipage}[c]{0.29\linewidth}
			\includegraphics[height=5.6cm]{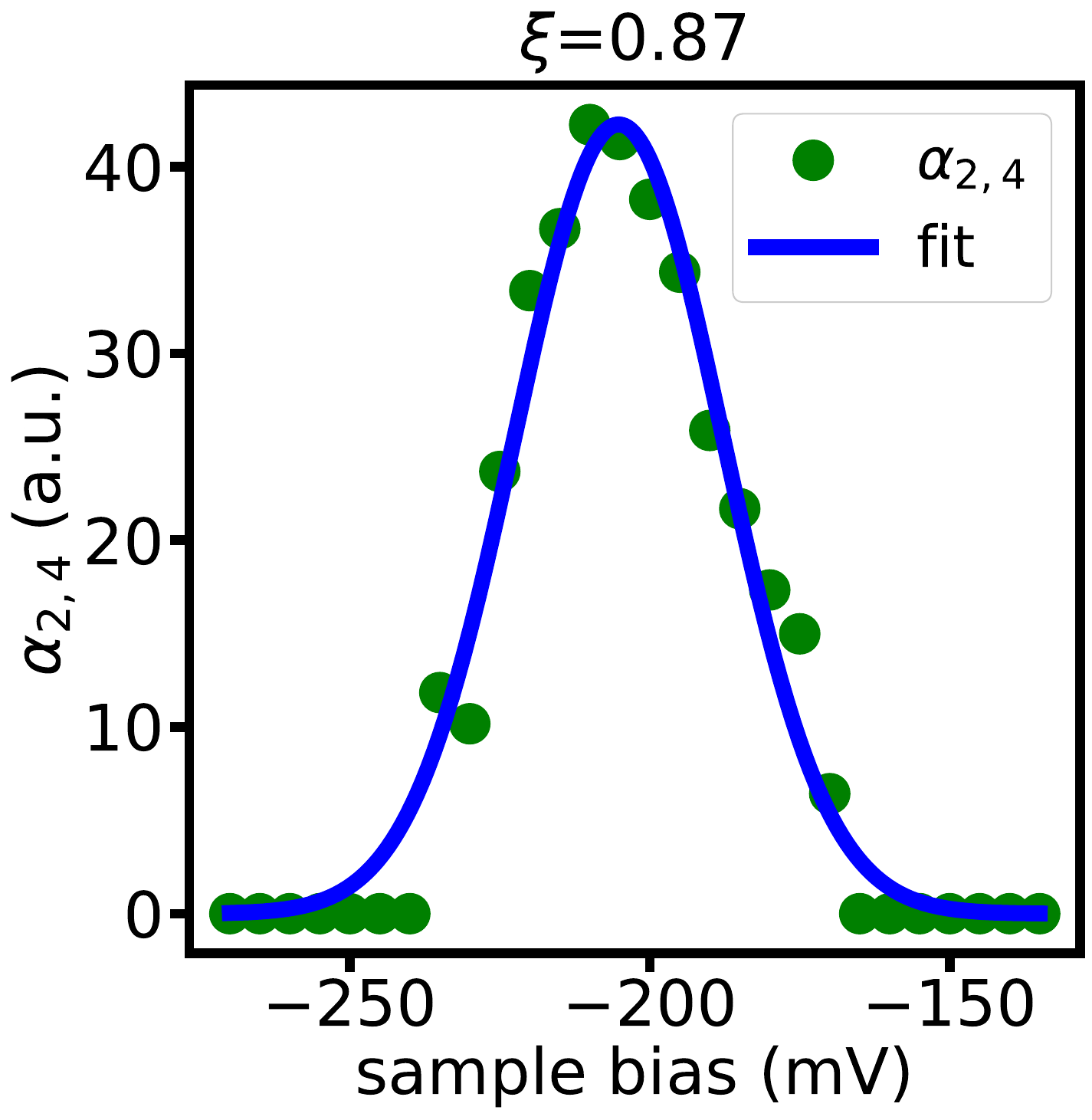}
		\end{minipage}
		\hfill
	\end{minipage}
	\caption{Reconstructed energy distributions for the $l = 4$ states (green dots). For the detailed line shape analysis equation (3) (main text) was fitted to the distributions. The fit is depicted with a full, blue line. The quality value $\xi$  can be found above each plot and was calculated as described in \ref{SM:roughness}.}
\end{figure}

\begin{table}[H]
	\begin{minipage}[c]{\linewidth}
		\centering
		\begin{minipage}[c]{0.325\linewidth}
			\begin{tabular}{|c|c|}\hline
				
				$n=4, l=4$ & fit (meV)  \\ \hline
				
				$\Gamma_\mathrm{wall}$ & $61 \pm 7$ \\
				
				$\Gamma_\mathrm{L}$ & $0 \varpm  10$ \\
				
				$E_\mathrm{4,4}$ &\textcolor{white}{$-$}$122 \pm 1$ \\ \hline
			\end{tabular}
		\end{minipage}
		\begin{minipage}[c]{0.325\linewidth}
			\begin{tabular}{|c|c|}\hline
				
				$n=3, l=4$ & fit (meV)  \\ \hline
				
				$\Gamma_\mathrm{wall}$ & $49 \pm 3$ \\
				
				$\Gamma_\mathrm{L}$ & $0 \varpm  4$ \\
				
				$E_\mathrm{3,4}$ & $-51.6 \pm 0.4$ \\ \hline
			\end{tabular}
		\end{minipage}
		\begin{minipage}[c]{0.325\linewidth}
			\begin{tabular}{|c|c|}\hline
				
				$n=2, l=4$ & fit (meV)  \\ \hline
				
				$\Gamma_\mathrm{wall}$ & $39 \pm 4$ \\
				
				$\Gamma_\mathrm{L}$ & $0 \varpm  1$ \\
				
				$E_\mathrm{2,4}$ & $-205.1  \pm 0.5$ \\ \hline
			\end{tabular}
		\end{minipage}
	\end{minipage}
	\caption{Results for fitting equation (3) of the main text to the reconstructed  $l=4$ energy distributions of the big corral.}
\end{table}

\newpage

\subsection{$l=5$, $l=6$ and $l=7$ states -  big corral}

\begin{figure}[H]
	\begin{minipage}[c]{\textwidth}
		\centering
		\begin{minipage}[c]{0.325\textwidth}
			\includegraphics[height=5.6cm]{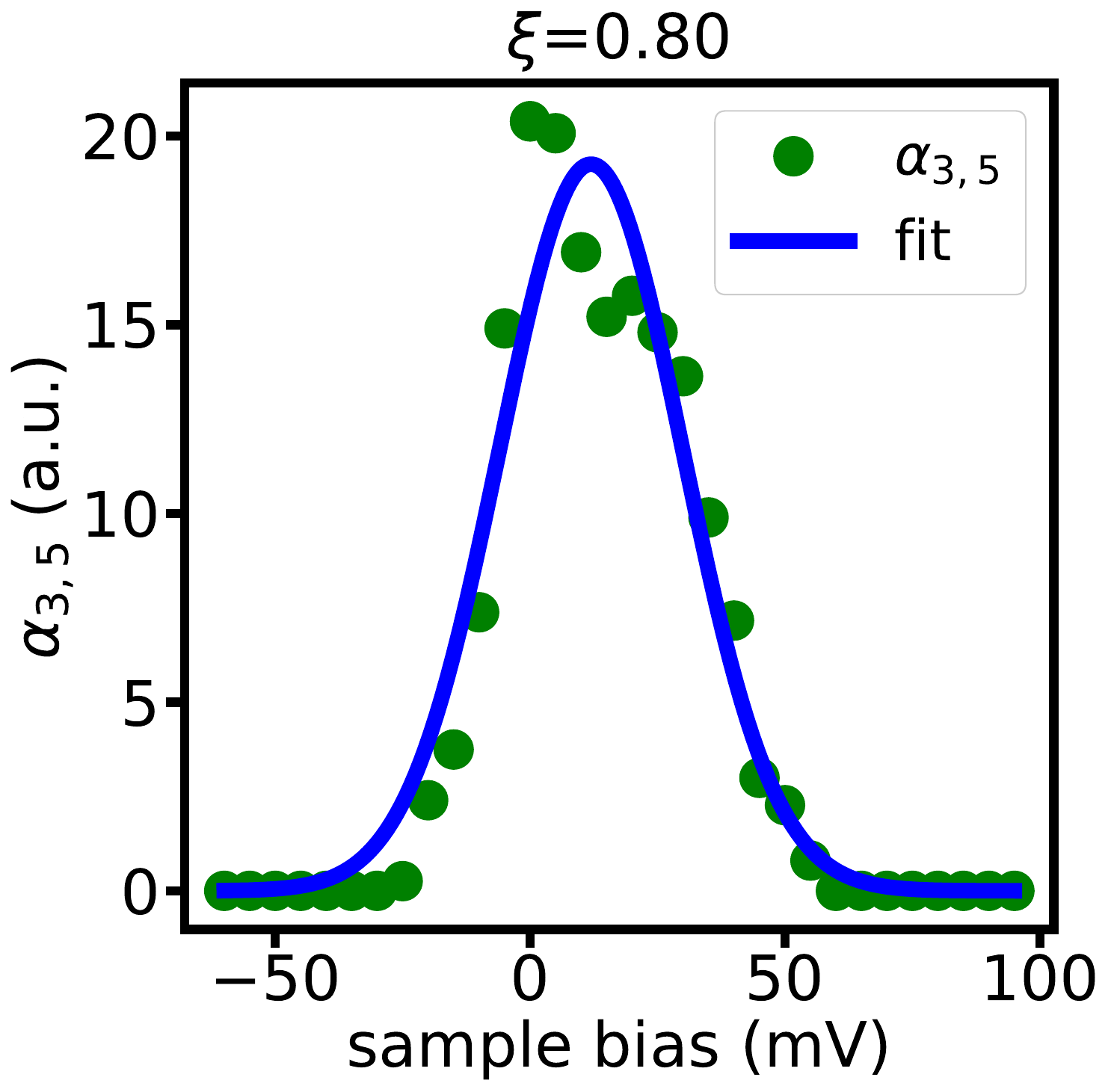}
		\end{minipage}
		\hfill
		\begin{minipage}[c]{0.325\textwidth}
			\includegraphics[height=5.6cm]{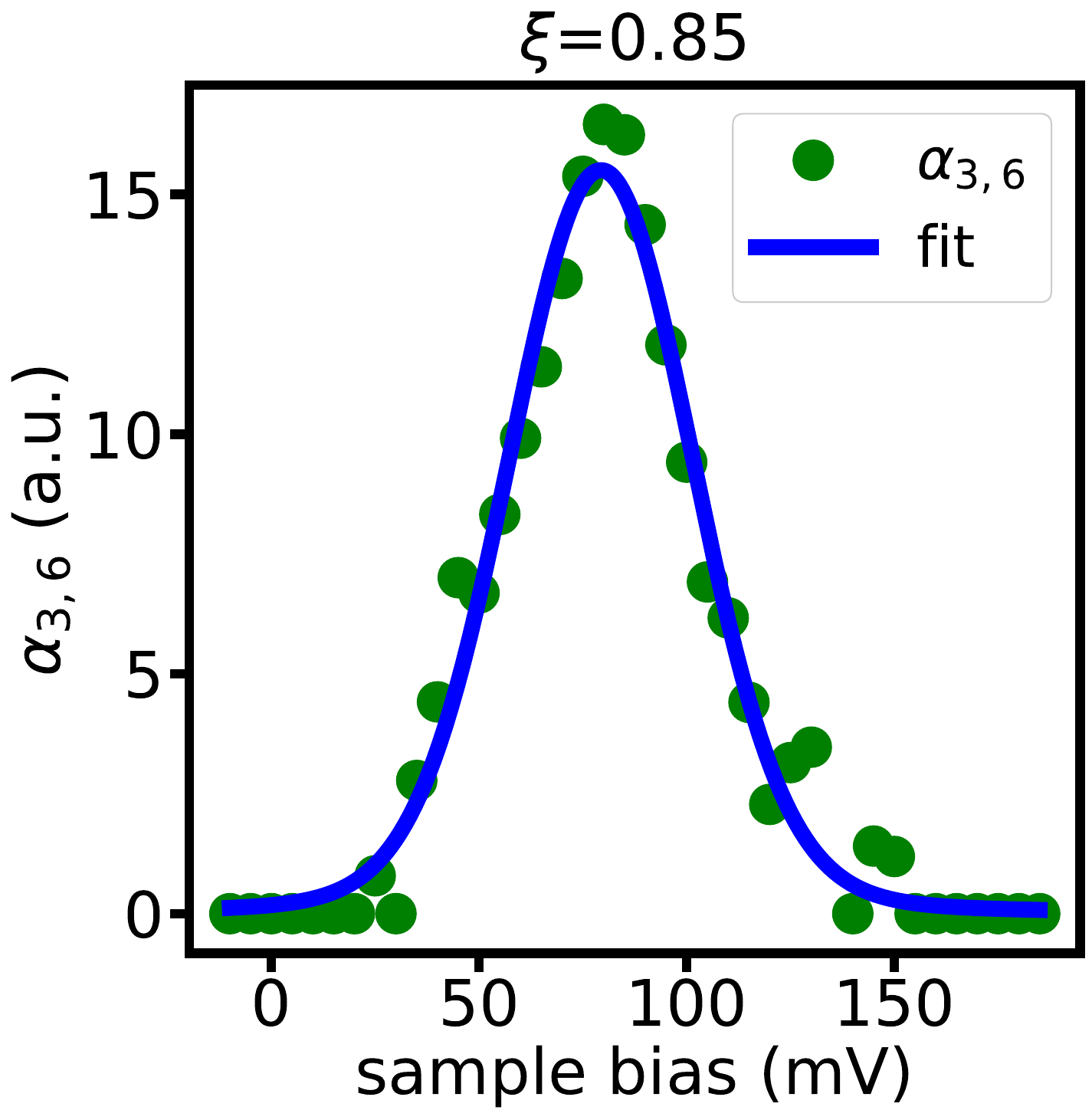}
		\end{minipage}
		\hfill
		\begin{minipage}[c]{0.325\textwidth}
			\includegraphics[height=5.6cm]{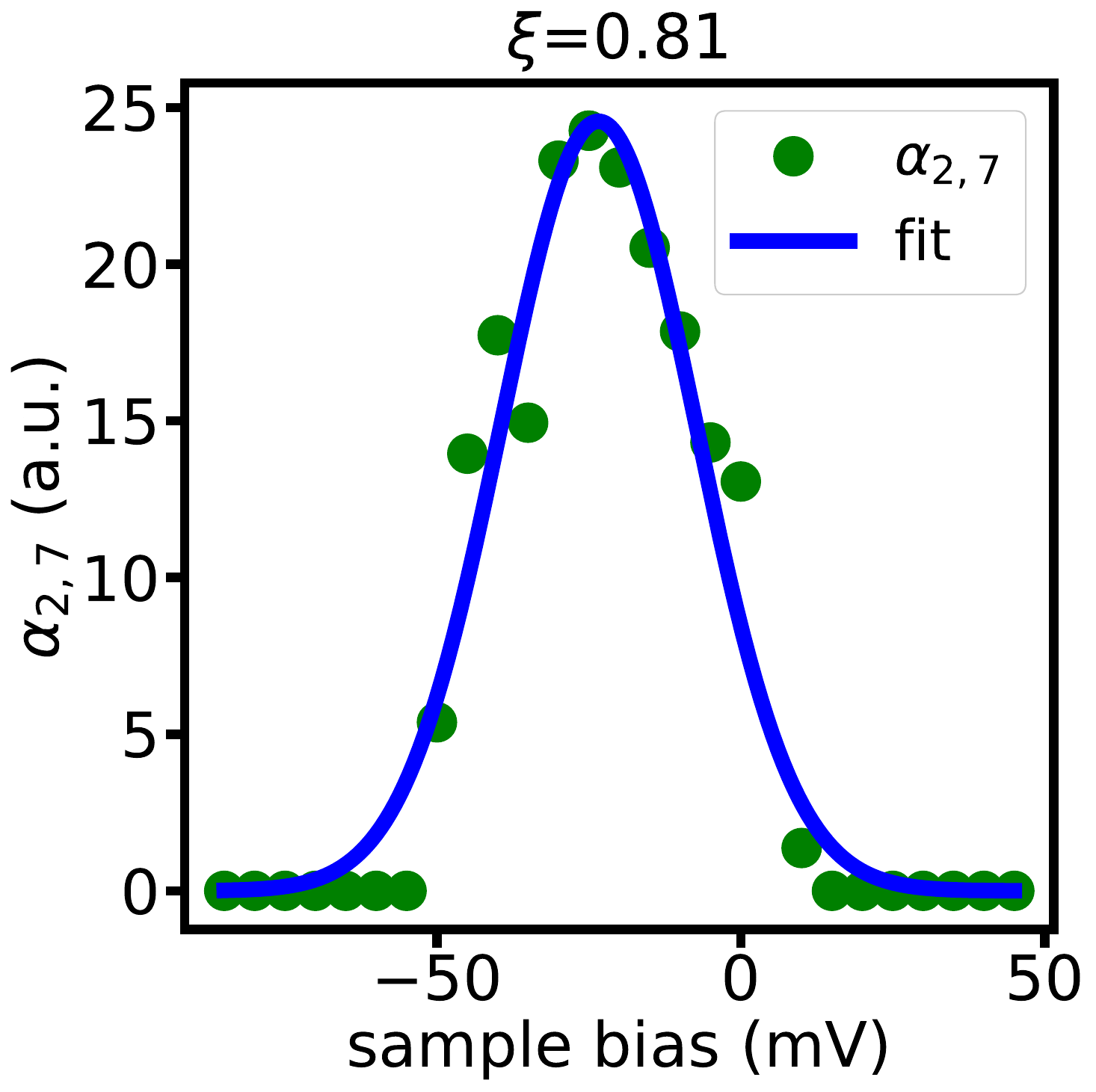}
		\end{minipage}
		\hfill
	\end{minipage}
	\caption{Reconstructed energy distributions for the $l = 5$, $l = 6$ and $l = 7$  states (green dots). For the detailed line shape analysis equation (3) (main text) was fitted to the distributions. The fit is depicted with a full, blue line. The quality value $\xi$  can be found above each plot and was calculated as described in \ref{SM:roughness}.}
\end{figure}

\begin{center}
	\begin{table}[H]
		\begin{minipage}[c]{\linewidth}
			\centering
			\begin{minipage}[c]{0.325\linewidth}
				\begin{tabular}{|c|c|}\hline
					
					$n=3, l=5$ & fit (meV)  \\ \hline
					
					$\Gamma_\mathrm{wall}$ &$41 \pm 5$ \\
					
					$\Gamma_\mathrm{L}$ & $0 \varpm  1$ \\
					
					$E_\mathrm{3,5}$ & $12.0 \pm 0.8$ \\ \hline
				\end{tabular}
			\end{minipage}
			\begin{minipage}[c]{0.325\linewidth}
				\begin{tabular}{|c|c|}\hline
					
					$n=3, l=6$ & fit (meV)  \\ \hline
					
					$\Gamma_\mathrm{wall}$ & $49 \pm 4$ \\
					
					$\Gamma_\mathrm{L}$ & $5 \pm 5$ \\
					
					$E_\mathrm{3,6}$ & $79.4 \pm 0.6$\textcolor{white}{$0$} \\ \hline
				\end{tabular}
			\end{minipage}
			\begin{minipage}[c]{0.325\linewidth}
				\begin{tabular}{|c|c|}\hline
					
					$n=2, l=7$ & fit (meV)  \\ \hline
					
					$\Gamma_\mathrm{wall}$ & $36\pm 5$ \\
					
					$\Gamma_\mathrm{L}$ & $0 \varpm 5$ \\
					
					$E_\mathrm{2,7}$ & $-23.4  \pm 0.7$ \\ \hline
				\end{tabular}
			\end{minipage}
		\end{minipage}
		\caption{Results for fitting equation (3) of the main text to the reconstructed  $l=5$, $l=6$ and $l=7$ energy distributions of the big corral.}
	\end{table}
\end{center}

\newpage

\subsection{$l=0$ states - small corral}
\begin{figure}[H]
	\begin{minipage}[c]{\textwidth}
		\centering
		\begin{minipage}[c]{0.29\linewidth}
			\includegraphics[height=5.6cm]{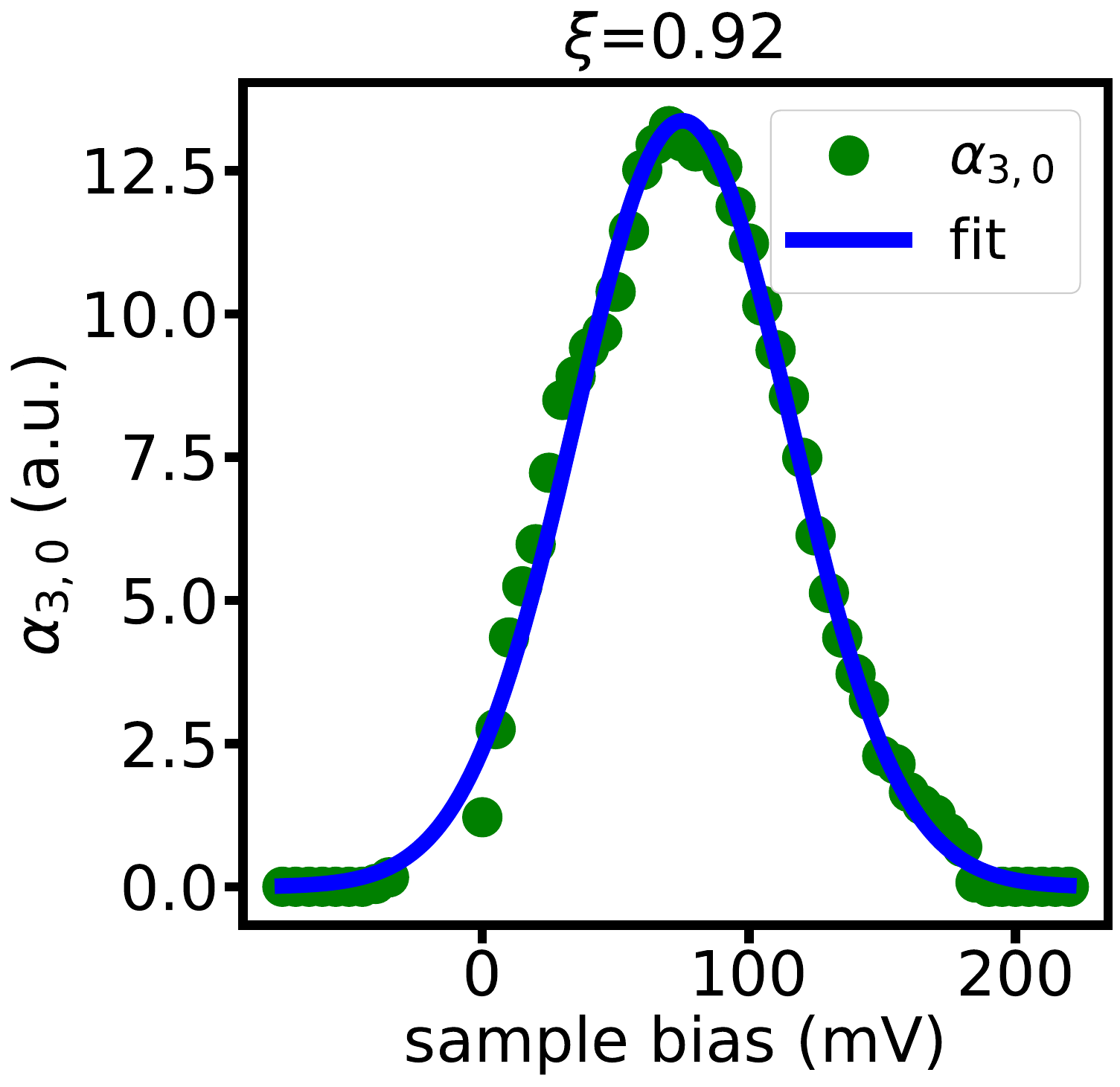}
		\end{minipage}
		\hfill
		\begin{minipage}[c]{0.29\linewidth}
			\includegraphics[height=5.6cm]{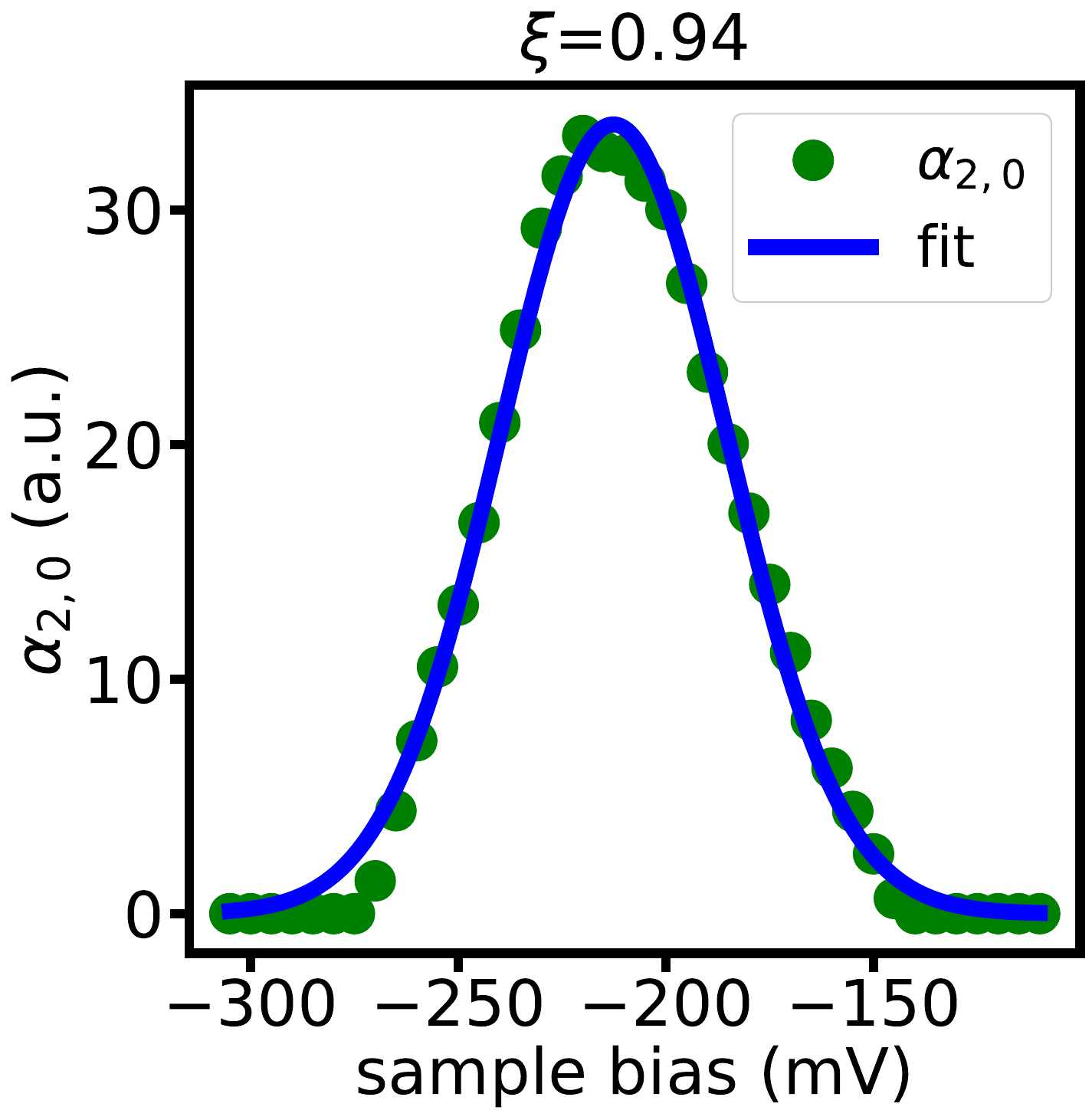}
		\end{minipage}
		\hfill
		\begin{minipage}[c]{0.29\linewidth}
			\includegraphics[height=5.6cm]{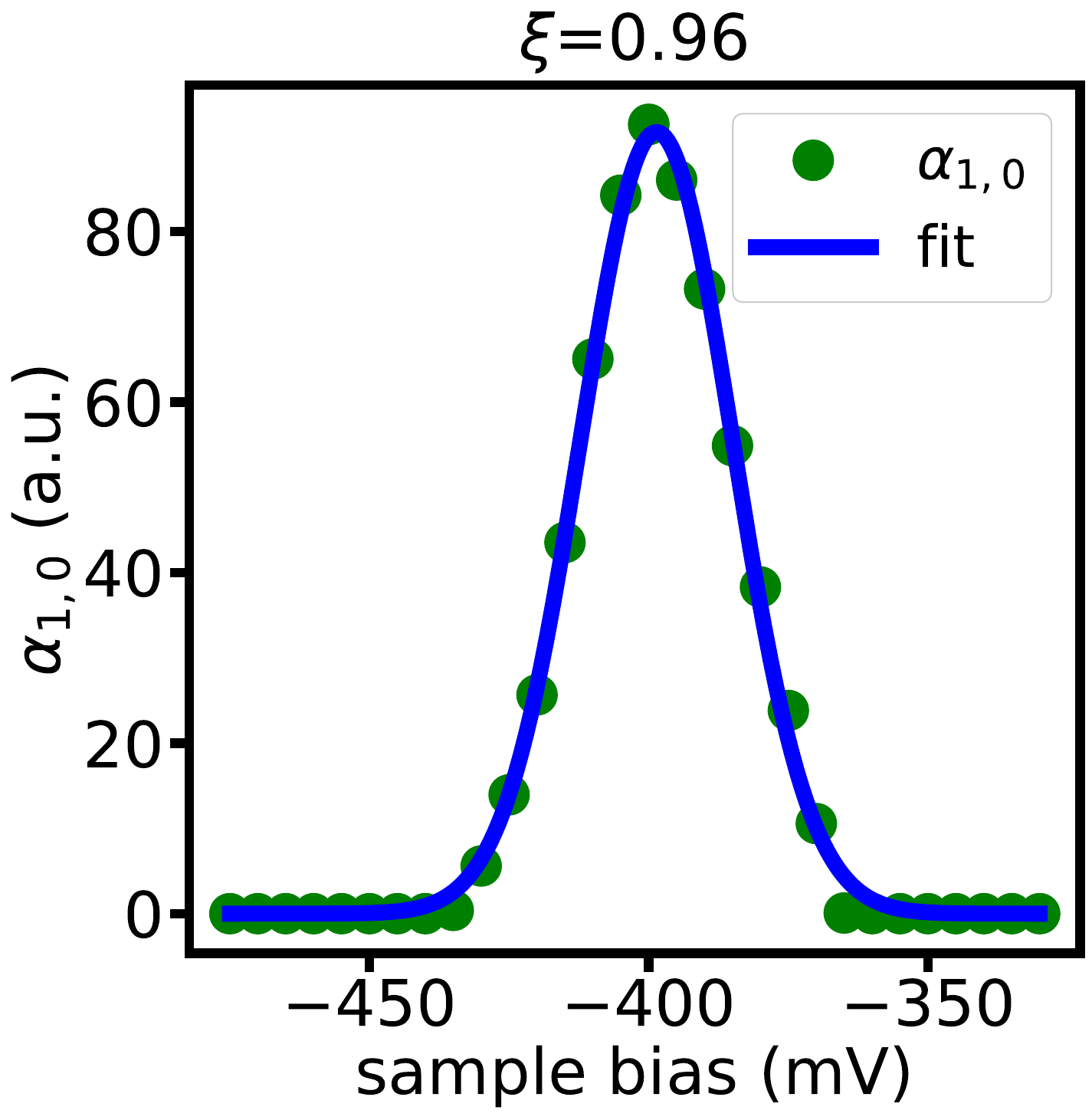}
		\end{minipage}
		\hfill
	\end{minipage}
	\caption{Reconstructed energy distributions for the $l = 0$ states (green dots). For the detailed line shape analysis equation (3) (main text) was fitted to the distributions. The fit is depicted with a full, blue line. The quality value $\xi$  can be found above each plot and was calculated as described in \ref{SM:roughness}.}
\end{figure}

\begin{center}
	\begin{table}[H]
		\begin{minipage}[c]{\linewidth}
			\centering
			\begin{minipage}[c]{0.325\linewidth}
				\begin{tabular}{|c|c|}\hline
					
					$n=3, l=0$ & fit (meV)  \\ \hline
					
					$\Gamma_\mathrm{wall}$ &$95 \pm 5$ \\
					
					$\Gamma_\mathrm{L}$ & $0 \varpm  7$ \\
					
					$E_\mathrm{3,0}$ &  \textcolor{white}{$0$} $75.1 \pm 0.8$ \textcolor{white}{$0$}\\ \hline
				\end{tabular}
			\end{minipage}
			\begin{minipage}[c]{0.325\linewidth}
				\begin{tabular}{|c|c|}\hline
					
					$n=2, l=0$ & fit (meV)  \\ \hline
					
					$\Gamma_\mathrm{wall}$ & $64 \pm 3$ \\
					
					$\Gamma_\mathrm{L}$ & $0 \varpm  1$ \\
					
					$E_\mathrm{2,0}$ & $-212.6  \pm 0.4$\\ \hline
				\end{tabular}
			\end{minipage}
			\begin{minipage}[c]{0.325\linewidth}
				\begin{tabular}{|c|c|}\hline
					
					$n=1, l=0$ & fit (meV)  \\ \hline
					
					$\Gamma_\mathrm{wall}$ & $30\pm 1$ \\
					
					$\Gamma_\mathrm{L}$ & $0 \varpm  1$ \\
					
					$E_\mathrm{1,0}$ & $-398.6  \pm 0.2$ \\ \hline
				\end{tabular}
			\end{minipage}
		\end{minipage}
		\caption{Results for fitting equation (3) of the main text to the reconstructed  $l=0$ energy distributions of the small corral.}
	\end{table}
\end{center}

\newpage
\subsection{$l=1$ states - small corral}
\begin{figure}[H]
	\begin{minipage}[c]{\textwidth}
		\centering
		\begin{minipage}[c]{0.325\linewidth}
			\includegraphics[height=5.6cm]{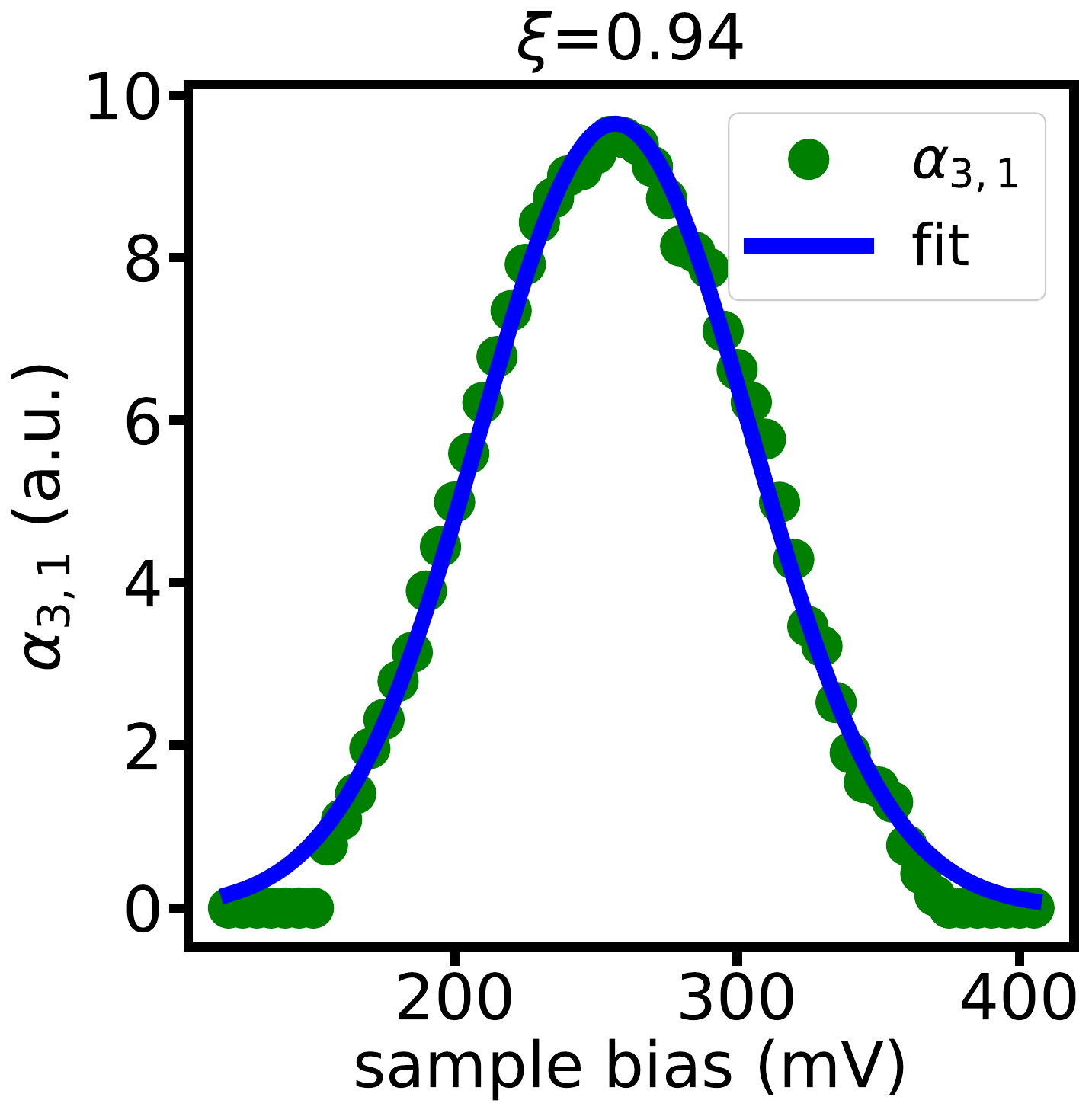}
		\end{minipage}
		\hfill
		\begin{minipage}[c]{0.325\linewidth}
			\includegraphics[height=5.6cm]{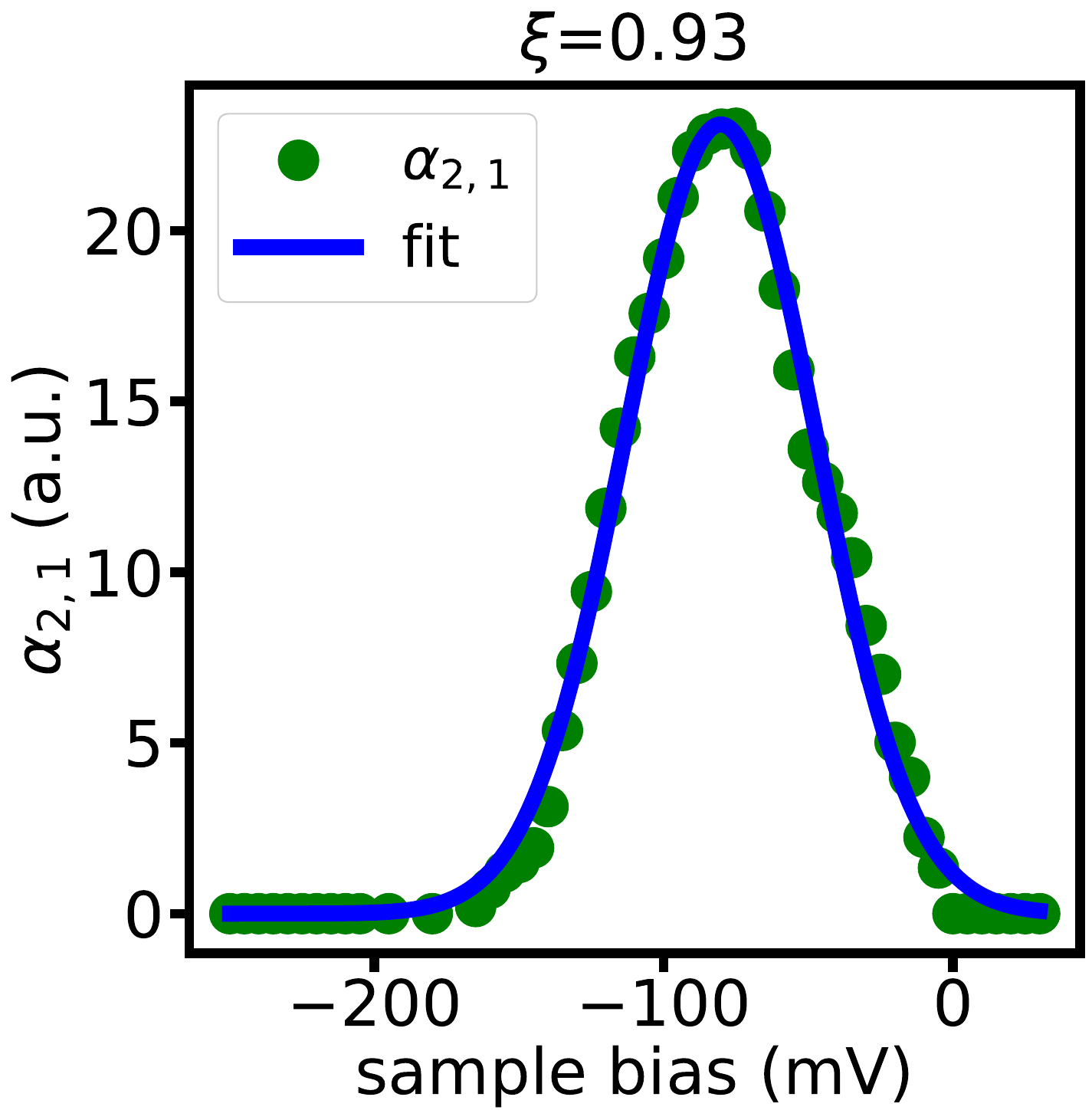}
		\end{minipage}
		\hfill
		\begin{minipage}[c]{0.325\linewidth}
			\includegraphics[height=5.6cm]{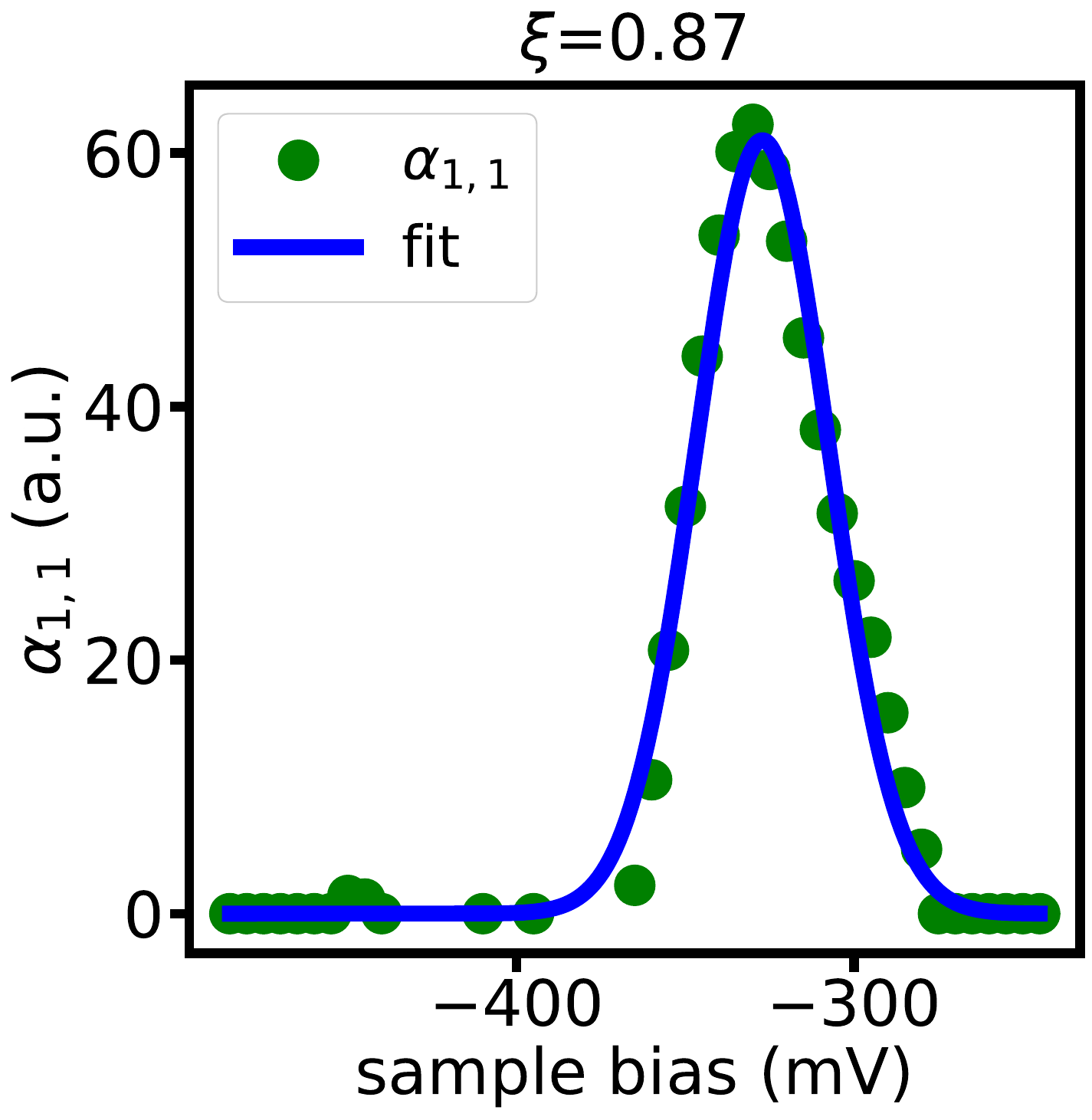}
		\end{minipage}
		\hfill
	\end{minipage}
	\caption{Reconstructed energy distributions for the $l = 1$ states (green dots). For the detailed line shape analysis equation (3) (main text) was fitted to the distributions. The fit is depicted with a full, blue line. The quality value $\xi$  can be found above each plot and was calculated as described in \ref{SM:roughness}.}
\end{figure}

\begin{center}
	\begin{table}[H]
		\begin{minipage}[c]{\linewidth}
			\centering
			\begin{minipage}[c]{0.325\linewidth}
				\begin{tabular}{|c|c|}\hline
					
					$n=3, l=1$ & fit (meV)  \\ \hline
					
					$\Gamma_\mathrm{wall}$ &$113 \pm 5$ \\
					
					$\Gamma_\mathrm{L}$ & $0 \varpm  7$ \\
					
					$E_\mathrm{3,1}$ & $256.8 \pm 0.6$ \\ \hline
				\end{tabular}
			\end{minipage}
			\begin{minipage}[c]{0.325\linewidth}
				\begin{tabular}{|c|c|}\hline
					
					$n=2, l=1$ & fit (meV)  \\ \hline
					
					$\Gamma_\mathrm{wall}$ & $77 \pm 3$ \\
					
					$\Gamma_\mathrm{L}$ & $0 \varpm  5$\\
					
					$E_\mathrm{2,1}$ & $-80.3  \pm 0.5$\\ \hline
				\end{tabular}
			\end{minipage}
			\begin{minipage}[c]{0.325\linewidth}
				\begin{tabular}{|c|c|}\hline
					
					$n=1, l=1$ & fit (meV)  \\ \hline
					
					$\Gamma_\mathrm{wall}$ & $45\pm 4$ \\
					
					$\Gamma_\mathrm{L}$ & $0 \varpm  1$ \\
					
					$E_\mathrm{1,1}$ & $-327.2   \pm 0.5$ \\ \hline
				\end{tabular}
			\end{minipage}
		\end{minipage}
		\caption{Results for fitting equation (3) of the main text to the reconstructed  $l=1$ energy distributions of the small corral.}
	\end{table}
\end{center}

\newpage

\subsection{$l=2$ and $l=3$ states - small corral}
\begin{figure}[H]
	\begin{minipage}[c]{\textwidth}
		\centering
		\begin{minipage}[c]{0.325\textwidth}
			\includegraphics[height=5.6cm]{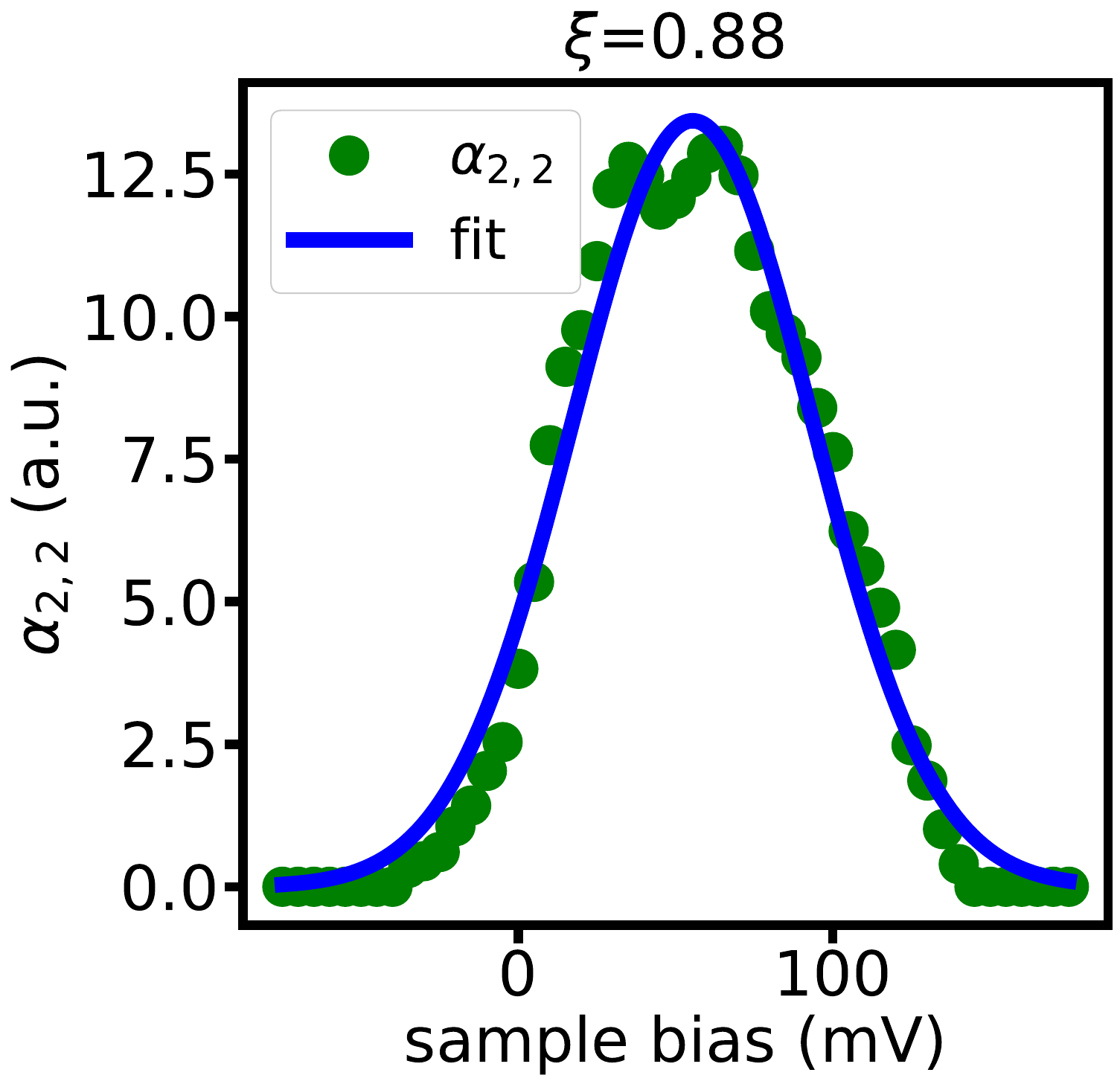}
		\end{minipage}
		\hspace{1cm}
		\begin{minipage}[c]{0.325\textwidth}
			\includegraphics[height=5.6cm]{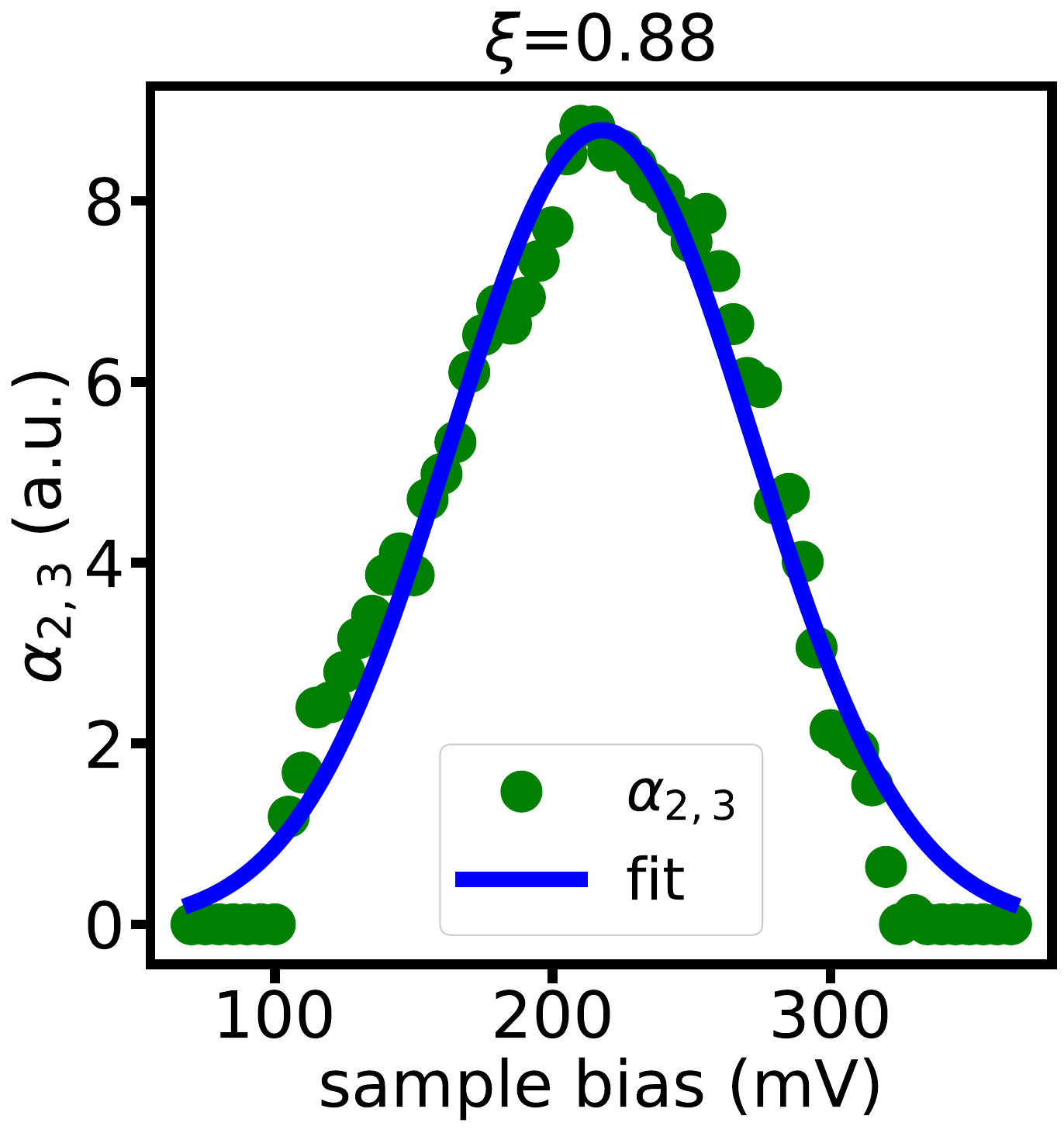}
		\end{minipage}
	\end{minipage}
	\caption{Reconstructed energy distributions for the $l=2$ and $l=3$  states (green dots). For the detailed line shape analysis equation (3) (main text) was fitted to the distributions. The fit is depicted with a full, blue line. The quality value $\xi$  can be found above each plot and was calculated as described in \ref{SM:roughness}.}
\end{figure}

\begin{center}
	\begin{table}[H]
		\begin{minipage}[c]{\linewidth}
			\centering
			\begin{minipage}[c]{0.325\linewidth}
				\begin{tabular}{|c|c|}\hline
					
					$n=2, l=2$ & fit (meV)  \\ \hline
					
					$\Gamma_\mathrm{wall}$ &$49 \pm 6$ \\
					
					$\Gamma_\mathrm{L}$ & $0 \varpm  1$ \\
					
					$E_\mathrm{2,2}$ & $55.4  \pm 0.9$ \\ \hline
				\end{tabular}
			\end{minipage}
			\begin{minipage}[c]{0.325\linewidth}
				\begin{tabular}{|c|c|}\hline
					
					$n=2, l=3$ & fit (meV)  \\ \hline
					
					$\Gamma_\mathrm{wall}$ & $128 \pm 10$ \\
					
					$\Gamma_\mathrm{L}$ & $0 \varpm  14$ \\
					
					$E_\mathrm{2,3}$ & $218   \pm 2$\\ \hline
				\end{tabular}
			\end{minipage}
		\end{minipage}
		\caption{Results for fitting equation (3) of the main text to the reconstructed  $l=2$ and $l=3$ energy distributions of the small corral.}
	\end{table}
\end{center}

\newpage

\section{Non-exponential decay of survival probability in quantum corrals}
In the main text, it was noted that up to now literature commonly employs Lorentzian curves to fit spectral peaks of quantum corrals. This implies an exponential decrease in the survival probability of electrons within the corral over time. An exponential decay in the time domain translates to a Lorentzian peak in the energy domain via Fourier transformation (time domain $\rightarrow$ FFT $\rightarrow$ energy domain). Such a decay assumes a uniform probability of decay for electrons at any given moment. However, our findings reveal that the primary determinant of electron lifetime in a quantum corral is predominantly their interaction with the wall, suggesting that electrons primarily "exit" the corral through the wall. 

\begin{figure}[H]
	\begin{minipage}[c]{\textwidth}
		\centering
		\begin{minipage}[c]{0.32\linewidth}
			\includegraphics[width=\linewidth]{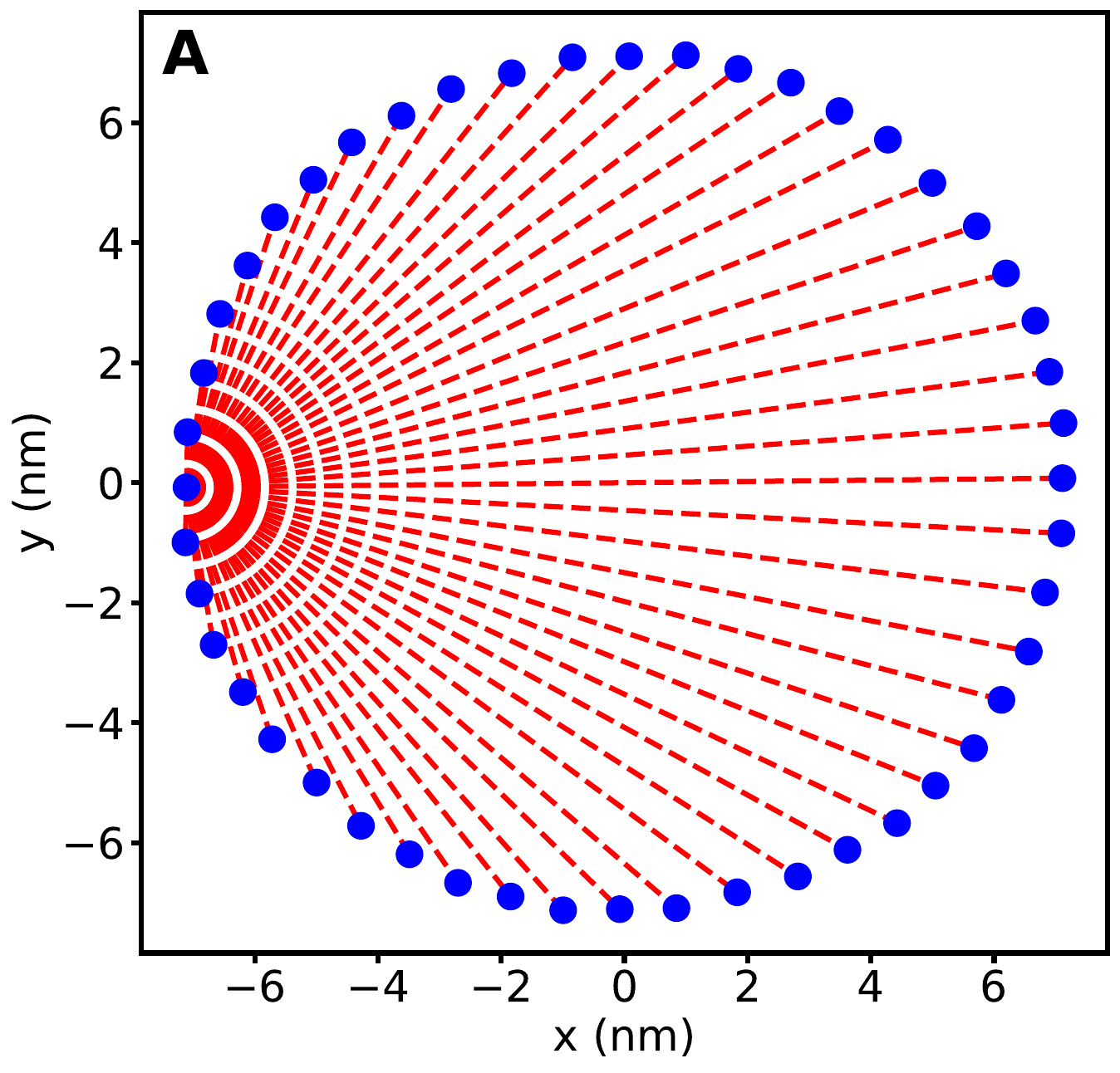}
		\end{minipage}
		\hfill
		\begin{minipage}[c]{0.32\linewidth}
			\includegraphics[width=\linewidth]{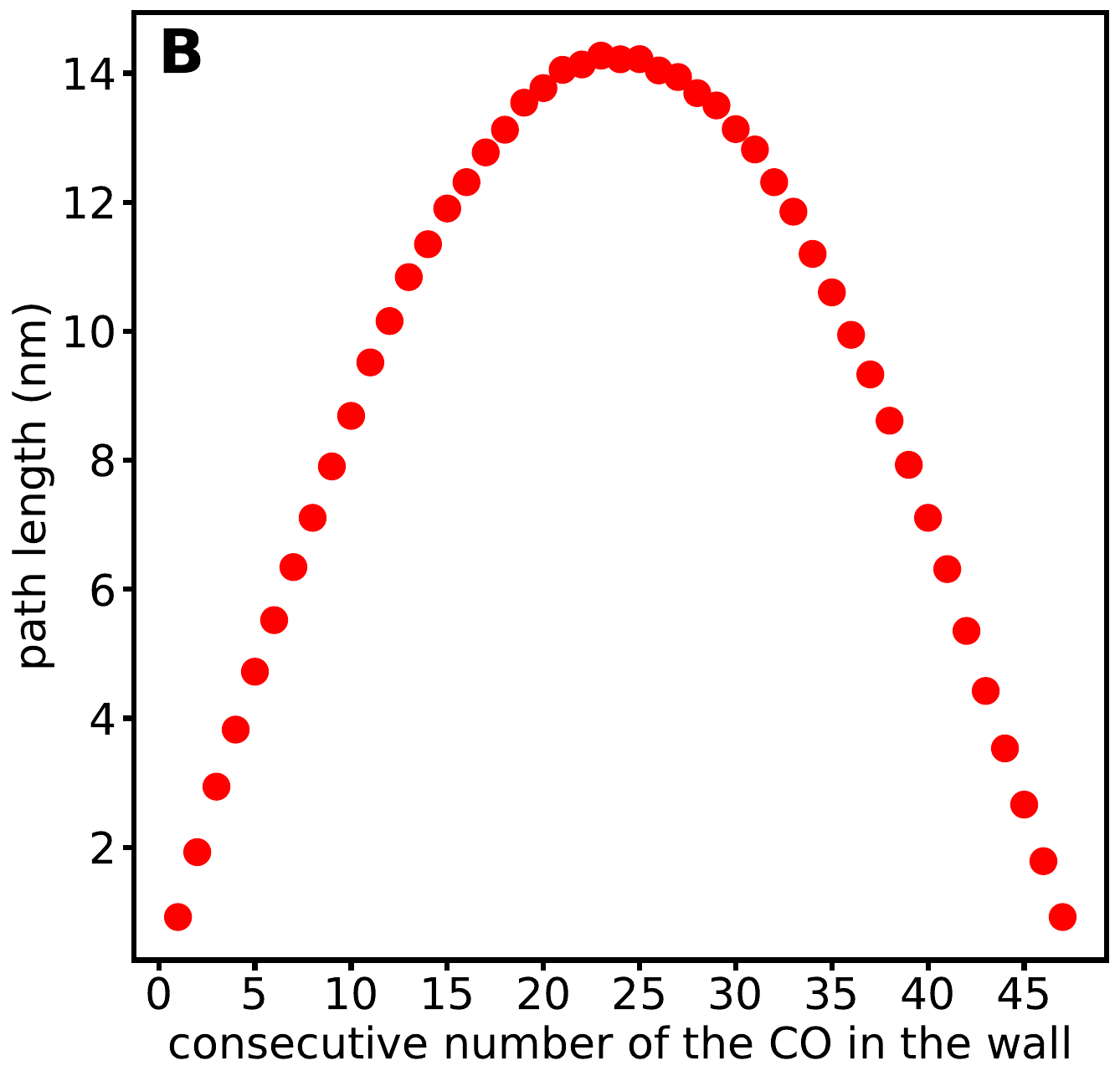}
		\end{minipage}
		\hfill
		\begin{minipage}[c]{0.32\linewidth}
			\includegraphics[width=\linewidth]{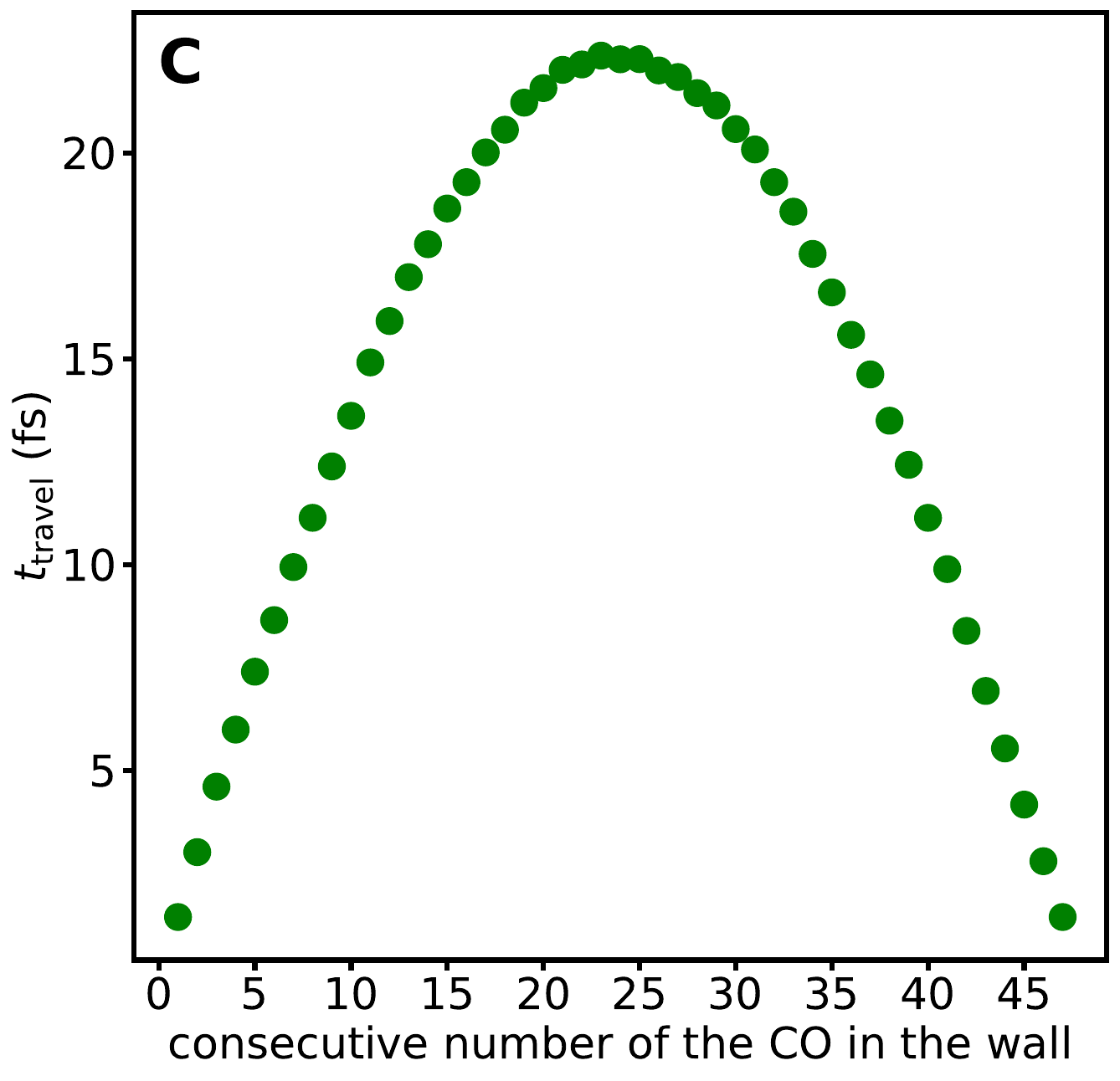}
		\end{minipage}
		
		\vspace{0,5cm}
		
		\hfill
		\begin{minipage}[c]{0.32\linewidth}
			\includegraphics[width=\linewidth]{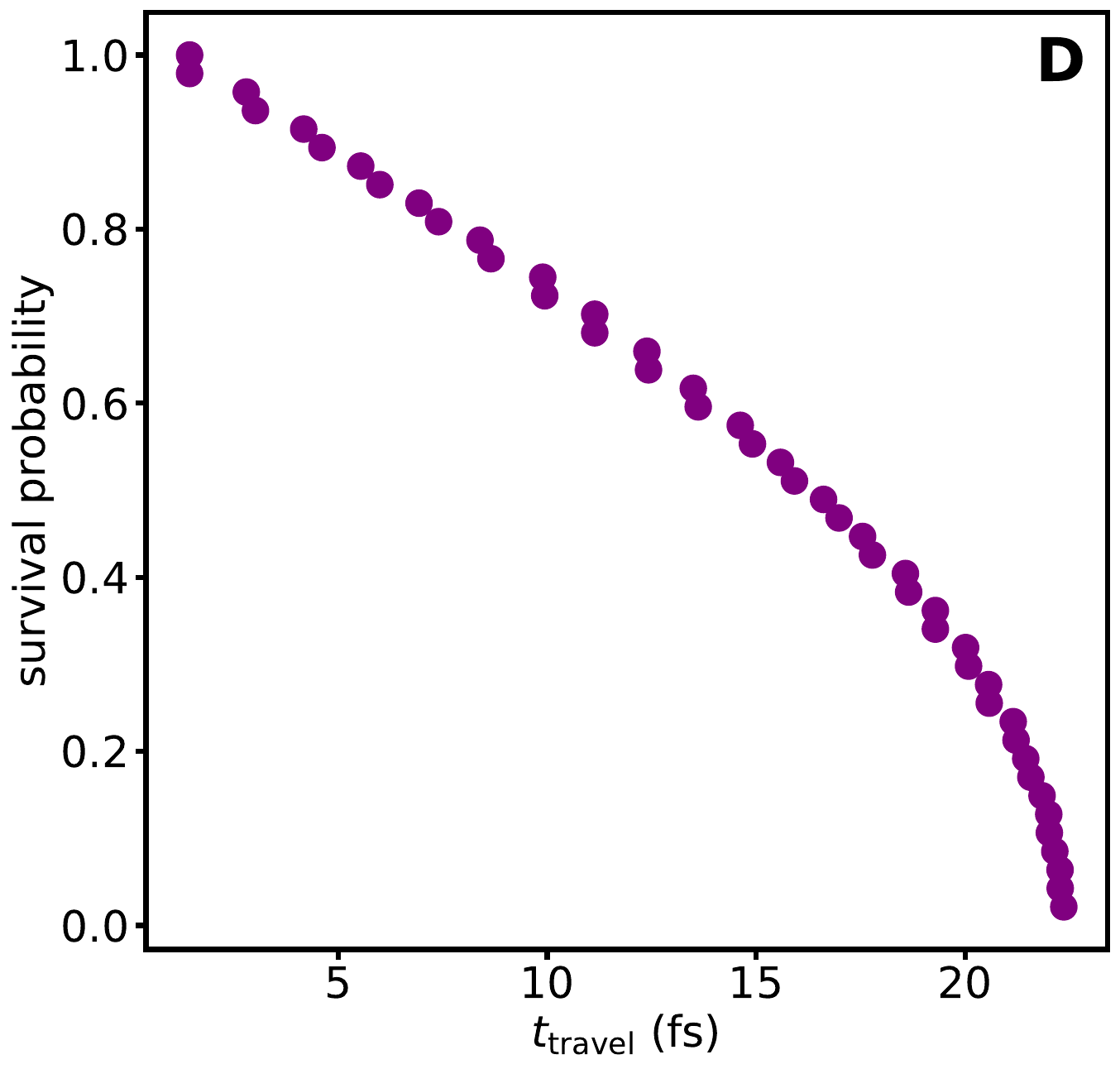}
		\end{minipage}
		\hfill
		\vspace{0,5cm}
	\end{minipage}
	\caption{\label{fig:FFT}
		\textbf{A:} The blue markers represent the positions of the CO molecules in a large 48-corral. The red dashed lines indicate the possible paths an electron, starting at the corral wall, can take. 
		\textbf{B:} This plot illustrates the path lengths from panel A. 
		\textbf{C:} Depicts the traveling times $t_\mathrm{travel}$ of an electron at the Fermi energy, corresponding to the path lengths given in panel B. 
		\textbf{D:} Shows the survival probability curve translated from panel C, considering only one scattering event. Due to the non-exponential shape in the time domain, a non-Lorentzian peak shape will result in the energy domain.}
\end{figure}

In our explanation, we adopt a single-particle approach, assuming an uninterrupted travel time for electrons through the corral. For clarity, we consider only one interaction event at the corral wall (without multiple scattering events). Our measurements, conducted with a maximum current of $\approx 500\,\mathrm{pA}$, correspond to tunneling one electron every $\approx 320\;$ps into or out of the corral. Despite this, lifetimes of electrons in quantum corrals of $\approx30\;$fs were measured, indicating that the measured electron lifetimes (proportional to inverse width of peaks) are negligibly influenced by the tip position and the tunneling electrons. Moreover, our analysis from the main text revealed that electron lifetimes in the corrals are predominantly governed by interactions of the electrons with the wall. Consequently, we use the corral wall as the reference point for the subsequent calculations. This provides the potential paths and path lengths for an electron within the corral for one interaction event, as depicted in Figure \ref{fig:FFT}A\&B. Figure \ref{fig:FFT}C then illustrates the distribution of traveling times for an electron at the Fermi energy.

Given that each electron path (see red dashed lines in Figure \ref{fig:FFT}A) carries the same probability, all must be considered. The resulting survival probability decay for one interaction event, after which an electron leaves the corral, is shown in Figure \ref{fig:FFT}D. This curve does clearly not exhibit an exponential decrease resulting in a non-Lorentzian peak in the energy domain after Fourier transformation.

Although the inclusion of multiple scattering events in this approach could produce peaks resembling a Gaussian curve in the energy domain, it falls outside the scope of this study and may be explored in future theoretical considerations.

\newpage

\newpage

\section{Spectral broadening due to breathing mode induced corral radius variation}

The energy positions of corral states can be approximated by the eigenenergies obtained from the hard wall model:

\begin{equation}
E_{n,l} = \dfrac{\hbar^2 \rho_{n,l}^2}{2 m^*} \cdot \dfrac{1}{r_\mathrm{C}^2}
\end{equation}
where $\rho_{n,l}$ represents the $n^\mathrm{th}$ zero point of a Bessel function of the first kind and order $l$. The effective mass of Shockley surface state electrons is denoted by $m^*$, and the corral radius is represented as $r_\mathrm{C}$. The energy axis starts at the onset of the band edge, which is $440\;$meV below the Fermi level.

According to this equation, a variation of the corral radius by $\mp\chi$ shifts the eigenenergies of the states up and down. The eigenenergy difference $\Delta E_\mathrm{breathing}$ at the extreme points $r_\mathrm{C}\mp\chi$ is then, generalized for all energies $E$ above the band edge, given by:

\begin{equation}\label{eq:E_breath}
\Delta{E}_{breathing} = E \cdot r_\mathrm{C}^2\cdot\left (\dfrac{1}{(r_\mathrm{C}-\chi)^2} -  \dfrac{1}{(r_\mathrm{C}+\chi)^2} \right).
\end{equation}

To account for the experimentally determined widths via the effects of the breathing mode on corral states, radial amplitudes of $200\;$pm would be necessary for both the large and small corral. This estimate is based on the measured peak widths around the Fermi level, which are $\approx 50\;$meV for the large corral and $\approx 100\;$meV for the small corral (see FIG. 3C in the main text). The calculated radial variation exceeds the radius of a copper atom ($\approx 127\;$pm).

An approximation of the uncertainty in the position of a single CO results in a Gaussian FWHM of about $33\;$pm \cite{Okabayashi2018}. This uncertainty can be transferred to the uncertainty of the corral radius (explained in the main text). Inserting the HWHM (half width at half maximum) into equation (\ref{eq:E_breath}) yields $\Delta{E}_{breathing} = 4.1\;$meV for states at the Fermi energy ($440\;$meV above the band edge) and $\Delta{E}_{breathing} = 7.7\;$meV for states $400\;$meV above the Fermi energy ($840\;$meV above the band edge). For this calculation the radius of the large corral was used ($7.13\;$nm).

\end{document}